\documentclass[usegraphicx,useAMS,usenatbib]{mn2e}
\graphicspath{{data/}{../images}{../../images/}{../../../images/}}

\usepackage{times}

\newcommand{\mum}{\,$\umu$m}

\newcommand{\nblaz}{143}
\newcommand{\nmeas}{10493}

\begin{document}

\title[Submillimetre observations of blazars]{Observations of flat-spectrum
  radio sources at $\lambda$ 850\mum\ from the James Clerk Maxwell Telescope II. April 2000 to June 2005}


\author[T.\ Jenness et al.]{ T.\ Jenness,$^1$ E.\ I.\ Robson$^2$ and J.\ A.\ Stevens$^3$ \\
$^1$ Joint Astronomy Centre, 660 N. A`oh\={o}k\={u} Place, University Park,
Hilo, Hawaii, 96720, USA \\
$^2$ UK Astronomy Technology Centre, Blackford Hill, Edinburgh, EH9 3HJ \\
$^3$ Centre for Astrophysics Research, University of Hertfordshire, College Lane, Hatfield, Hertfordshire, AL10 9AB}
\date{Accepted 2009 September 14.  Received 2009 September 9; in original form 2009 August 5}

\maketitle

\begin{abstract}
Calibrated data for \nblaz\ flat-spectrum extragalactic radio
sources are presented at a wavelength of 850\mum\ covering a
five-year period from April 2000. The data, obtained at the James
Clerk Maxwell Telescope using the SCUBA camera in pointing mode,
were analysed using an automated pipeline process based on the
Observatory Reduction and Acquisition Control - Data Reduction
(ORAC-DR) system. This paper describes the techniques used to
analyse and calibrate the data, and presents the database of results
along with a representative sample of the better-sampled
lightcurves. A re-analysis of previously published data from 1997 to
2000 is also presented.  The combined catalogue, comprising \nmeas\
flux density measurements, provides a unique and valuable resource
for studies of extragalactic radio sources.
\end{abstract}


\section{Introduction}

In Paper 1 \citep*{2001MNRAS.327..751R} we described a series of
monitoring observations of flat-spectrum radio-sources at a wavelength
of 850\mum\ using the Submillimetre Common-User Bolometer Array,
SCUBA \citep{1999MNRAS.303..659H}, on the 15-m diameter James Clerk
Maxwell Telescope on Mauna Kea, Hawaii. The data reported in this
paper extend this work and comprise \nmeas\ flux density measurements
from pointing observations of \nblaz\ flat-spectrum extragalactic
radio sources taken over the lifetime of SCUBA between 1997 April 5
and 2005 June 4. The observations have been reduced using the
automatic SCUBA data reduction pipeline \cite{2002MNRAS.336...14J}
encompassing on-line atmospheric extinction correction using
quasi-continuous monitoring data from the Caltech Submillimetre
Observatory (CSO) \citep{2002MNRAS.336....1A}.

The radio through submillimetre emission from this class of source is
synchrotron emission from a relativistic jet emanating from the region
of the supermassive black hole.  These objects represent some of the
most variable sources of emission we know, and understanding the
emission processes requires multifrequency, multi-epoch observing.
Indeed, one of the biggest limiting factors to progress is the lack of
such data; therefore, this dataset will be a valuable database for
such studies.  For example, data presented in Paper 1 have been
incorporated into multifrequency variability studies of individual
objects such as 3C~273 \citep{2008A&A...486..411S}, 3C~279 \citep{2006A&A...456..895L}
and NRAO~530 \citep{2006A&A...456...97F} but have also been used for calibration
purposes \citep[e.g.][]{2004MNRAS.354.1239S} and in statistical studies of
large samples of objects \citep[e.g.][]{2004A&A...421..129S}.

As discussed in Paper I, the observations presented in this paper were
not part of a systematic blazar monitoring programme but were present
in the JCMT pointing catalogue either because they were very bright,
close to interesting parts of the sky or they are located where there
are very few submillimetre point-sources suitable for pointing
calibration. This means that the light curves are sometimes
well-sampled and sometimes very sparsely sampled. In some cases only a
few observations were taken, usually because the target was found to
be too weak. Those are presented for completeness. A comprehensive
discussion of the observation technique using jiggle mapping is
detailed in Paper I and in \cite{1999MNRAS.303..659H}.  As an archive
resource, all data presented in paper 1 have been re-analyzed and are
presented alongside the new data for completeness and uniformity of
approach.

\begin{figure*}
\includegraphics[angle=-90,width=\textwidth]{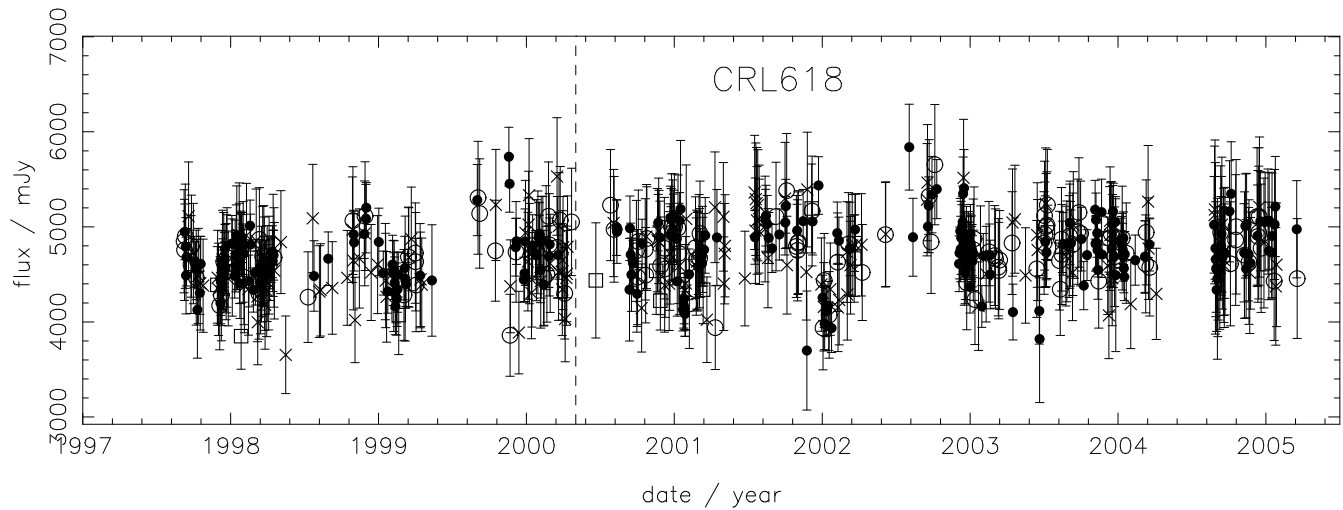}
\caption{Light-curve of CRL\,618 obtained using the same techniques as
applied to the blazar data. The calibration is stable to better than 7
per cent.}
\label{fig:fcf}
\end{figure*}

\section{Observation selection criteria}
\label{sect:obssel}

Paper I discusses the specific issues associated with processing
blazar pointing data while the paper by \cite{2002MNRAS.336...14J}
discusses the general case of the pipeline process of data extraction
and reduction in much greater detail. Only the new techniques are
discussed in this paper.

The main selection criterion is that all data are from pointing
observations of blazars present in the JCMT pointing catalogue but
other criteria must be included in order to generate an observation
list suitable for automated processing.

The presence of the polarimeter in the beam \citep{2003MNRAS.340..353G}
during a pointing observation affects the calibration factor by
approximately a factor of 2 due to attenuation by the waveplate, and
additionally some blazars are themselves highly polarized at
submillimetre wavelengths \citep{1996ApJ...462L..23S,1998MNRAS.297..667N,2007AJ....134..799J}
to the extent that it becomes
difficult to accurately correct the data for this effect. All
observations with the polarimeter in the beam have thus been
discarded. In data taken since 2002 August 8 the data headers indicate
whether the polarimeter is connected and these data are easily
discarded. Between 1999 July 6 and 2002 August 8, the presence of the
polarimeter must be inferred by using other metadata.  In cases where
it is inferred that the polarimeter has been used all pointing data
from the night are discarded. This does leave open the rare
possibility that the polarimeter is in the beam for a setup pointing
and then removed prior to doing any polarimeter observations and this
case can only be tested by examining the resulting flux density
data. For observations taken during the polarimeter commissioning
period (1998 May 13 to 1999 July 6) the presence of the polarimeter is
harder to infer and can depend on observation log entries made by the
observer.  This approach is fairly inaccurate and results in manual
removal of observations for this period.

For nights on which the Moon was observed
\citep[see e.g.,][]{2000JCMTNews15NJ}, pointing observations are ignored because
the SCUBA sensitivity must be adjusted in order to observe this bright
source. That adjustment is not present in the data headers and it is
easiest for the archive search to remove them. This only accounts for
18 nights of data during the period covered by this paper.
  
Although the data reduction attempts to be insensitive to small focus
changes, large focus shifts can still be problematic especially given
that a pointing observation is always done prior to a focus (and also
prior to the first focus of the evening). The archive extraction
routine disgards all pointings that are taken before the first focus
of the night, all pointings that are taken more than 1.5 hours after
the last focus of the night, and all pointings that are followed by a
focus that changes by more than 0.2 mm unless they are closer to the
previous focus than they are to the next.

Some nights have known problems with the secondary mirror (such as a
failure of one axis) or problems with the dish shape resulting in very
large variations in beam quality and calibration factors. These nights
can not be detected automatically but are removed using a look-up
table. Less than 10 nights were affected by these problems.  Finally,
all observations with a zenith sky opacity at 225 GHz greater than
0.30 are discarded \citep[see e.g.,][]{2002MNRAS.336....1A}.

\section{Calibration}

Calibration is based on long-term observations of Uranus and, to a
lesser extent Mars, thereby producing a time-dependent flux
calibration factor (FCF) for SCUBA. These have been very stable over
the period, changes due to upgrades of the instrument are clearly seen
and reflected in the changing FCF. The values determined for Paper I
were checked before re-processing the data and it was determined that
the accuracy for the old narrow-band filter could be improved by
calculating the FCF over shorter periods of a few months at a time
rather than taking yearly averages. These calibration changes mean
that for some periods the newly calculated flux densities can differ
by up to 10 per cent from the results previously published in Paper
I. In most cases the calibration difference is no more than 5 per
cent.  The calibration accuracy is shown in Fig.\ \ref{fig:fcf}, where
data for the best secondary calibrator, CRL\,618, have been processed
using the same recipes used to process the pointing data. The
light-curve is flat with a Gaussian of $4730\pm 330$\,mJy fitting the
distribution, corresponding to a calibration accuracy of $\pm7$ per
cent over the 8 year period, agreeing with the accuracy demonstrated
in \cite{2002MNRAS.336...14J}.  Data uncertainties were calculated as
described in Paper I.

\section{Post pipeline processing}
\label{sect:post}

The output data from the pipeline for any source are first averaged
over an individual night; there has been no attempt to determine
variability within a single night. The nightly averaged data are first
viewed to determine whether there are any obviously erroneous
points. While this is easy to accomplish for a calibration source, or
a source that is not variable, for these variable extragalactic
radio-sources, this can introduce a level of subjectivity. As
described in section \ref{sect:obssel} early polarimeter observations
are problematic and must be removed by inspection of the light
curves. In general the light curves were inspected and in cases where
a point was obviously discrepant or isolated, the processed map and
observing log were inspected to decide whether the point was valid.

Manual removal of points mainly occurred for the following reasons:

\begin{itemize}
\item SCUBA occassionally suffered from what was thought to be a film
of helium migrating across the array. In some cases this film obscured
the central source.
\item Some observations placed the target near to an extremely noisy
bolometer that would affect the aperture photometry.
\item Some images are extremely non-Gaussian 'pancakes'.
\item Sometimes a fault in the system would result in the source being
split into 2.
\item There are occurrences of very out of focus images (probably
tests of the focus system).
\end{itemize}

In total, approximately 50 observations (0.5 per cent) were removed manually from the data set.


\begin{table*}
\caption{The sources observed for this paper along with the J2000 RA/dec, the date of the first and last observation, the minimum and maximum flux densities measured for the period (in mJy), the total number of observations for that target and the number of nights during that period on which the source was observed}
\begin{tabular}{llcrccrrrr}
 & & RA(J2000) & Dec(J2000) & UT(min) & UT(max) & Flux(min) & Flux(max) & Nobs & Nnights\\
\hline
0003$-$066 &  & 00 06 13.9 & $-$06 23 35.33 & 19970808 & 20050601 & 765.9 & 1578.1 &  120 &  60\\
  0016+731 &  & 00 19 45.8 &  73 27 30.02 & 20000922 & 20020426 & 176.4 & 358.4 &    7 &   6\\
0048$-$097 &  & 00 50 41.3 & $-$09 29 05.21 & 19970909 & 20030828 & 374.2 & 1032.7 &   24 &  12\\
  0106+013 &  & 01 08 38.8 &  01 35 00.32 & 19970705 & 20050604 & 187.1 & 1333.1 &  154 &  66\\
  0119+041 &  & 01 21 56.9 &  04 22 24.73 & 20000804 & 20000804 & 158.4 & 158.4 &    1 &   1\\
  0133+476 &  & 01 36 58.6 &  47 51 29.10 & 19970405 & 20050531 & 960.7 & 2805.0 &  227 &  90\\
0135$-$247 &  & 01 37 38.3 & $-$24 30 53.89 & 20000809 & 20050603 & 220.0 & 735.9 &   35 &  23\\
0138$-$097 &  & 01 41 25.8 & $-$09 28 43.67 & 20040111 & 20040111 & 208.2 & 208.2 &    2 &   1\\
  0149+218 &  & 01 52 18.1 &  22 07 07.70 & 19980228 & 20020917 & 372.8 & 531.6 &    9 &   5\\
  0202+319 &  & 02 05 04.9 &  32 12 30.10 & 20020805 & 20040904 & 213.7 & 395.0 &    4 &   4\\
  0215+015 &  & 02 17 49.0 &  01 44 49.70 & 19971208 & 20011212 & 352.2 & 567.5 &    8 &   6\\
  0212+735 &  & 02 17 30.8 &  73 49 32.62 & 20010108 & 20041007 & 168.4 & 512.3 &    5 &   4\\
  0219+428 & (3C\,66A) & 02 22 39.6 &  43 02 07.80 & 19970910 & 19970910 & 719.0 & 719.0 &    2 &   1\\
  0221+067 &  & 02 24 28.4 &  06 59 23.34 & 19971002 & 20020304 & 166.9 & 461.3 &   39 &  23\\
  0224+671 &  & 02 28 50.1 &  67 21 03.03 & 19980215 & 20041019 & 259.6 & 685.1 &   23 &  13\\
  0234+285 &  & 02 37 52.4 &  28 48 08.99 & 19971006 & 20041216 & 761.8 & 1810.8 &   68 &  29\\
  0235+164 &  & 02 38 38.9 &  16 36 59.27 & 19970808 & 20050601 & 457.9 & 2697.2 &   78 &  41\\
  0300+471 &  & 03 03 35.2 &  47 16 16.28 & 19981028 & 20031216 & 333.2 & 586.6 &    4 &   4\\
  0306+102 &  & 03 09 03.6 &  10 29 16.34 & 20010114 & 20010114 & 274.5 & 274.5 &    1 &   1\\
  0316+413 & (3C\,84) & 03 19 48.2 &  41 30 42.10 & 19970816 & 20041216 & 1095.5 & 2427.5 &  243 &  87\\
  0333+321 &  & 03 36 30.1 &  32 18 29.34 & 20010909 & 20010909 & 451.9 & 451.9 &    3 &   1\\
0336$-$019 &  & 03 39 30.9 & $-$01 46 35.80 & 19970705 & 20050601 & 419.1 & 1724.1 &  197 &  90\\
0338$-$214 &  & 03 40 35.6 & $-$21 19 31.17 & 20010207 & 20050106 & 409.4 & 599.6 &   67 &  21\\
  0355+508 &  & 03 59 29.7 &  50 57 50.16 & 19980105 & 20041216 & 1364.0 & 3519.1 &   28 &  18\\
  0415+379 & (3C\,111) & 04 18 21.3 &  38 01 36.07 & 19970706 & 20040831 & 414.7 & 2410.4 &   71 &  39\\
0420$-$014 &  & 04 23 15.8 & $-$01 20 33.07 & 19970810 & 20050601 & 956.0 & 8168.3 &  244 & 104\\
  0422+004 &  & 04 24 46.8 &  00 36 06.33 & 20031216 & 20031216 & 1026.8 & 1026.8 &    1 &   1\\
  0430+052 & (3C\,120) & 04 33 11.1 &  05 21 15.62 & 19980310 & 20031216 & 821.5 & 1821.3 &    7 &   4\\
0454$-$234 &  & 04 57 03.2 & $-$23 24 52.02 & 20020211 & 20041216 & 506.3 & 2238.3 &   30 &  14\\
0458$-$020 &  & 05 01 12.8 & $-$01 59 14.26 & 20000129 & 20050126 & 215.2 & 476.1 &   17 &   9\\
0521$-$365 &  & 05 22 58.0 & $-$36 27 30.85 & 19971013 & 20030130 & 960.3 & 2973.9 &   38 &  22\\
  0528+134 &  & 05 30 56.4 &  13 31 55.15 & 19970406 & 20041216 & 564.1 & 1560.6 &  129 &  62\\
  0529+075 &  & 05 32 39.0 &  07 32 43.35 & 19971217 & 20040201 & 277.6 & 683.8 &   25 &  14\\
0537$-$441 & (PKS\,0537) & 05 38 50.4 & $-$44 05 08.94 & 19971013 & 20031220 & 993.6 & 7161.3 &   45 &  22\\
  0552+398 &  & 05 55 30.8 &  39 48 49.17 & 19970909 & 20041108 & 188.6 & 668.3 &   82 &  43\\
0605$-$085 &  & 06 07 59.7 & $-$08 34 49.98 & 19970908 & 20031113 & 281.5 & 767.2 &   28 &  19\\
0607$-$157 &  & 06 09 41.0 & $-$15 42 40.67 & 19970410 & 20031216 & 772.4 & 3378.2 &   82 &  39\\
  0642+449 &  & 06 46 32.0 &  44 51 16.59 & 19970410 & 20041007 & 326.8 & 653.2 &   97 &  37\\
  0716+714 &  & 07 21 53.4 &  71 20 36.36 & 19970910 & 20041220 & 454.9 & 2930.8 &   76 &  33\\
0723$-$008 &  & 07 25 50.6 & $-$00 54 56.54 & 20010207 & 20031223 & 462.6 & 609.8 &    4 &   4\\
0727$-$115 &  & 07 30 19.1 & $-$11 41 12.60 & 19970912 & 20040311 & 434.5 & 1239.3 &   20 &  17\\
  0735+178 &  & 07 38 07.4 &  17 42 19.00 & 19971208 & 20040311 & 243.9 & 852.7 &   33 &  19\\
  0736+017 &  & 07 39 18.0 &  01 37 04.62 & 19970410 & 20041108 & 499.5 & 3732.6 &   76 &  40\\
  0745+241 &  & 07 48 36.1 &  24 00 24.11 & 19970410 & 20040311 & 203.1 & 527.0 &   69 &  30\\
  0748+126 &  & 07 50 52.0 &  12 31 04.83 & 19980316 & 20010207 & 235.1 & 682.0 &    6 &   4\\
  0754+100 &  & 07 57 06.6 &  09 56 34.85 & 19970410 & 20030203 & 630.7 & 1141.8 &    6 &   6\\
  0814+425 &  & 08 18 16.0 &  42 22 45.41 & 20000325 & 20041007 & 214.5 & 559.5 &   39 &  17\\
0818$-$128 &  & 08 20 57.4 & $-$12 58 59.17 & 20011005 & 20011005 & 225.6 & 225.6 &    5 &   1\\
  0823+033 &  & 08 25 50.3 &  03 09 24.52 & 20001119 & 20040110 & 262.2 & 671.0 &   22 &  11\\
  0829+046 &  & 08 31 48.9 &  04 29 39.09 & 19970410 & 20040915 & 345.0 & 936.3 &   35 &  21\\
  0836+710 &  & 08 41 24.4 &  70 53 42.17 & 19970410 & 20041124 & 343.1 & 1337.2 &   95 &  54\\
  0851+202 & (OJ\,287) & 08 54 48.9 &  20 06 30.64 & 19970410 & 20040311 & 747.0 & 4540.0 &   82 &  46\\
  0859+470 &  & 09 03 04.0 &  46 51 04.14 & 20010416 & 20010416 & 164.1 & 164.1 &    1 &   1\\
  0917+449 &  & 09 20 58.5 &  44 41 53.99 & 19970410 & 20020121 & 168.7 & 324.7 &   17 &  12\\
  0917+624 &  & 09 21 36.2 &  62 15 52.18 & 20010207 & 20010207 & 200.4 & 200.4 &    1 &   1\\
  0923+392 &  & 09 27 03.0 &  39 02 20.85 & 19970405 & 20050323 & 835.0 & 2338.1 & 1179 & 354\\
0925$-$203 &  & 09 27 51.8 & $-$20 34 51.23 & 20030428 & 20030428 & 301.6 & 301.6 &    1 &   1\\
  0954+556 &  & 09 57 38.2 &  55 22 57.61 & 19970410 & 20050115 & 188.4 & 288.8 &   21 &   9\\
  0954+658 &  & 09 58 47.2 &  65 33 54.82 & 19970410 & 20050604 & 210.6 & 1105.4 &  163 &  69\\
  1012+232 &  & 10 14 47.1 &  23 01 16.57 & 20001209 & 20040118 & 124.8 & 453.9 &   15 &  11\\
1034$-$293 &  & 10 37 16.1 & $-$29 34 02.81 & 19970410 & 20041019 & 325.4 & 1094.0 &  115 &  35\\
\hline
\end{tabular}
\label{tab:data}
\end{table*}
\begin{table*}
\contcaption{}
\begin{tabular}{llcrccrrrr}
 & & RA(J2000) & Dec(J2000) & UT(min) & UT(max) & Flux(min) & Flux(max) & Nobs & Nnights\\
\hline
  1044+719 &  & 10 48 27.6 &  71 43 35.94 & 19970410 & 20050513 & 218.6 & 1162.1 &  193 &  85\\
  1053+815 &  & 10 58 11.5 &  81 14 32.68 & 20010428 & 20030304 & 415.1 & 450.4 &    3 &   3\\
  1055+018 &  & 10 58 29.6 &  01 33 58.82 & 19970410 & 20050319 & 388.0 & 3000.4 &  278 & 118\\
  1116+128 &  & 11 18 57.3 &  12 34 41.72 & 20000812 & 20050128 & 226.5 & 345.5 &   15 &  10\\
1124$-$186 &  & 11 27 04.4 & $-$18 57 17.44 & 20000418 & 20030529 & 219.7 & 710.8 &   47 &  16\\
  1128+385 &  & 11 30 53.3 &  38 15 18.55 & 19991230 & 20040122 & 253.3 & 373.3 &    3 &   3\\
  1144+402 &  & 11 46 58.3 &  39 58 34.30 & 20001209 & 20050116 & 97.2 & 670.6 &   37 &  20\\
  1147+245 &  & 11 50 19.2 &  24 17 53.84 & 19980130 & 20040113 & 307.8 & 555.3 &   34 &  22\\
  1156+295 &  & 11 59 31.8 &  29 14 43.83 & 19970410 & 20041230 & 245.9 & 1999.8 &  176 &  77\\
1213$-$172 &  & 12 15 46.8 & $-$17 31 45.40 & 19980319 & 20030316 & 233.3 & 728.5 &   27 &  10\\
  1216+487 &  & 12 19 06.4 &  48 29 56.16 & 20001226 & 20050604 & 117.0 & 228.2 &   24 &  15\\
  1219+285 &  & 12 21 31.7 &  28 13 58.30 & 19970410 & 20011031 & 277.2 & 759.6 &   99 &  50\\
  1226+023 & (3C\,273) & 12 29 06.7 &  02 03 08.60 & 19970405 & 20050319 & 2268.2 & 16209.7 &  508 & 210\\
1243$-$072 &  & 12 46 04.2 & $-$07 30 46.57 & 20020218 & 20020218 & 177.1 & 177.1 &    2 &   1\\
1244$-$255 &  & 12 46 46.8 & $-$25 47 49.29 & 20000213 & 20030213 & 702.0 & 2293.0 &   29 &  11\\
1253$-$055 & (3C\,279) & 12 56 11.2 & $-$05 47 21.52 & 19970407 & 20050319 & 3658.4 & 17011.2 &  599 & 234\\
  1308+326 &  & 13 10 28.7 &  32 20 43.78 & 19970410 & 20050125 & 170.3 & 1507.3 &  696 & 216\\
1313$-$333 &  & 13 16 08.0 & $-$33 38 59.17 & 19970410 & 20050529 & 327.7 & 1640.9 &   38 &  24\\
  1328+307 &  & 13 31 08.3 &  30 30 32.96 & 20000213 & 20050125 & 217.4 & 271.7 &    7 &   5\\
1334$-$127 &  & 13 37 39.8 & $-$12 57 24.69 & 19970410 & 20041220 & 936.7 & 5019.7 &  124 &  56\\
  1354+195 &  & 13 57 04.4 &  19 19 07.37 & 20000323 & 20030127 & 205.1 & 384.2 &    9 &   6\\
  1413+135 &  & 14 15 58.8 &  13 20 23.71 & 19970410 & 20040308 & 213.8 & 1830.8 &  102 &  56\\
  1418+546 &  & 14 19 46.6 &  54 23 14.78 & 19970604 & 20050602 & 157.9 & 995.5 &  302 & 122\\
  1502+106 &  & 15 04 25.0 &  10 29 39.20 & 20010105 & 20041215 & 134.1 & 866.4 &   30 &  14\\
1510$-$089 &  & 15 12 50.5 & $-$09 05 59.83 & 19970606 & 20041220 & 269.4 & 1135.5 &   84 &  43\\
1514$-$241 &  & 15 17 41.8 & $-$24 22 19.48 & 19970410 & 20030507 & 838.8 & 1875.8 &   39 &  25\\
1519$-$273 &  & 15 22 37.7 & $-$27 30 10.79 & 20010206 & 20010206 & 185.1 & 185.1 &    1 &   1\\
  1538+149 &  & 15 40 49.5 &  14 47 45.88 & 19970706 & 20000317 & 201.5 & 324.4 &    6 &   5\\
  1548+056 &  & 15 50 35.3 &  05 27 10.45 & 19980219 & 20041214 & 533.7 & 984.3 &   13 &   6\\
  1600+335 &  & 16 02 07.3 &  33 26 53.07 & 20020322 & 20020322 & 173.4 & 173.4 &    1 &   1\\
  1606+106 &  & 16 08 46.2 &  10 29 07.78 & 19971006 & 20040317 & 210.2 & 2056.1 &   22 &  12\\
  1611+343 &  & 16 13 41.1 &  34 12 47.91 & 19970405 & 20041227 & 528.3 & 1043.8 &  272 & 119\\
1622$-$253 &  & 16 25 46.9 & $-$25 27 38.33 & 20020425 & 20020425 & 473.2 & 473.2 &    1 &   1\\
1622$-$297 &  & 16 26 06.0 & $-$29 51 26.97 & 20020425 & 20020425 & 277.9 & 277.9 &    1 &   1\\
  1633+382 &  & 16 35 15.5 &  38 08 04.50 & 19970405 & 20050318 & 497.0 & 5565.0 &   84 &  42\\
  1637+574 &  & 16 38 13.5 &  57 20 23.98 & 20000205 & 20050601 & 311.1 & 809.3 &  123 &  44\\
  1638+398 &  & 16 40 29.6 &  39 46 46.03 & 20000603 & 20000603 & 236.5 & 236.5 &    1 &   1\\
  1642+690 &  & 16 42 07.8 &  68 56 39.76 & 20000809 & 20040118 & 427.8 & 597.4 &   63 &  24\\
  1641+399 & (3C\,345) & 16 42 58.8 &  39 48 36.99 & 19970602 & 20050509 & 1203.6 & 4078.4 &  451 & 150\\
  1656+477 &  & 16 58 02.8 &  47 37 49.23 & 20010303 & 20020303 & 160.5 & 221.7 &    7 &   4\\
  1655+077 &  & 16 58 09.0 &  07 41 27.54 & 20000604 & 20020606 & 309.9 & 502.0 &    9 &   8\\
1657$-$261 &  & 17 00 53.2 & $-$26 10 51.72 & 19970701 & 20030324 & 228.7 & 791.2 &   22 &  15\\
  1716+686 &  & 17 16 13.9 &  68 36 38.68 & 19980317 & 19980322 & 277.8 & 341.2 &   43 &   5\\
  1717+178 &  & 17 19 13.0 &  17 45 06.44 & 20020422 & 20030529 & 364.3 & 1954.3 &    3 &   2\\
1730$-$130 &  & 17 33 02.7 & $-$13 04 49.55 & 19970405 & 20050531 & 731.7 & 4084.5 &  121 &  59\\
  1739+522 &  & 17 40 37.0 &  52 11 43.41 & 19970405 & 20030320 & 101.4 & 799.6 &   63 &  25\\
1741$-$038 &  & 17 43 58.9 & $-$03 50 04.62 & 19970701 & 20050601 & 713.0 & 2028.0 &   17 &  13\\
  1749+701 &  & 17 48 32.8 &  70 05 50.77 & 20010522 & 20010522 & 346.1 & 346.1 &    1 &   1\\
  1749+096 &  & 17 51 32.8 &  09 39 00.73 & 19970410 & 20020604 & 896.1 & 2705.0 &   46 &  25\\
  1803+784 &  & 18 00 45.7 &  78 28 04.02 & 19970405 & 20030529 & 604.9 & 1269.8 &   20 &  15\\
  1758+388 &  & 18 00 24.8 &  38 48 30.70 & 20000605 & 20020323 & 116.9 & 170.7 &   11 &   6\\
  1800+440 &  & 18 01 32.3 &  44 04 21.90 & 20000321 & 20030705 & 427.4 & 1396.6 &   32 &  17\\
  1807+698 &  & 18 06 50.7 &  69 49 28.11 & 20010423 & 20050528 & 764.8 & 1020.1 &    6 &   6\\
  1823+568 &  & 18 24 07.1 &  56 51 01.49 & 19970405 & 20050529 & 386.3 & 1209.6 &  143 &  66\\
1908$-$202 &  & 19 11 09.7 & $-$20 06 55.11 & 19970701 & 20020917 & 717.5 & 2116.6 &   25 &  19\\
1921$-$293 &  & 19 24 51.1 & $-$29 14 30.12 & 19970703 & 20050531 & 2282.2 & 6020.9 &   88 &  44\\
  1923+210 &  & 19 25 59.6 &  21 06 26.16 & 19981027 & 20030625 & 467.8 & 2030.0 &   17 &  13\\
  1928+738 &  & 19 27 48.5 &  73 58 01.57 & 19971005 & 20010721 & 329.1 & 1025.3 &   22 &   9\\
  1954+513 &  & 19 55 42.7 &  51 31 48.55 & 20000824 & 20050124 & 235.3 & 405.0 &    3 &   3\\
1958$-$179 &  & 20 00 57.1 & $-$17 48 57.67 & 19970701 & 20021230 & 410.7 & 2164.8 &   45 &  18\\
  2007+776 &  & 20 05 31.0 &  77 52 43.25 & 19970701 & 20011002 & 302.5 & 592.3 &   26 &  16\\
\hline
\end{tabular}
\label{tab:data2}
\end{table*}
\begin{table*}
\contcaption{}
\begin{tabular}{llcrccrrrr}
 & & RA(J2000) & Dec(J2000) & UT(min) & UT(max) & Flux(min) & Flux(max) & Nobs & Nnights\\
\hline
  2005+403 &  & 20 07 44.9 &  40 29 48.60 & 20010721 & 20010721 & 347.2 & 347.2 &    2 &   1\\
2008$-$159 &  & 20 11 15.7 & $-$15 46 40.25 & 20010721 & 20030905 & 267.4 & 561.4 &    7 &   4\\
  2021+317 &  & 20 23 19.0 &  31 53 02.31 & 19970405 & 20010612 & 307.3 & 463.0 &    8 &   6\\
  2037+511 &  & 20 38 37.0 &  51 19 12.66 & 19970405 & 20040829 & 383.0 & 516.8 &   18 &  12\\
  2059+034 &  & 21 01 38.8 &  03 41 31.32 & 19980422 & 20010721 & 206.1 & 536.0 &    3 &   3\\
  2121+053 &  & 21 23 44.5 &  05 35 22.09 & 20000812 & 20011003 & 420.1 & 722.4 &   11 &   7\\
  2134+004 &  & 21 36 38.6 &  00 41 54.21 & 20010721 & 20030614 & 340.6 & 624.2 &    5 &   4\\
  2145+067 &  & 21 48 05.5 &  06 57 38.60 & 19970810 & 20041216 & 1289.4 & 3831.5 &  122 &  59\\
2155$-$152 &  & 21 58 06.3 & $-$15 01 09.33 & 20010721 & 20041120 & 481.1 & 1025.2 &    9 &   5\\
  2200+420 & (BL\,Lac) & 22 02 43.3 &  42 16 39.98 & 19970405 & 20041216 & 1024.2 & 5159.4 &  201 &  93\\
  2201+315 &  & 22 03 15.0 &  31 45 38.27 & 19970405 & 20041003 & 452.5 & 1330.0 &   19 &  11\\
2223$-$052 & (3C\,446) & 22 25 47.3 & $-$04 57 01.39 & 19970703 & 20050531 & 794.6 & 4032.8 &  144 &  66\\
2227$-$088 &  & 22 29 40.1 & $-$08 32 54.44 & 19970920 & 20040826 & 311.6 & 2332.3 &   12 &   9\\
  2230+114 &  & 22 32 36.4 &  11 43 50.90 & 19970528 & 20011002 & 1182.5 & 5541.5 &   13 &   9\\
  2234+282 &  & 22 36 22.5 &  28 28 57.41 & 20000530 & 20040827 & 263.6 & 834.3 &   21 &  13\\
2243$-$123 &  & 22 46 18.2 & $-$12 06 51.28 & 19981028 & 20011002 & 416.2 & 505.4 &    4 &   3\\
  2251+158 & (3C\,454.3) & 22 53 57.7 &  16 08 53.56 & 19970606 & 20041216 & 1316.4 & 26346.7 &  125 &  66\\
2255$-$282 &  & 22 58 06.0 & $-$27 58 21.26 & 19970808 & 20041114 & 617.1 & 6556.6 &   51 &  27\\
  2318+049 &  & 23 20 44.9 &  05 13 49.95 & 19971203 & 20050108 & 149.3 & 589.5 &   67 &  37\\
2345$-$167 &  & 23 48 02.6 & $-$16 31 12.02 & 19971005 & 20011002 & 769.3 & 1041.4 &    9 &   4\\
  2351+456 &  & 23 54 21.7 &  45 53 04.24 & 20050531 & 20050531 & 297.7 & 297.7 &    2 &   1\\
\hline
\end{tabular}
\label{tab:data3}
\end{table*}

\section{The data}

The observed sources are presented in Table \ref{tab:data} which, for
each source, lists the date of the first and last observation, the
total number of observations and the number of nights on which the
source was observed. The full set of results are available
electronically, with a subset presented in Table \ref{tab:tseries} and the
lightcurves for the sources with data from more than one night (filtering out 19 sources) are
presented in Fig.\ \ref{fig:light-curves}.

\begin{table}
\caption{These are the flux density measurements for a single source, 1908$-$202, provided as an example. Modified Julian Date (MJD) is defined as Julian Date $-$ 2 400 000.5. The number of measurements taken on each night is given in the last column. The data for all sources are available in the online version of the article (see Supporting Information)}
\label{tab:tseries}
\begin{center}
\begin{tabular}{llrrr}
\hline
Date & Date & Flux & Error & \# \\
MJD  &  UT   & / mJy & / mJy & \\
\hline
50630 & 19970701 &  1883.3 &  211 & 1\\
50917 & 19980414 &  1646.2 &  184 & 1\\
50918 & 19980415 &  1521.9 &  170 & 1\\
50919 & 19980416 &  1614.9 &  181 & 1\\
50920 & 19980417 &  1617.5 &  181 & 2\\
50925 & 19980422 &  1560.1 &  174 & 2\\
50927 & 19980424 &  1627.8 &  154 & 3\\
51311 & 19990513 &  1094.7 &  122 & 1\\
51360 & 19990701 &   942.9  &  105 & 1\\
51362 & 19990703 &   966.9  &  108 & 1\\
51780 & 20000824 & 1122.3  & 125  & 1\\
52027 & 20010428 &   800.4  &   89  & 1\\
52056 & 20010527 &   921.1  &  103 & 1\\
52109 & 20010719 &   780.1  &   87  & 1\\
52111 & 20010721 &   812.2  &   91  &1\\
52133 & 20010812 &   804.0  &   63  &3\\
52184 & 20011002 &   717.5  &   80  &1\\
52428 & 20020603 &   899.7  &  101 &1\\
52534 & 20020917 &  2116.6 &  237 &1\\
\hline
\end{tabular}
\end{center}
\end{table}

\section{Conclusions}

This paper presents the second data release of SCUBA pointing data for
flat-spectrum radio sources. In total we have now catalogued \nmeas\
flux density measurements for \nblaz\ sources at 850\mum. Some of
these sources have very sparsely sampled data but our observations
provide perhaps the only measurements of these objects at
submillimetre wavelengths. Other sources have well-sampled light
curves that can be used as part of multifrequency studies. We have
shown that with care, the data can be calibrated in an automated
manner to an accuracy of $<10$ per cent.  Furthermore, the development
and refinement of the pipeline process means that it is a relatively
simple task to extract and calibrate all these data and make them
available to the scientific community.

\begin{figure*}
\begin{tabular}{cc}
\includegraphics[angle=-90,width=76mm]{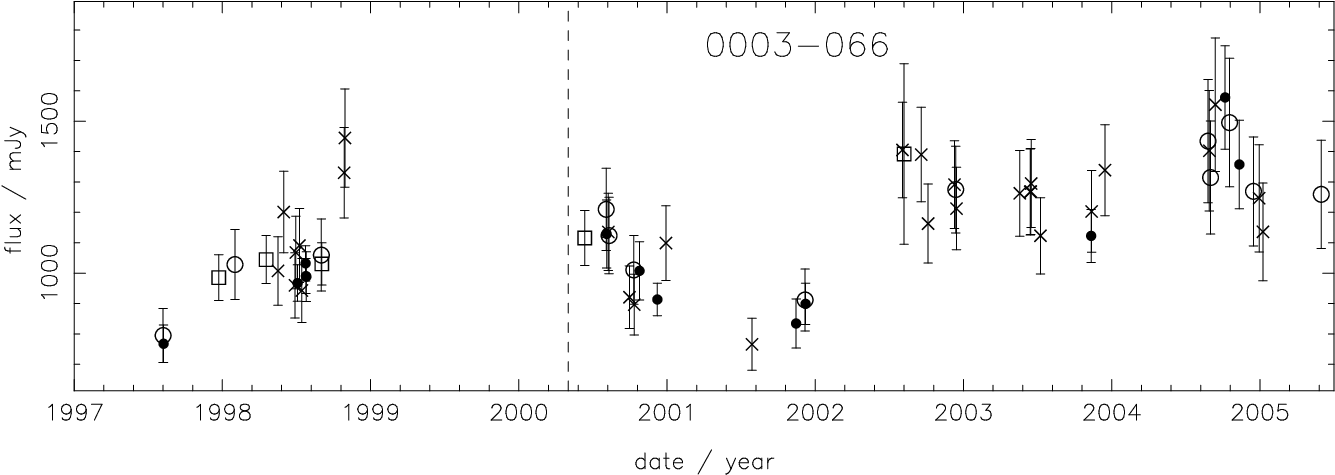}
&
\includegraphics[angle=-90,width=76mm]{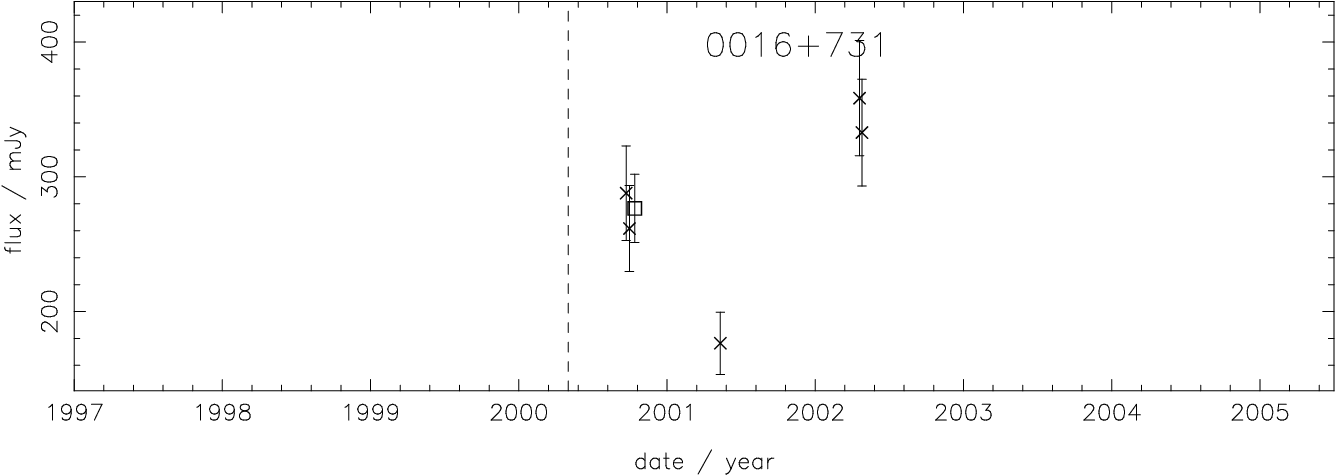}
\\
\includegraphics[angle=-90,width=76mm]{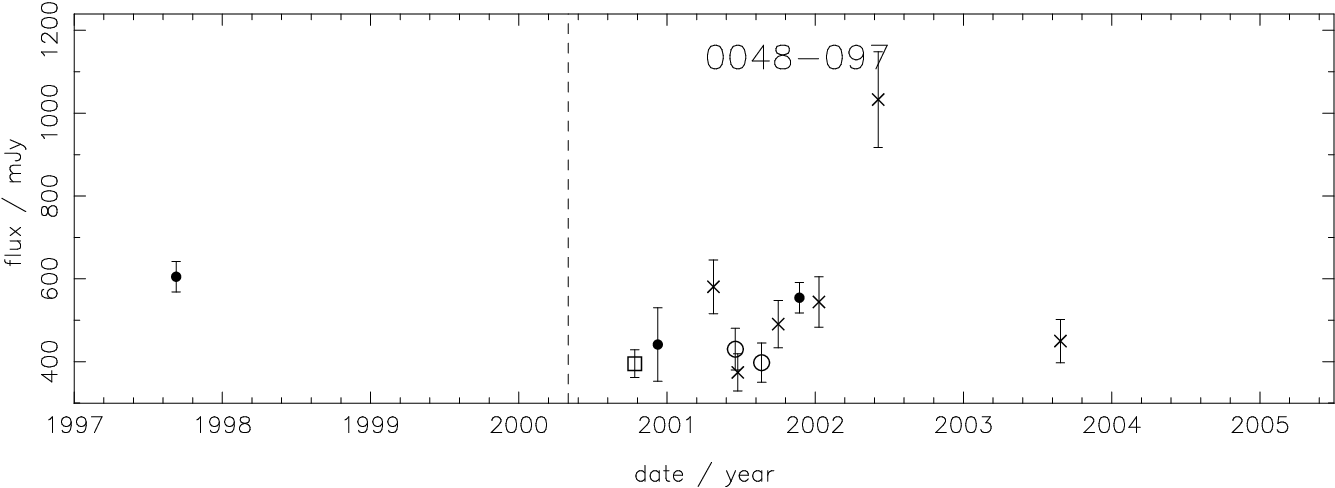}
&
\includegraphics[angle=-90,width=76mm]{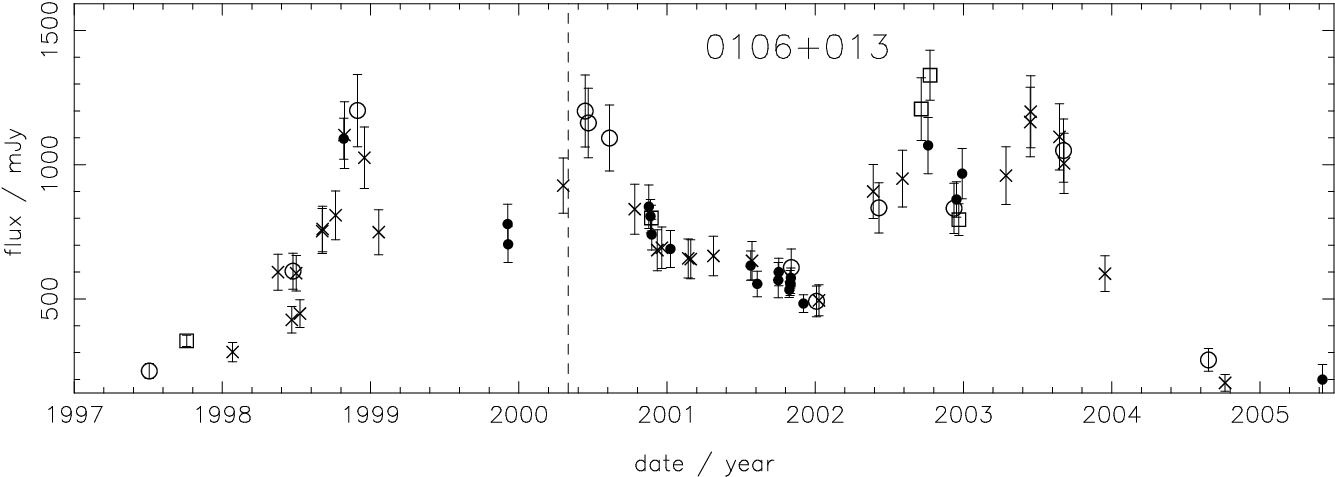}
\\
\includegraphics[angle=-90,width=76mm]{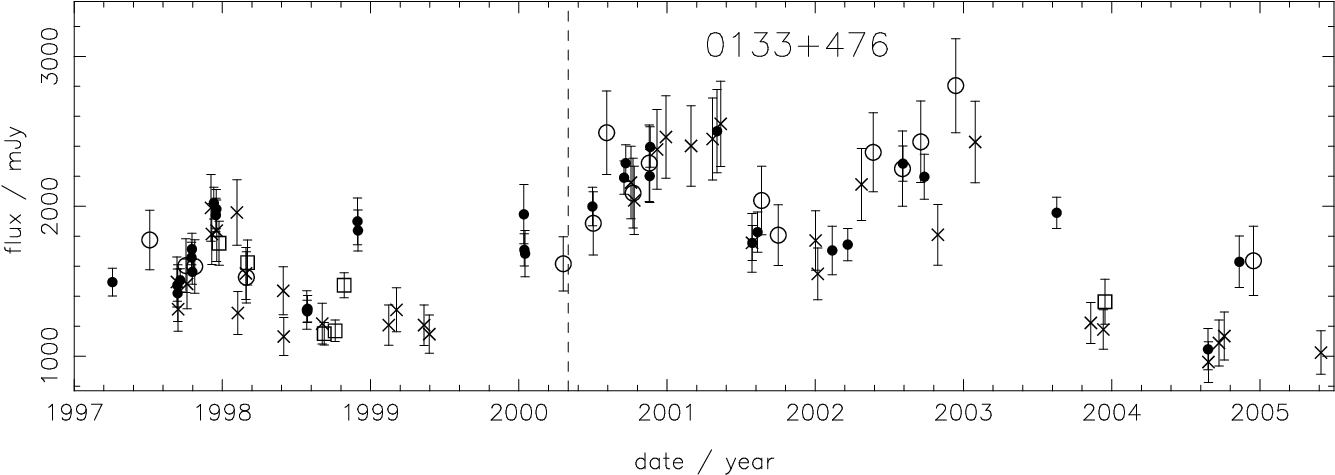}
&
\includegraphics[angle=-90,width=76mm]{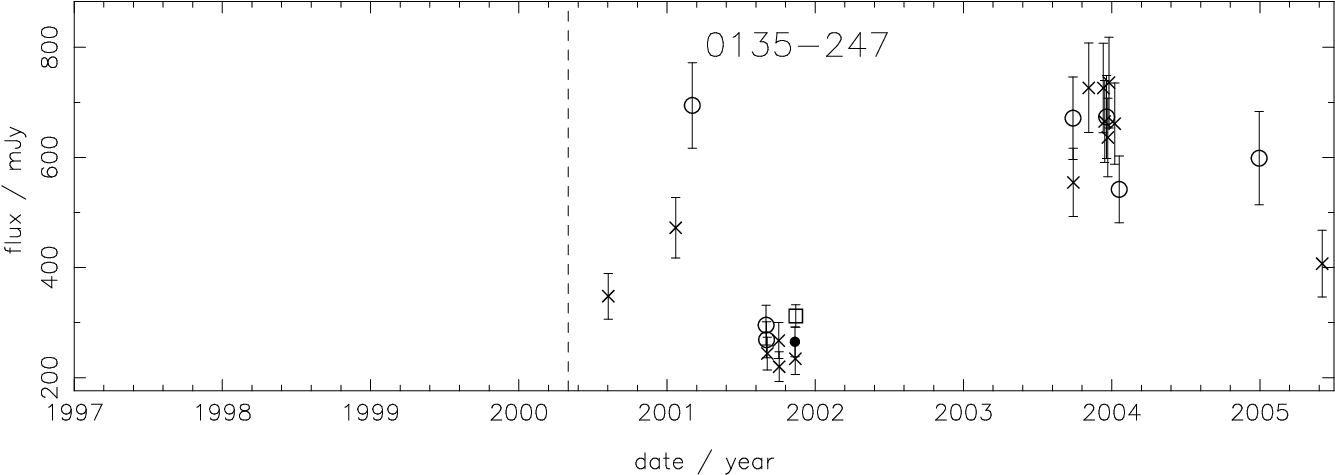}
\\
\includegraphics[angle=-90,width=76mm]{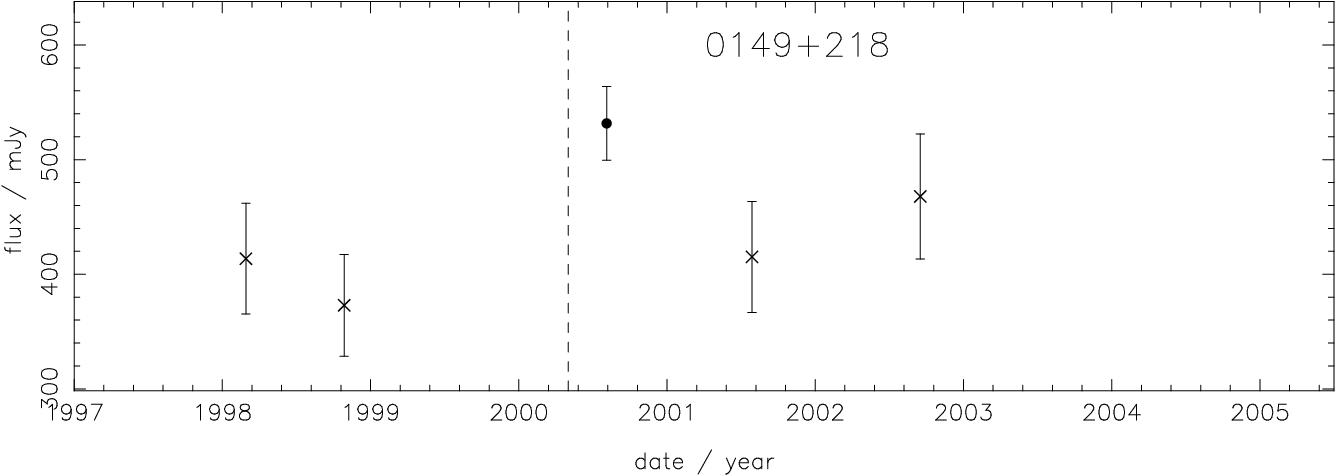}
&
\includegraphics[angle=-90,width=76mm]{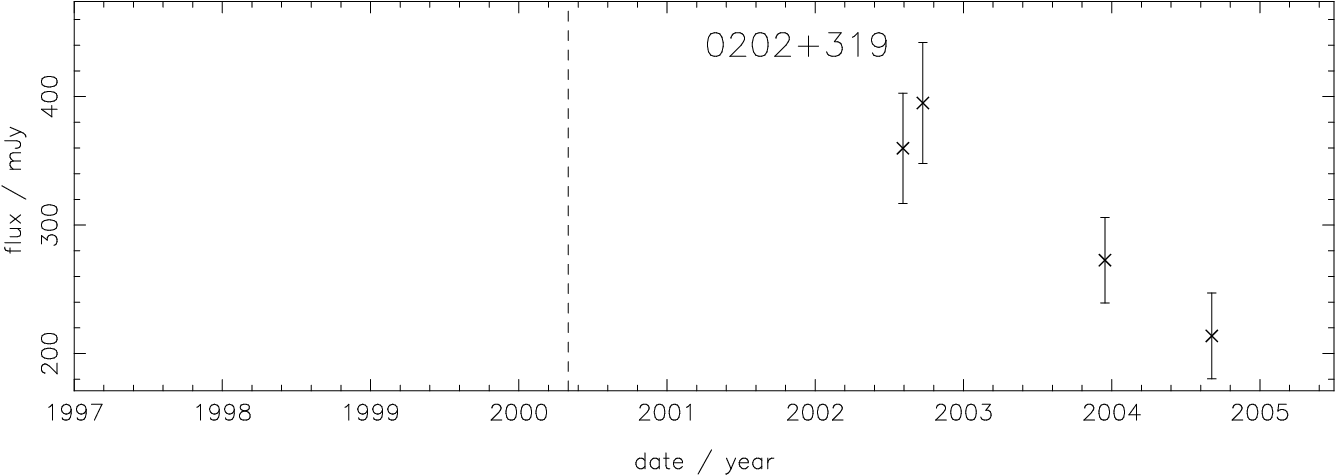}
\\
\includegraphics[angle=-90,width=76mm]{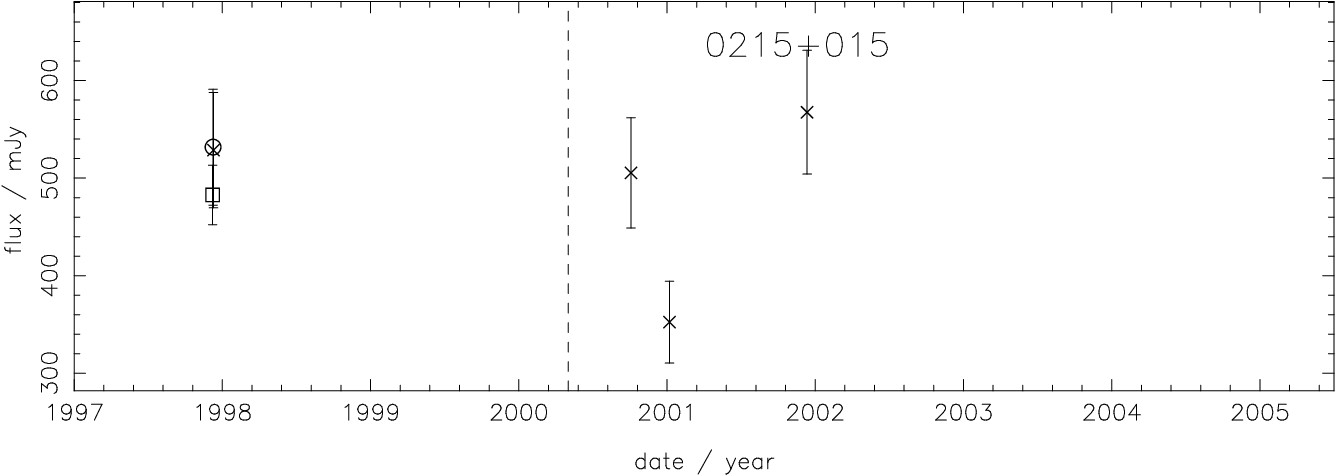}
&
\includegraphics[angle=-90,width=76mm]{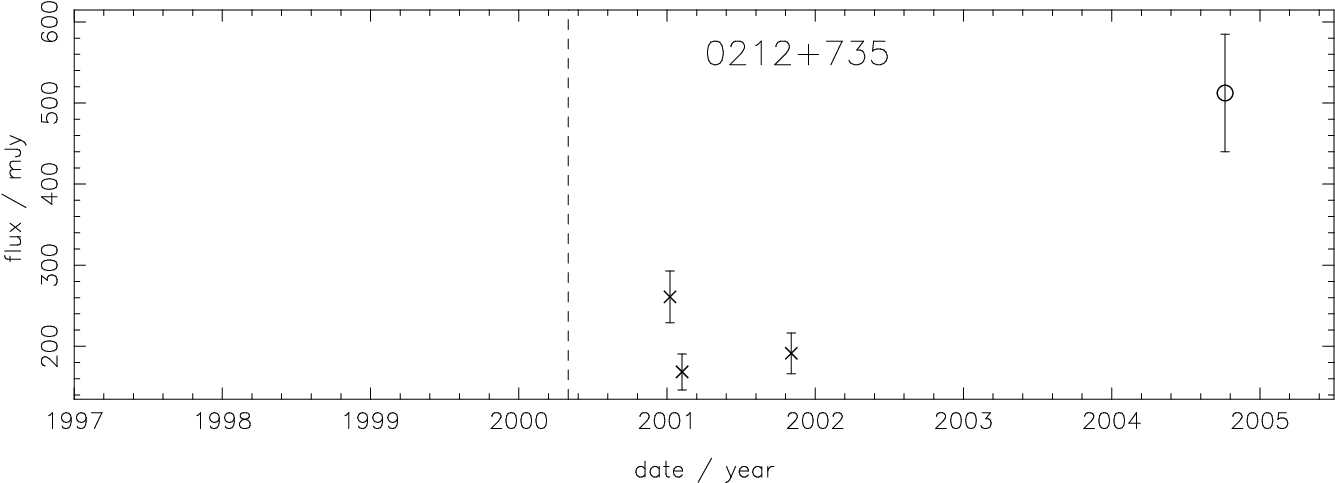}
\\
\includegraphics[angle=-90,width=76mm]{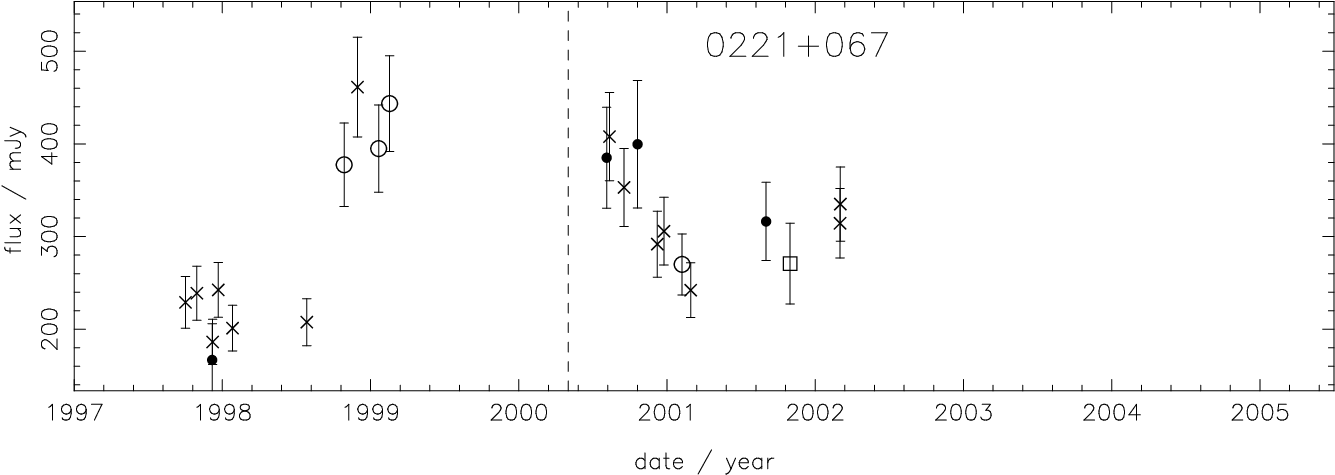}
&
\includegraphics[angle=-90,width=76mm]{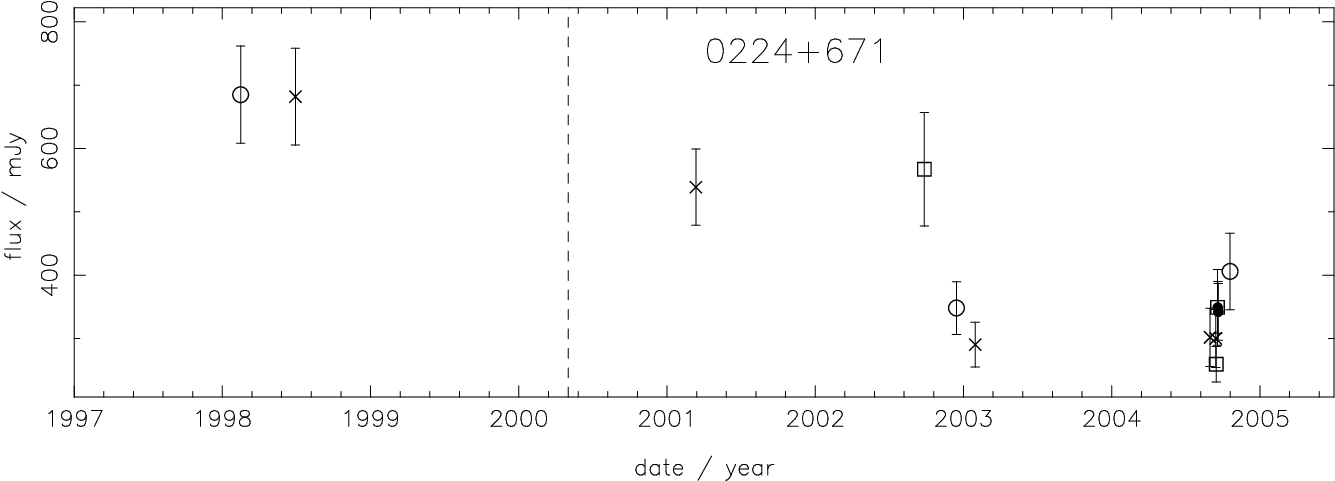}
\\
\includegraphics[angle=-90,width=76mm]{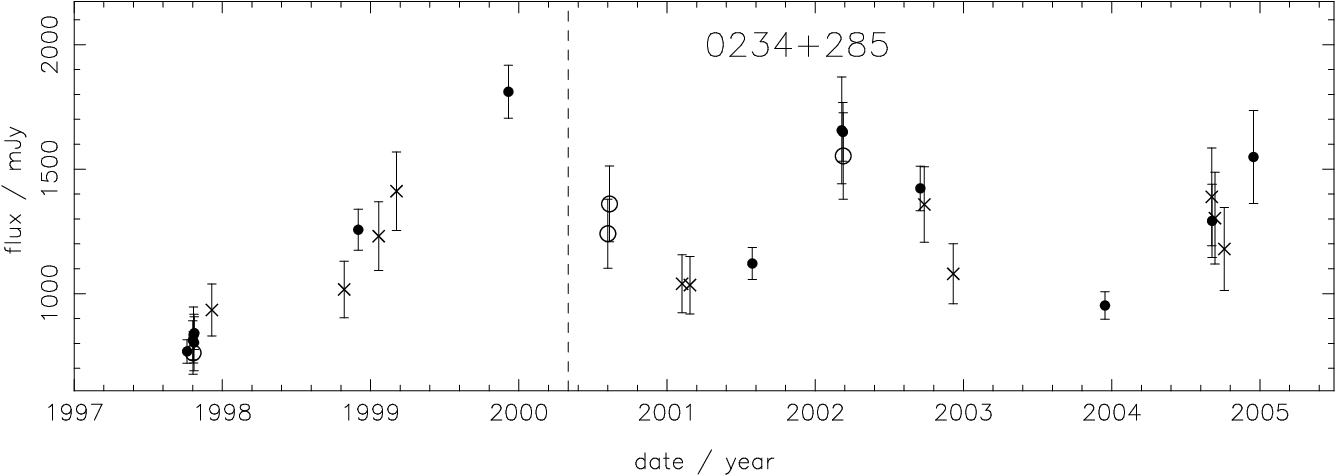}
&
\includegraphics[angle=-90,width=76mm]{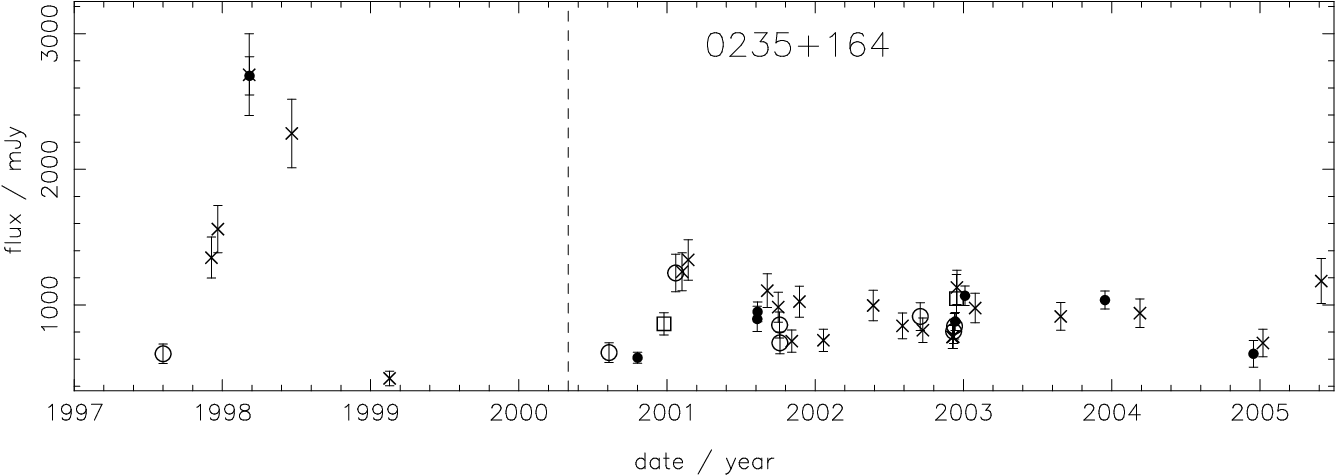}
\\
\end{tabular}
\caption{Light curves for well-sampled data sets. The vertical dashed line indicates the break between this paper and Paper I (although all data have been reanalyzed). Solid circles indicate 3 or more measurements, open circle indicates 2 measurements, open box indicates 2 measurements but the error was derived by the separation of the two points and cross indicates a single mesurement.}
\label{fig:light-curves}
\end{figure*}
\clearpage
\begin{figure*}
\begin{tabular}{cc}
\includegraphics[angle=-90,width=76mm]{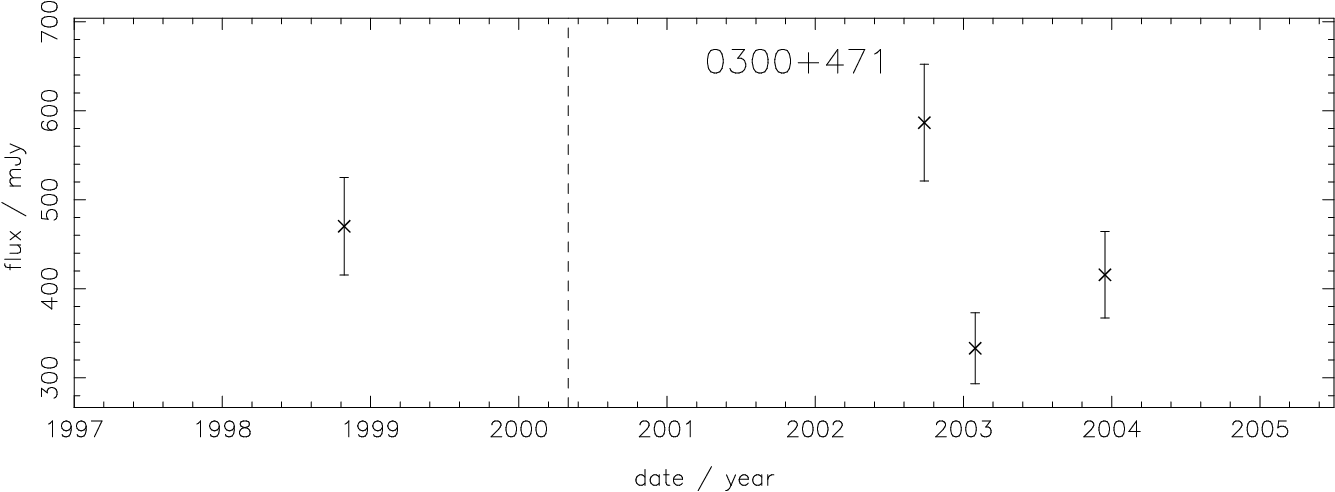}
&
\includegraphics[angle=-90,width=76mm]{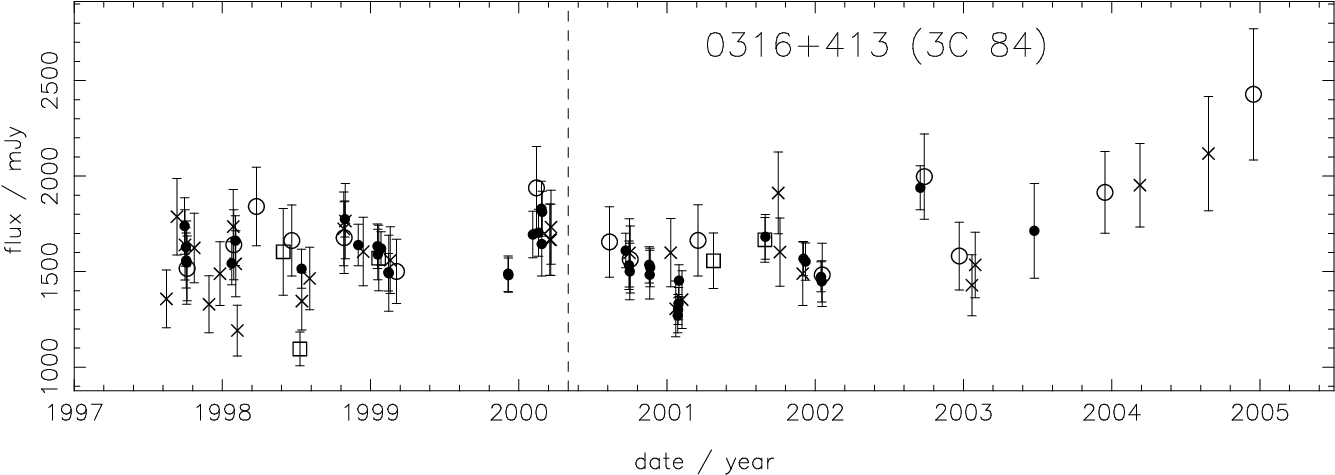}
\\
\includegraphics[angle=-90,width=76mm]{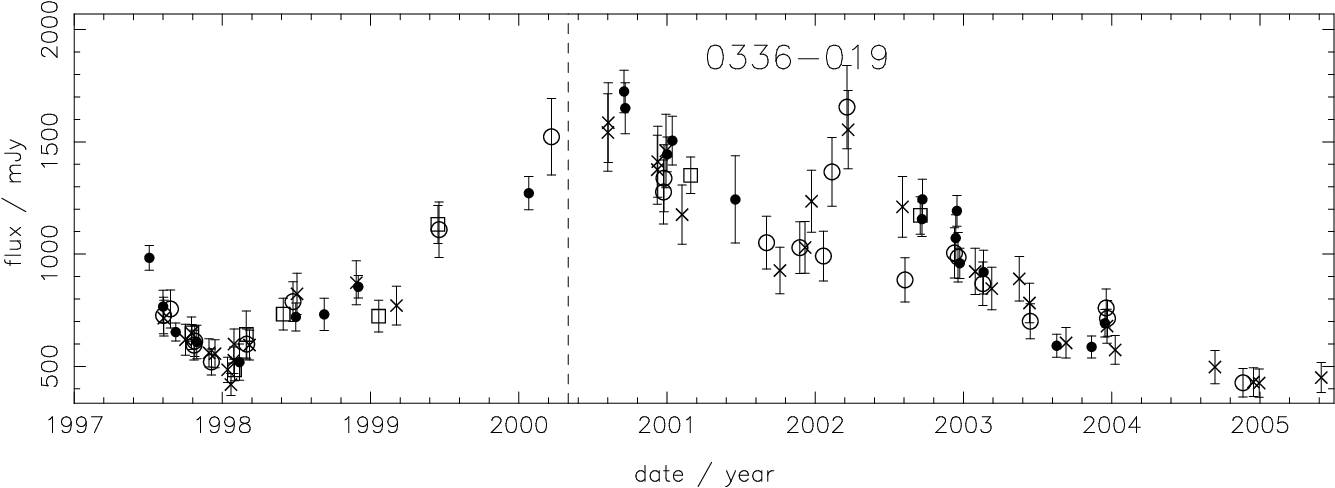}
&
\includegraphics[angle=-90,width=76mm]{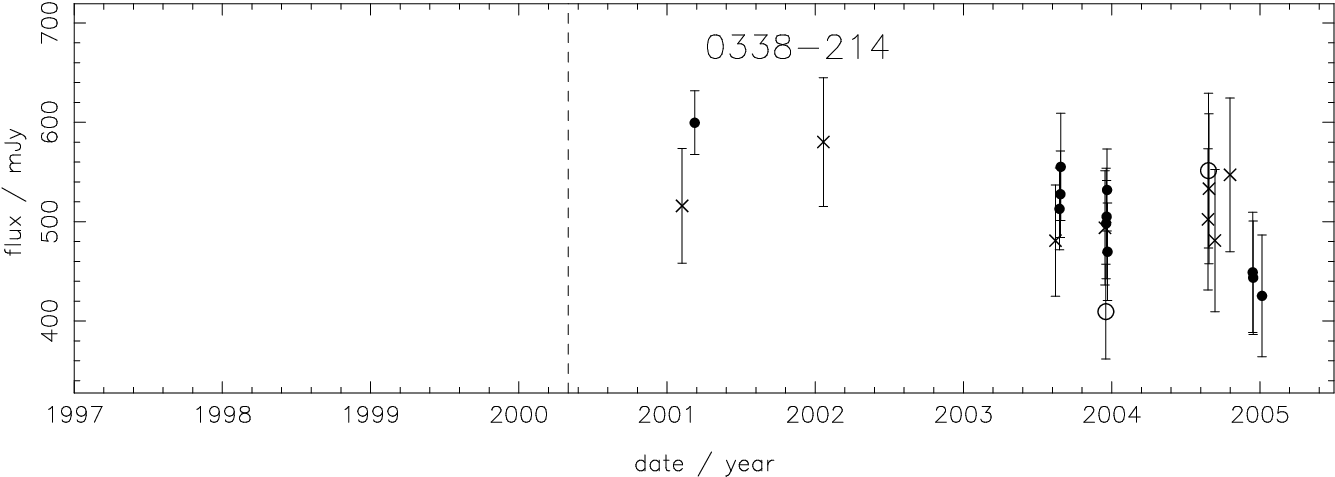}
\\
\includegraphics[angle=-90,width=76mm]{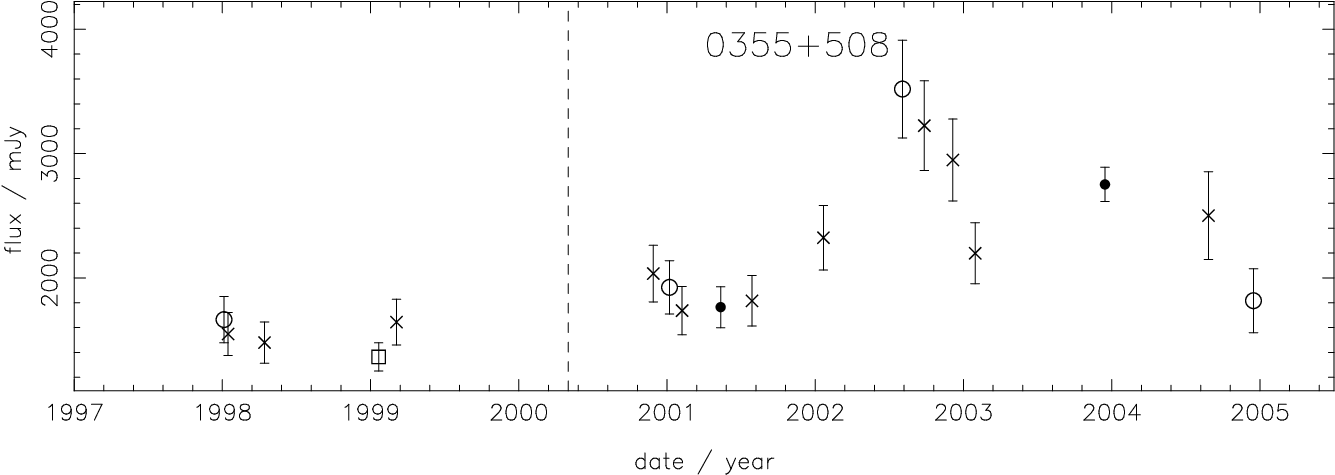}
&
\includegraphics[angle=-90,width=76mm]{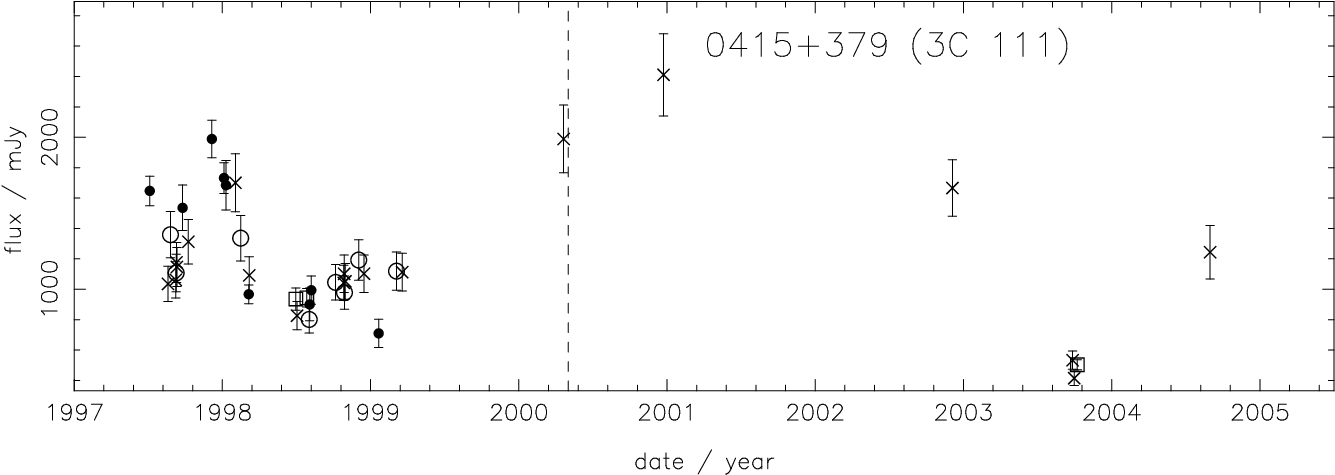}
\\
\includegraphics[angle=-90,width=76mm]{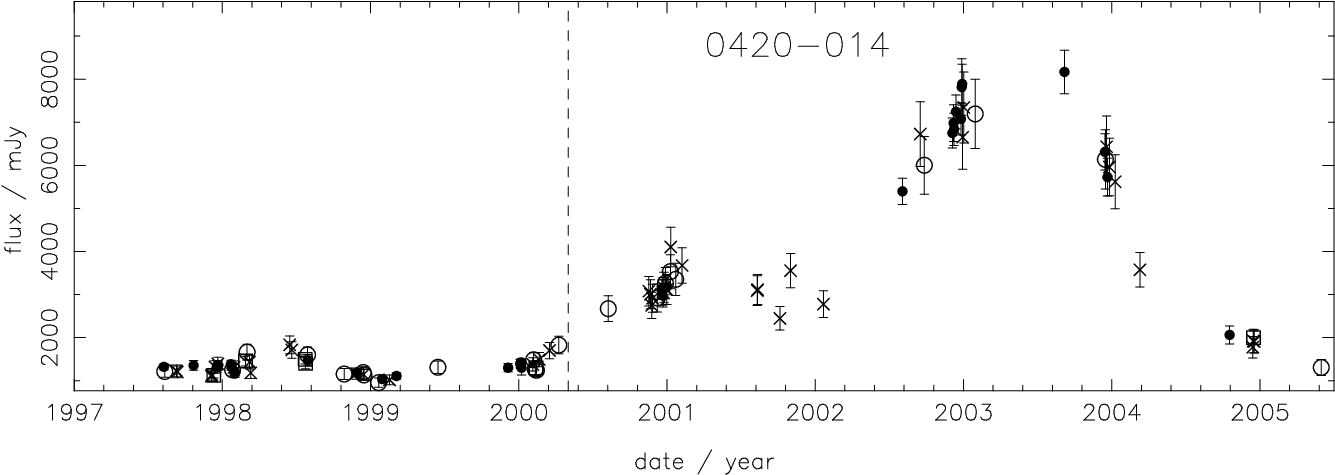}
&
\includegraphics[angle=-90,width=76mm]{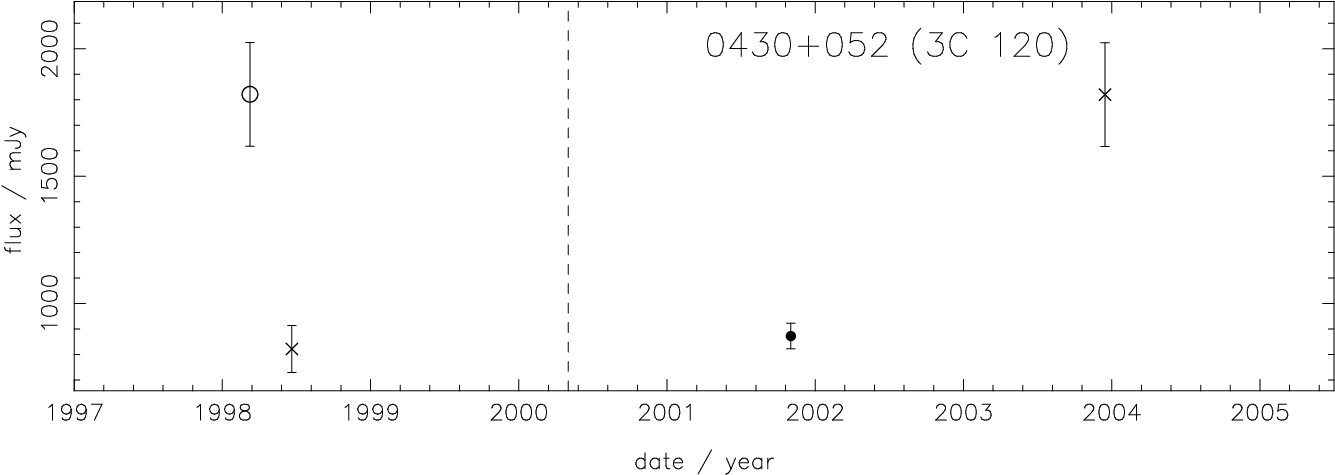}
\\
\includegraphics[angle=-90,width=76mm]{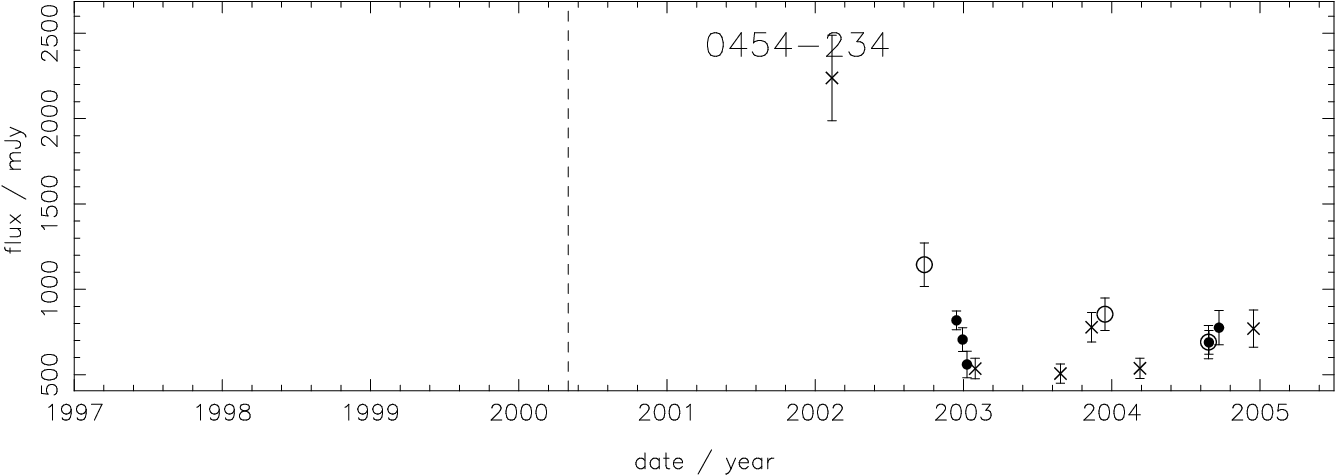}
&
\includegraphics[angle=-90,width=76mm]{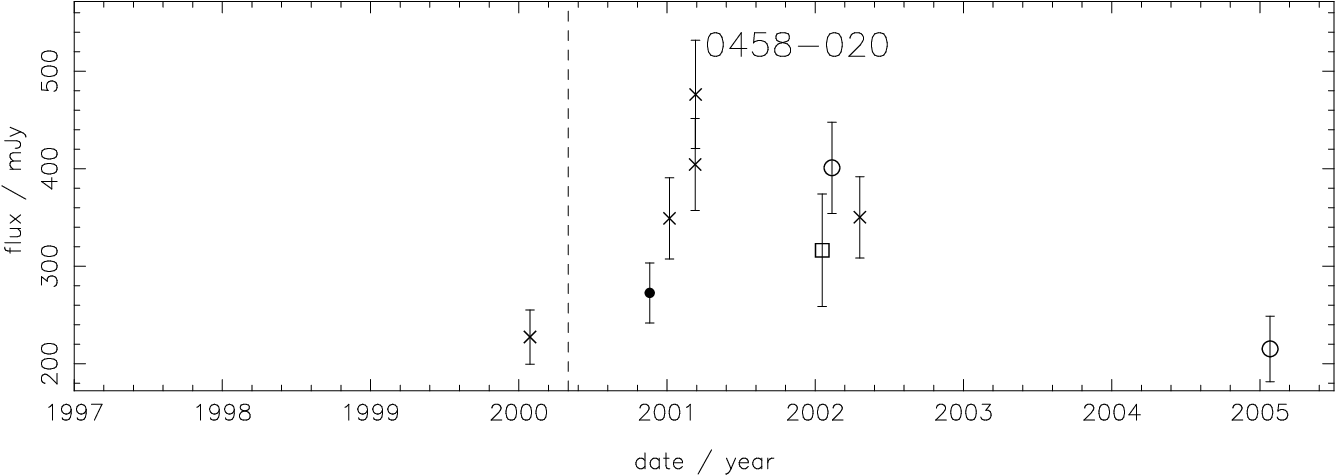}
\\
\includegraphics[angle=-90,width=76mm]{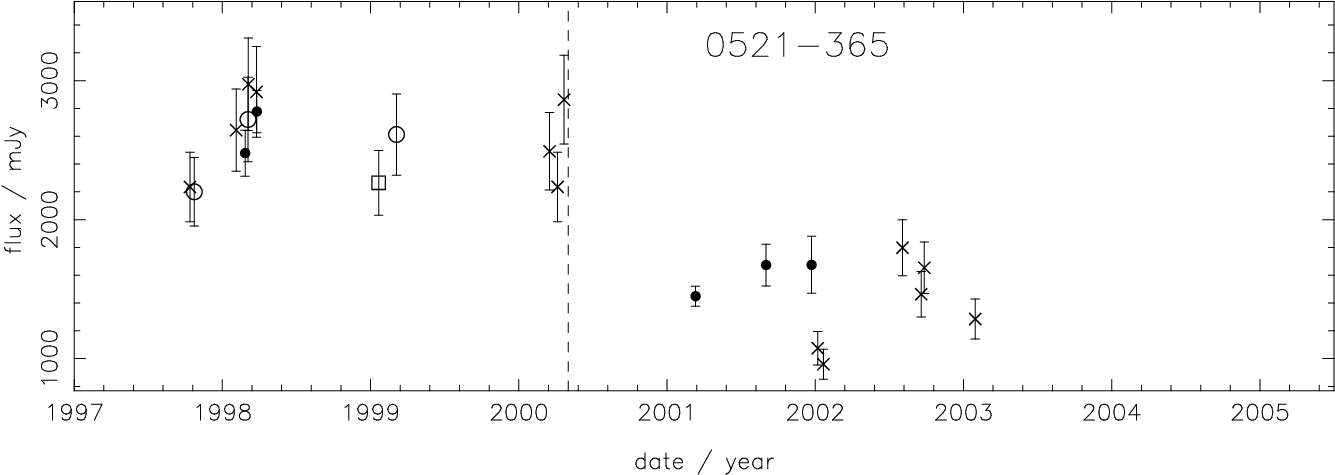}
&
\includegraphics[angle=-90,width=76mm]{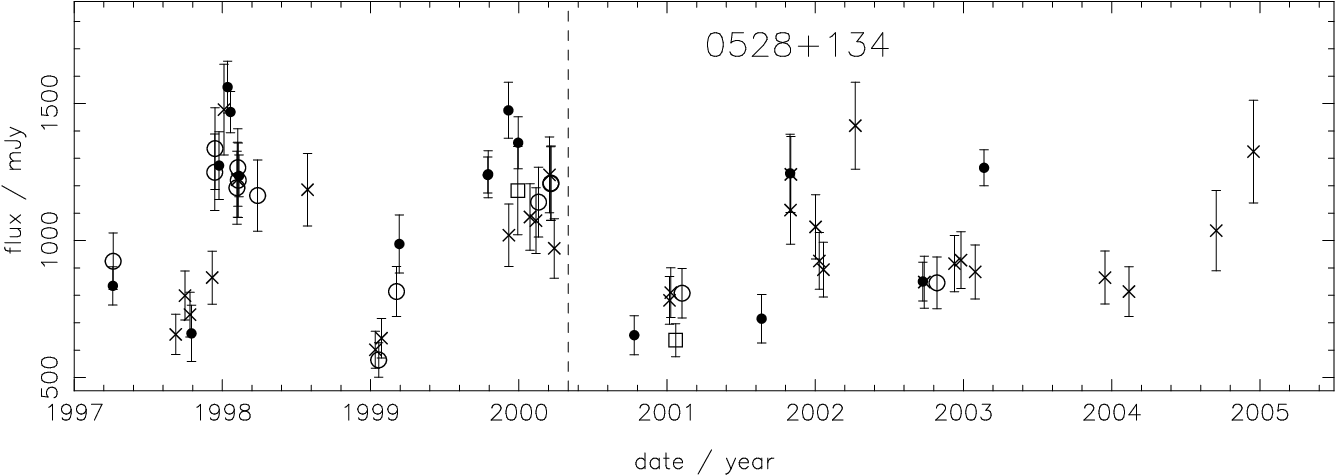}
\\
\includegraphics[angle=-90,width=76mm]{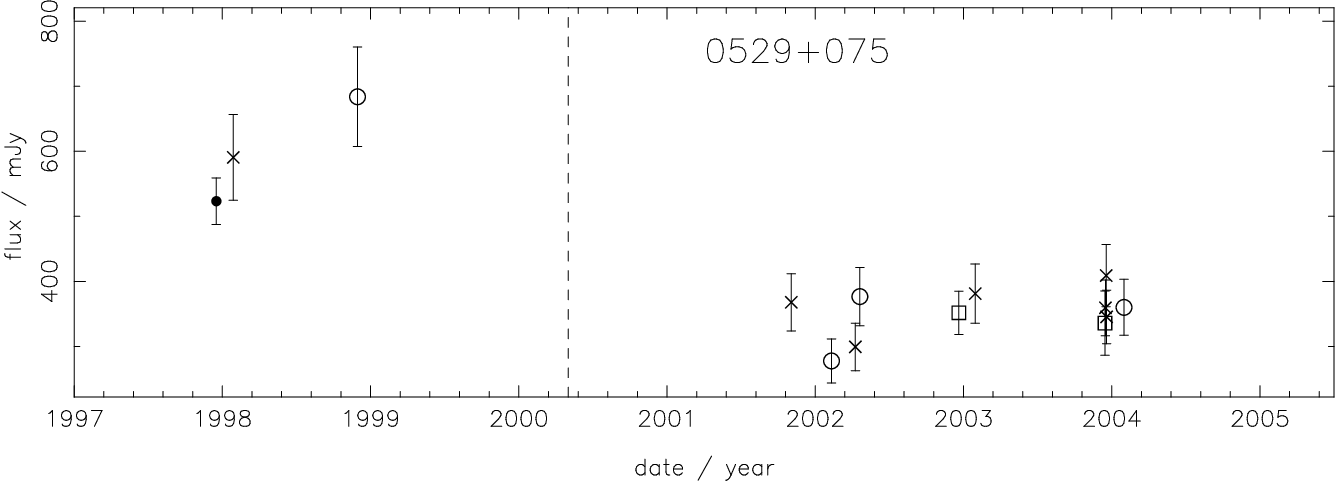}
&
\includegraphics[angle=-90,width=76mm]{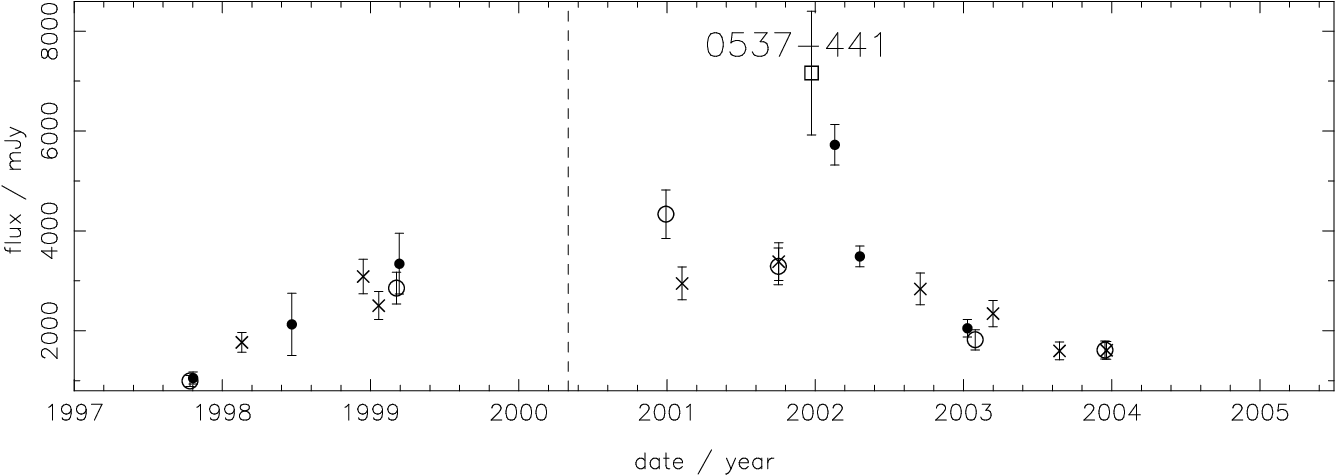}
\\
\end{tabular}
\contcaption{}
\label{fig:light-curves2}
\end{figure*}
\clearpage
\begin{figure*}
\begin{tabular}{cc}
\includegraphics[angle=-90,width=76mm]{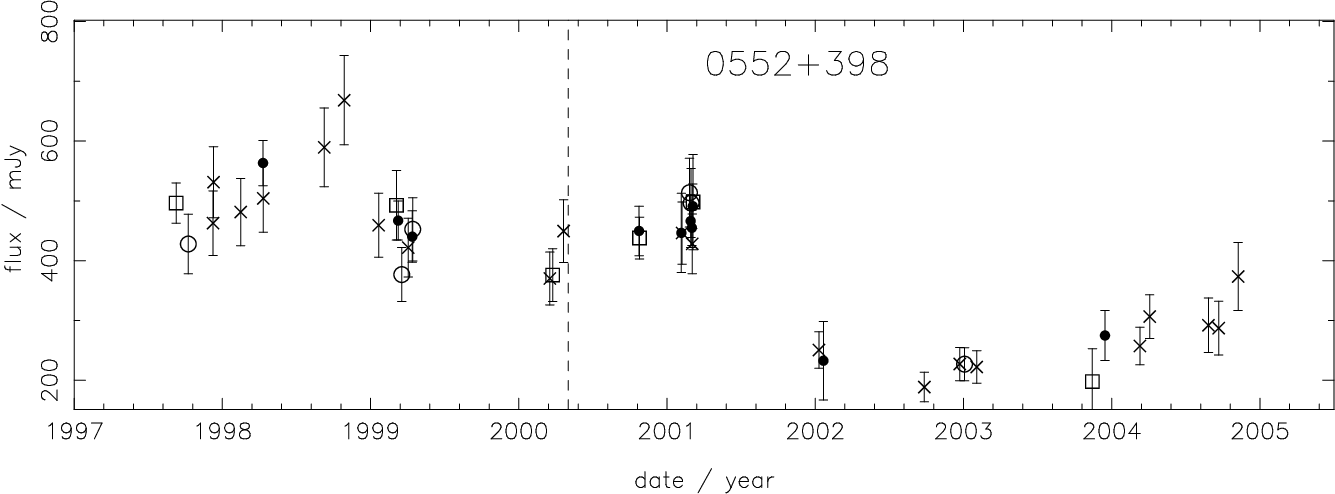}
&
\includegraphics[angle=-90,width=76mm]{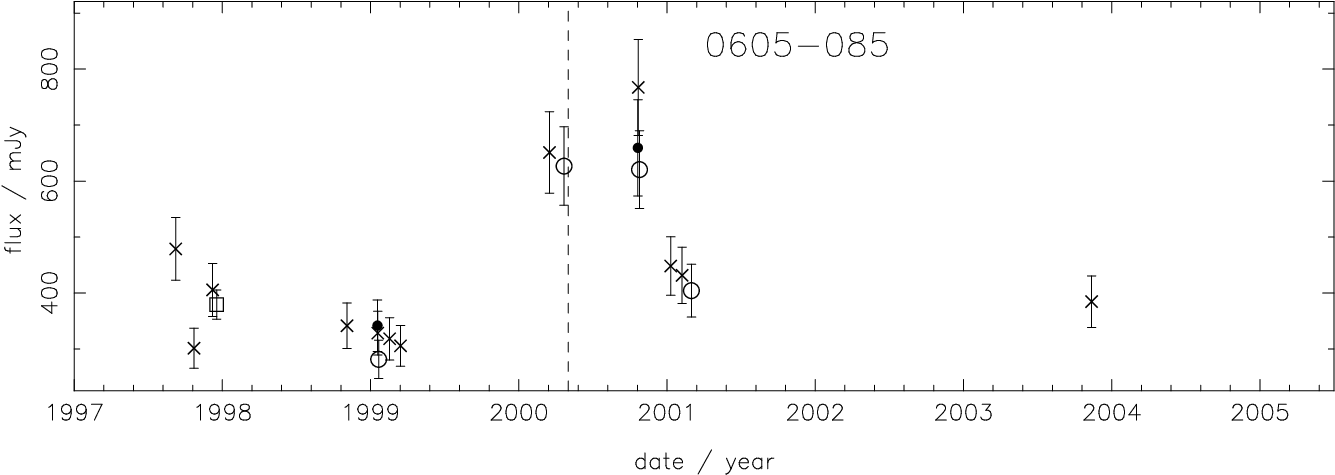}
\\
\includegraphics[angle=-90,width=76mm]{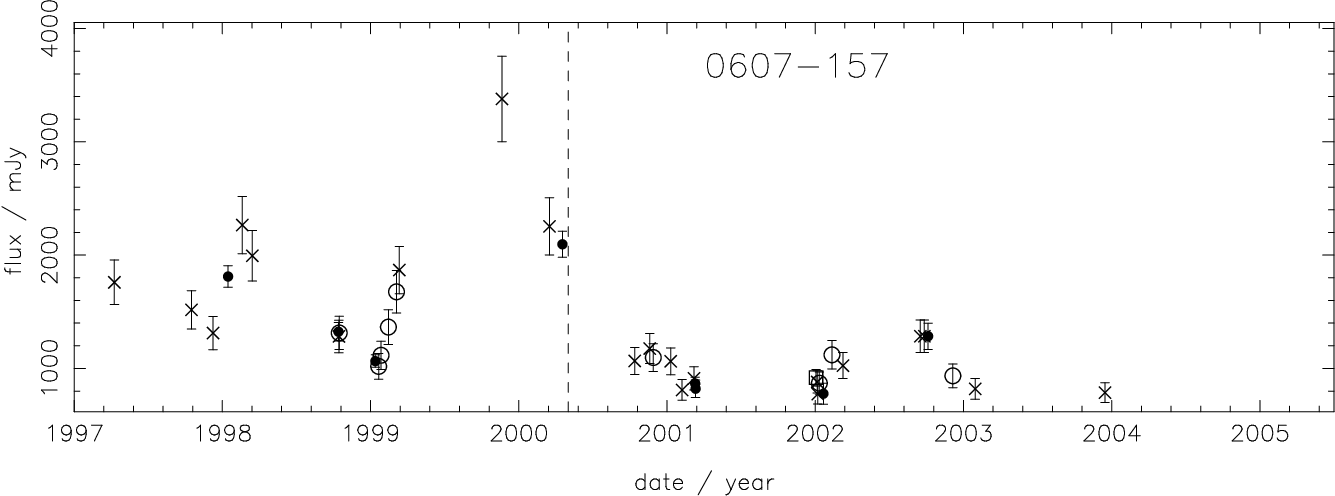}
&
\includegraphics[angle=-90,width=76mm]{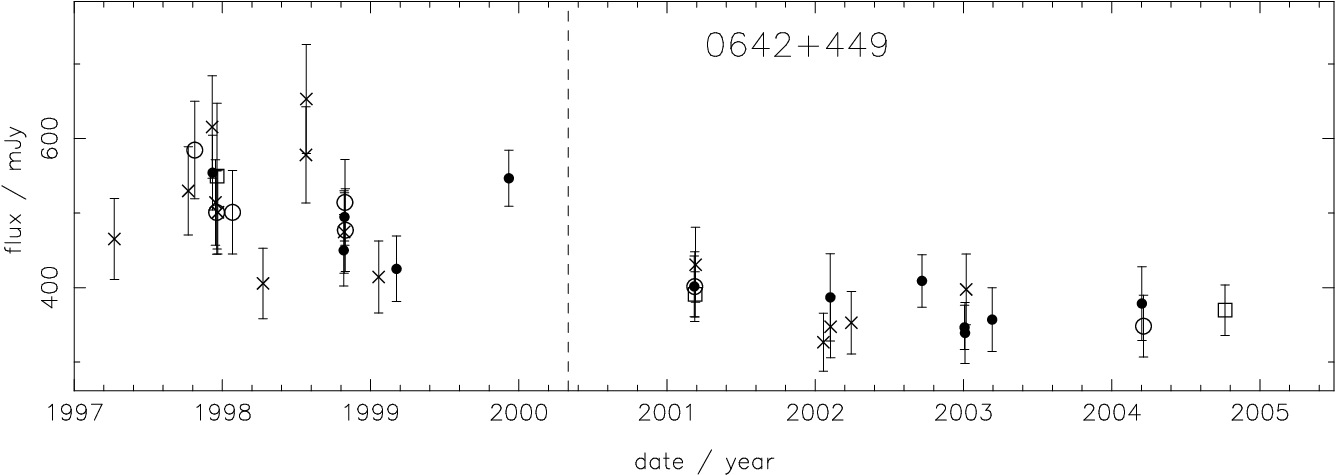}
\\
\includegraphics[angle=-90,width=76mm]{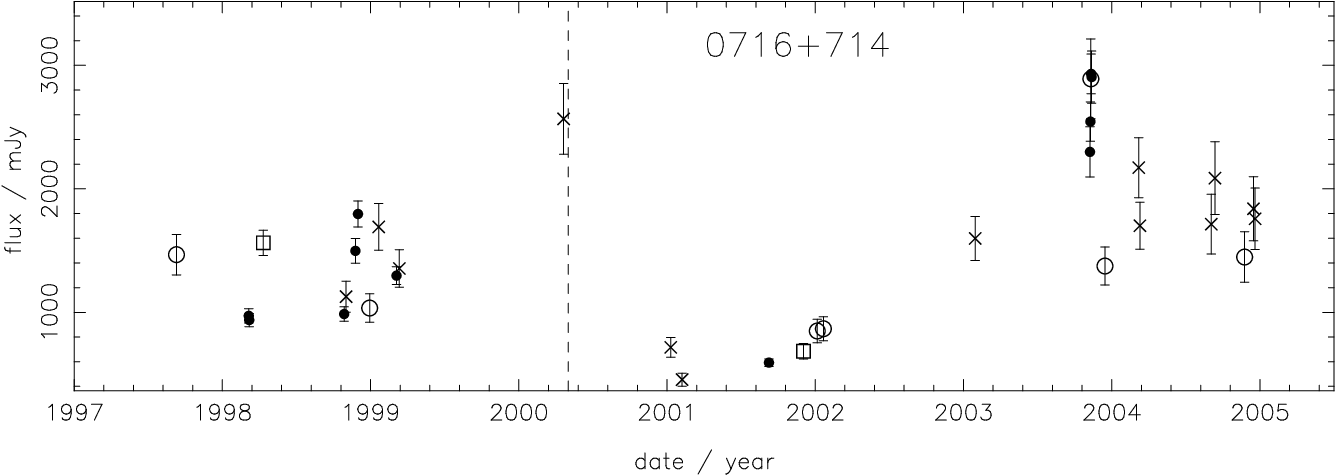}
&
\includegraphics[angle=-90,width=76mm]{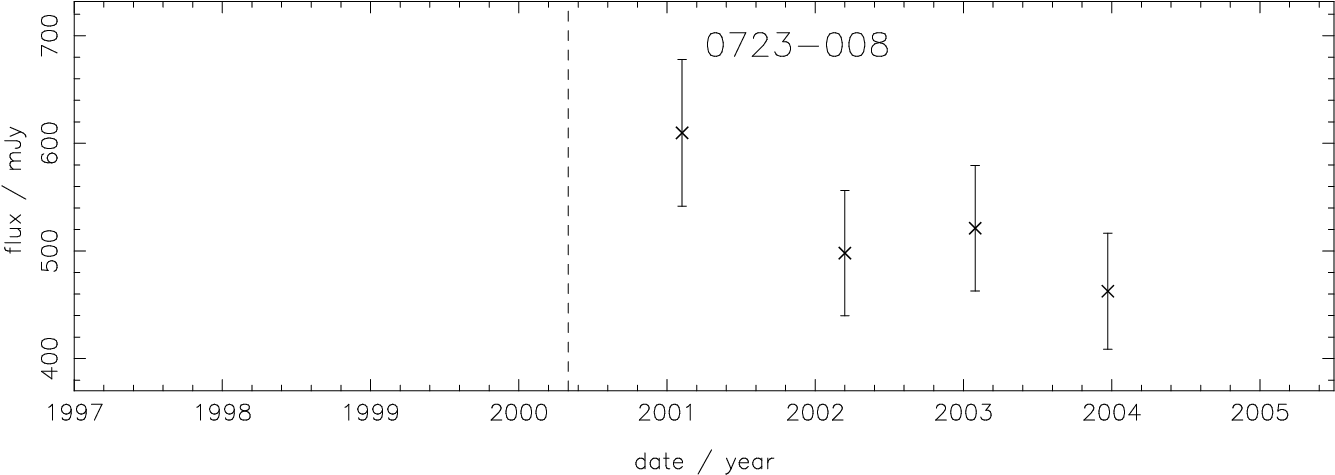}
\\
\includegraphics[angle=-90,width=76mm]{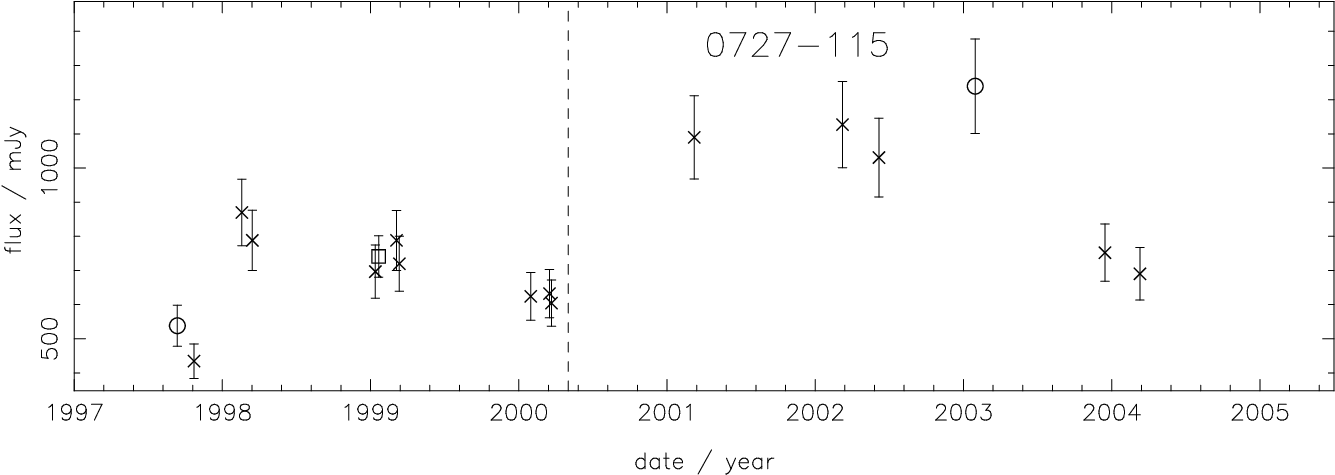}
&
\includegraphics[angle=-90,width=76mm]{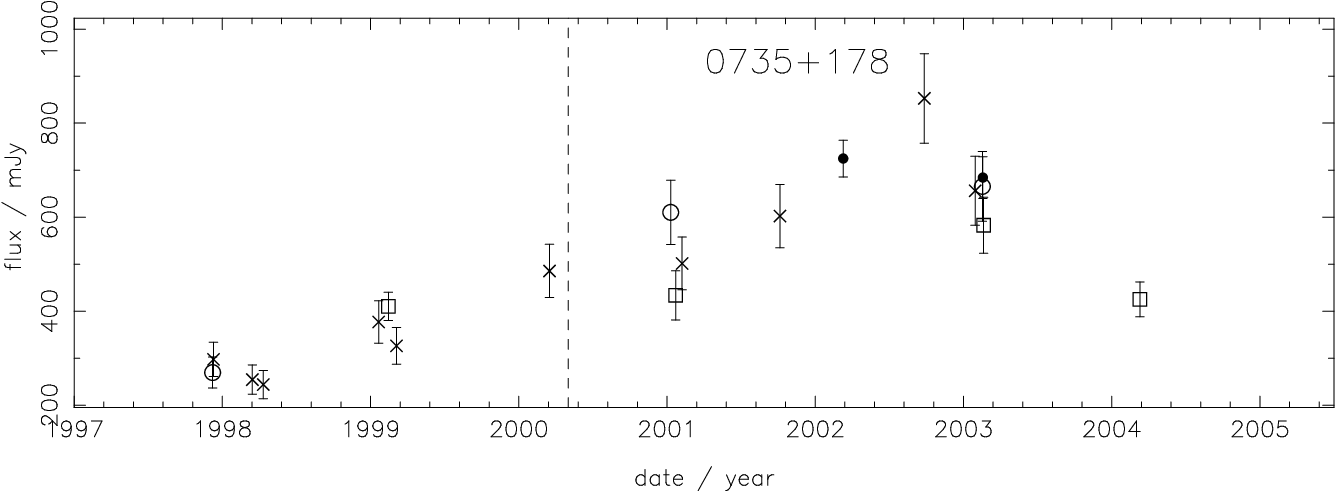}
\\
\includegraphics[angle=-90,width=76mm]{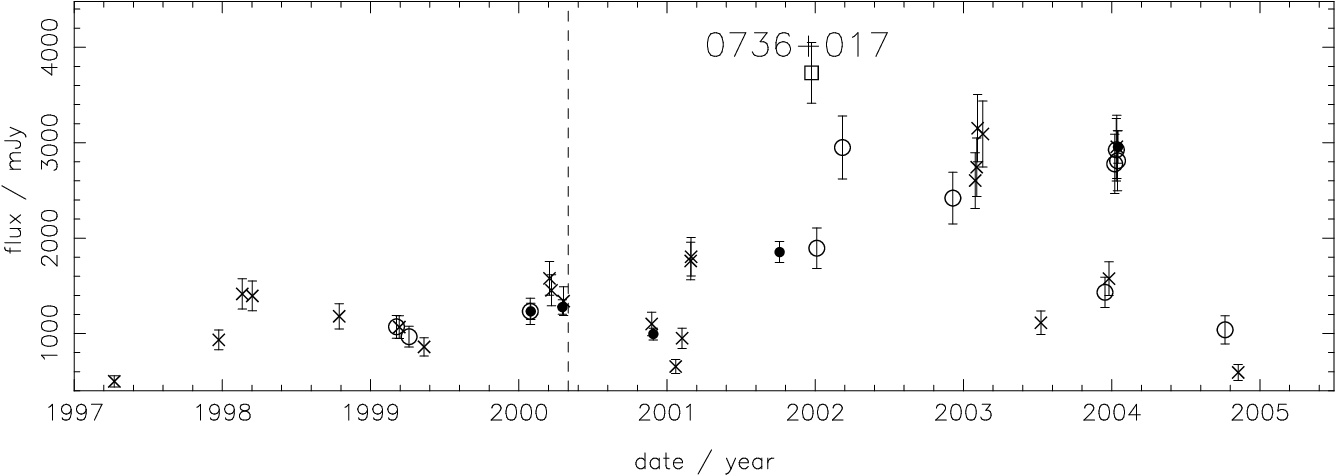}
&
\includegraphics[angle=-90,width=76mm]{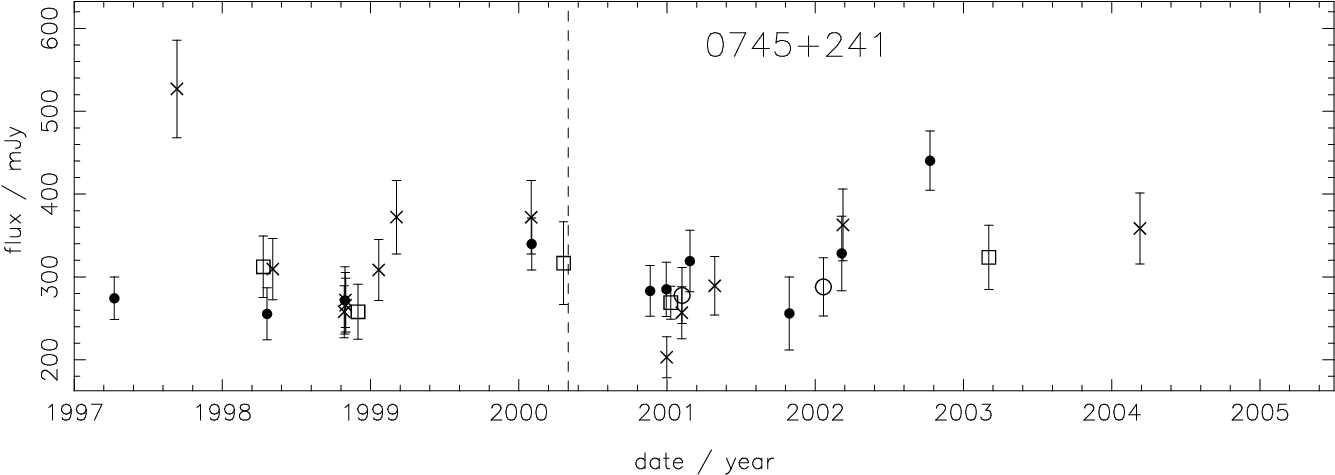}
\\
\includegraphics[angle=-90,width=76mm]{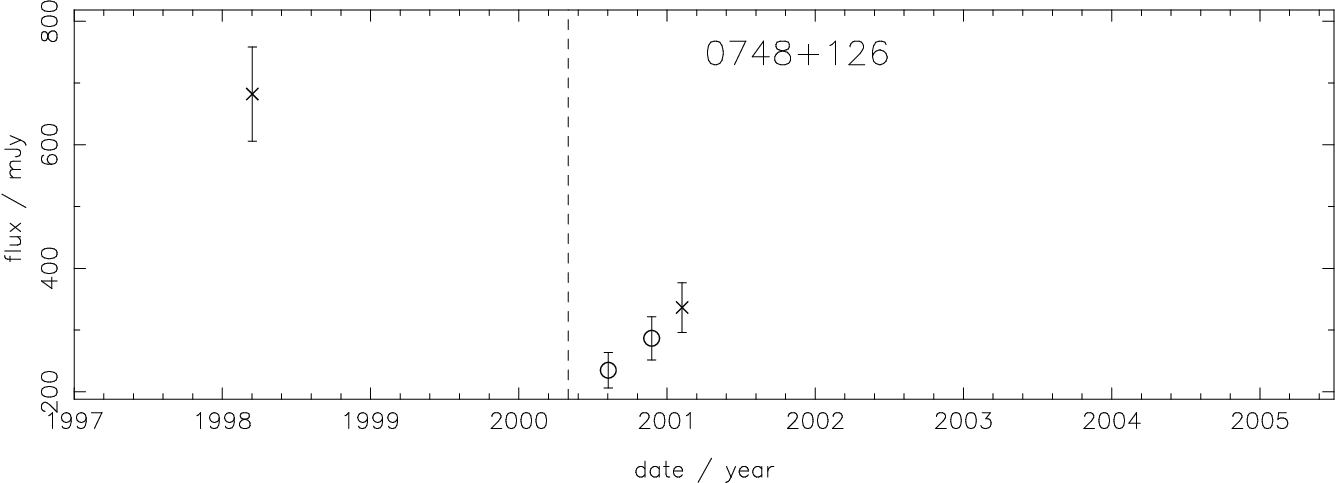}
&
\includegraphics[angle=-90,width=76mm]{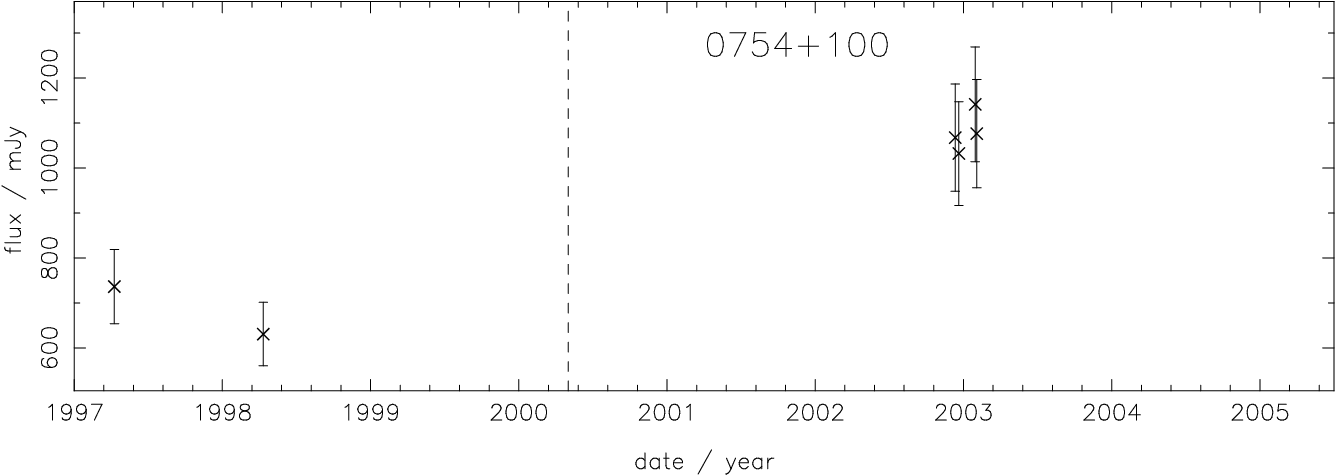}
\\
\includegraphics[angle=-90,width=76mm]{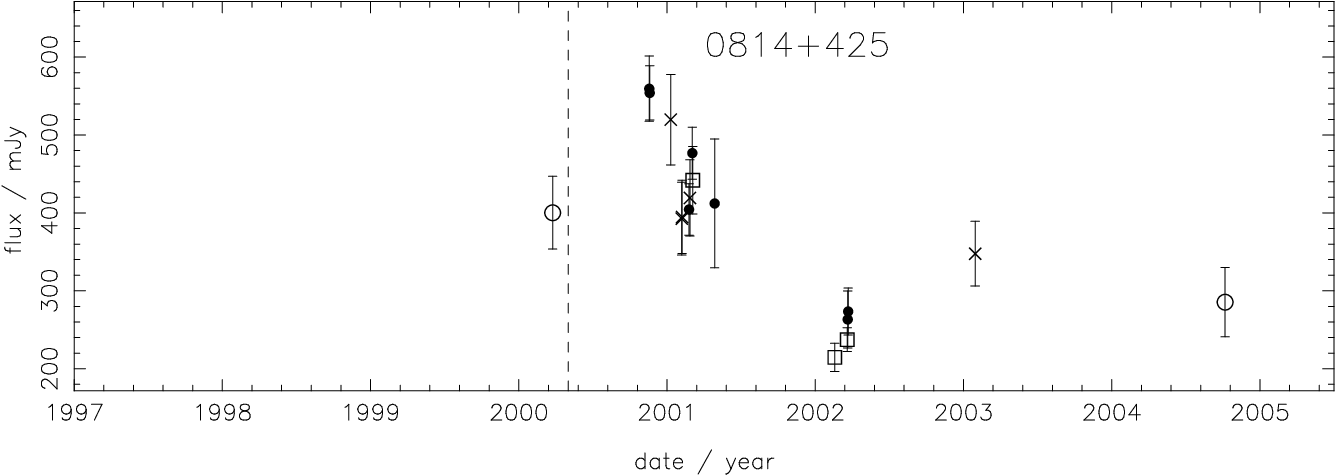}
&
\includegraphics[angle=-90,width=76mm]{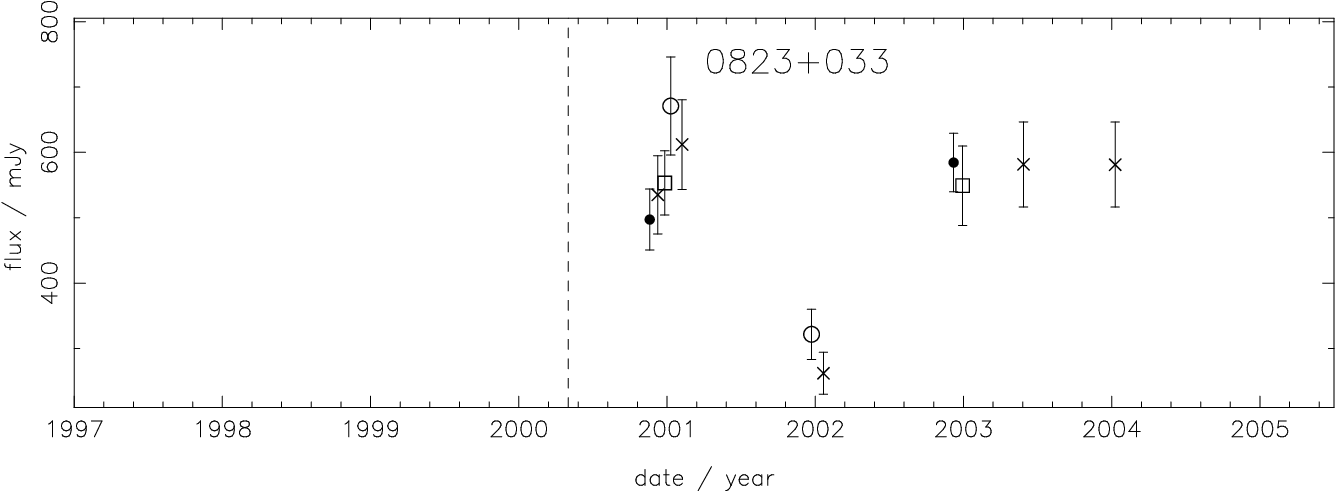}
\\
\end{tabular}
\contcaption{}
\label{fig:light-curves3}
\end{figure*}
\clearpage
\begin{figure*}
\begin{tabular}{cc}
\includegraphics[angle=-90,width=76mm]{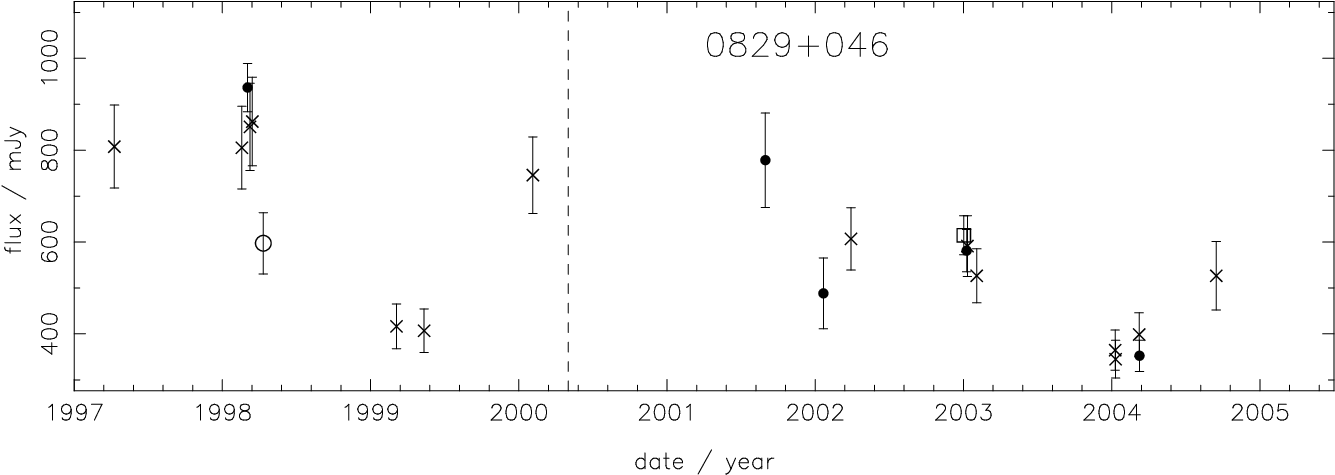}
&
\includegraphics[angle=-90,width=76mm]{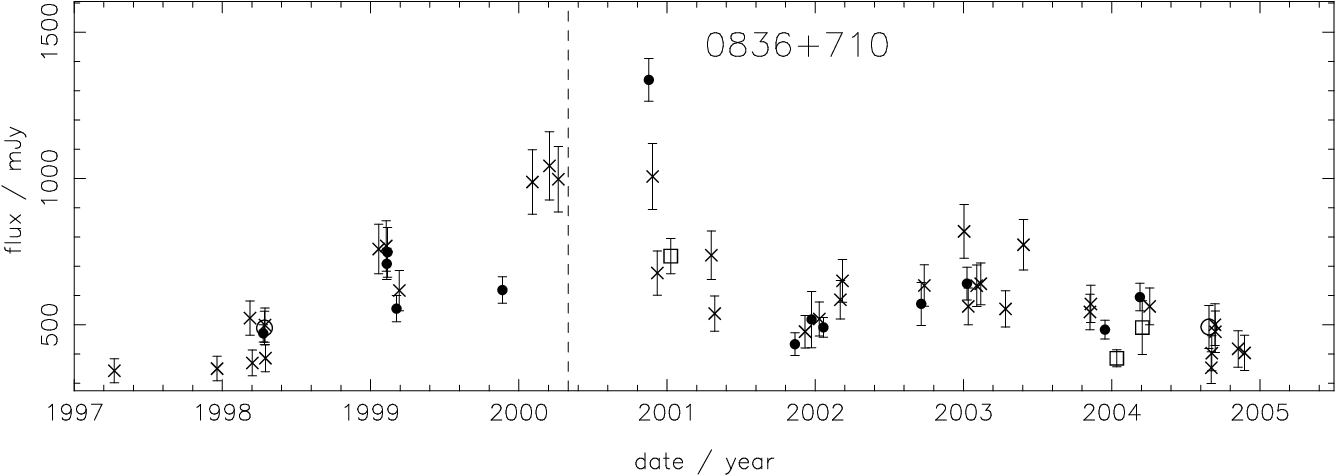}
\\
\includegraphics[angle=-90,width=76mm]{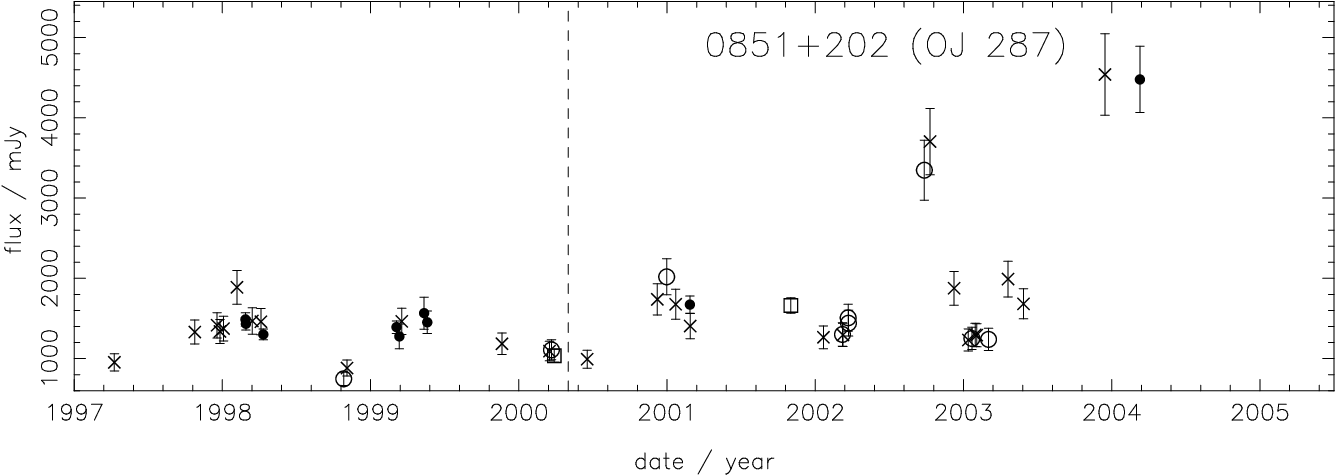}
&
\includegraphics[angle=-90,width=76mm]{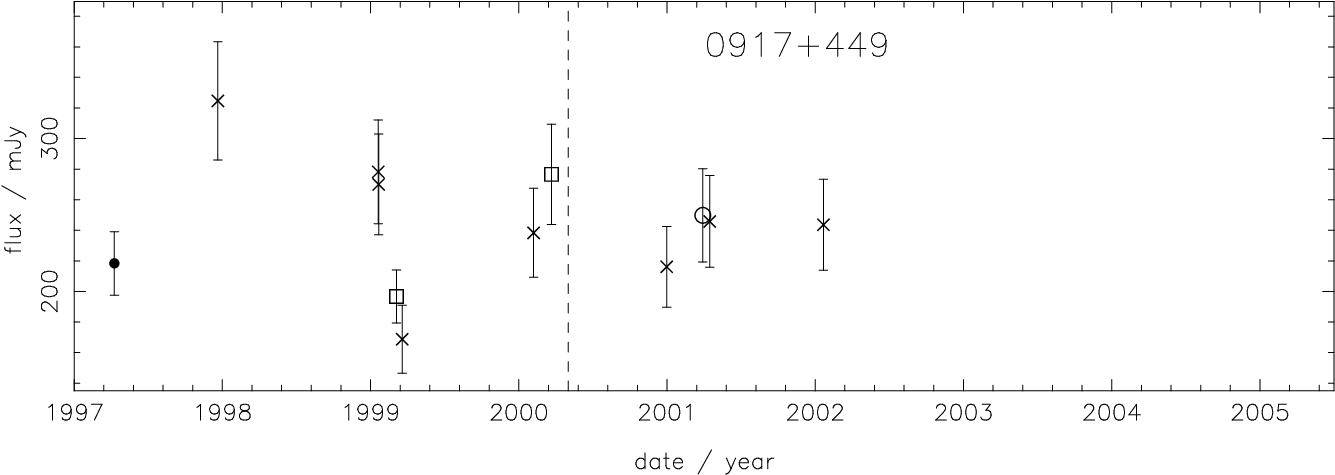}
\\
\includegraphics[angle=-90,width=76mm]{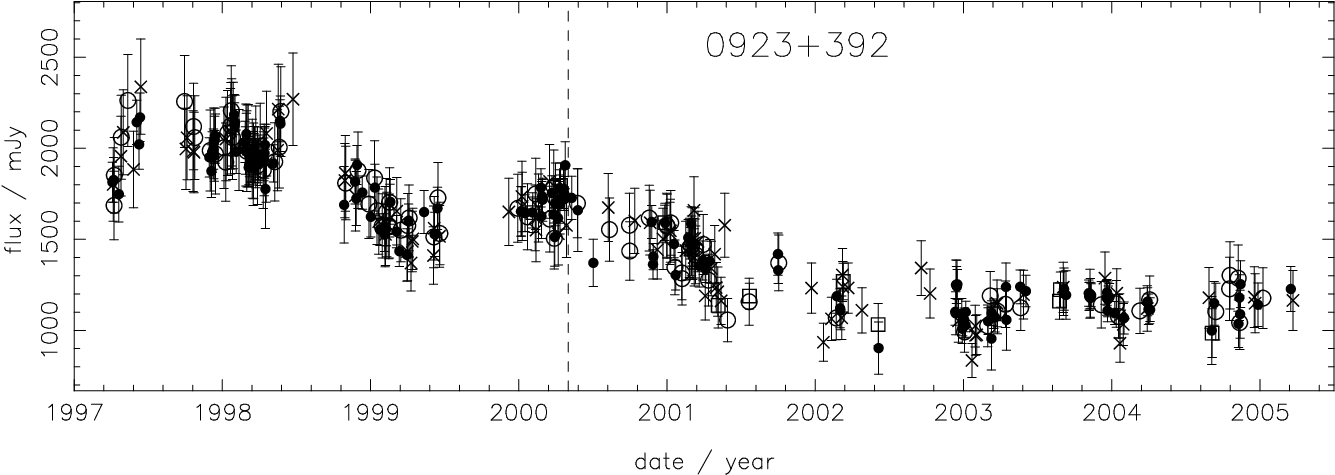}
&
\includegraphics[angle=-90,width=76mm]{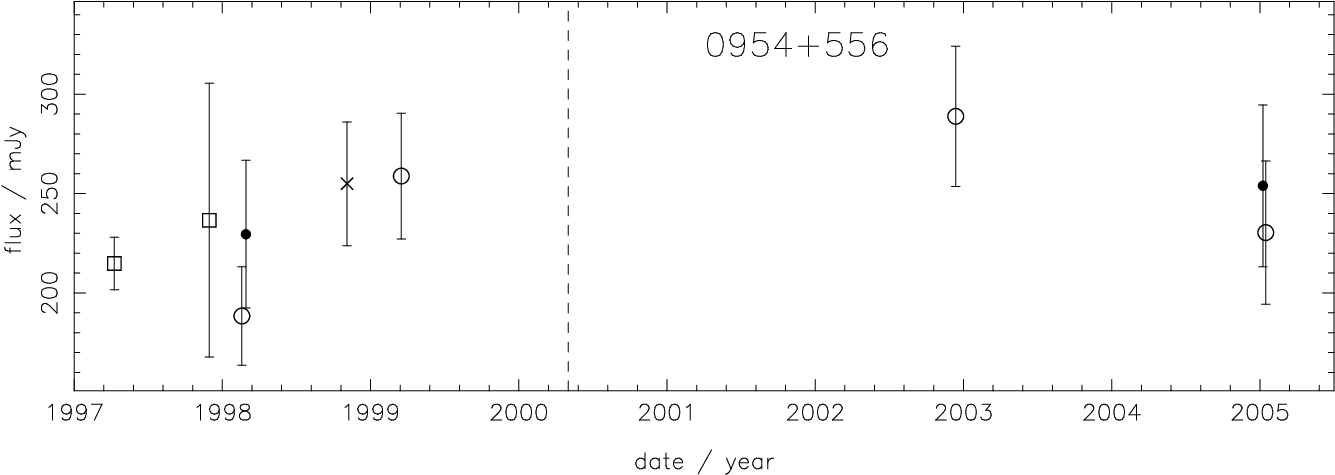}
\\
\includegraphics[angle=-90,width=76mm]{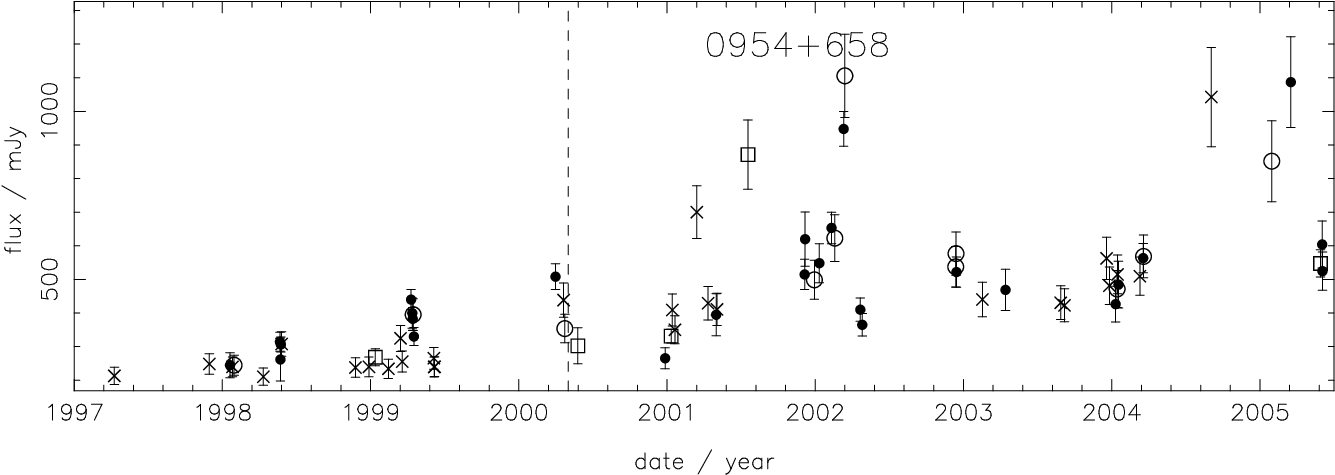}
&
\includegraphics[angle=-90,width=76mm]{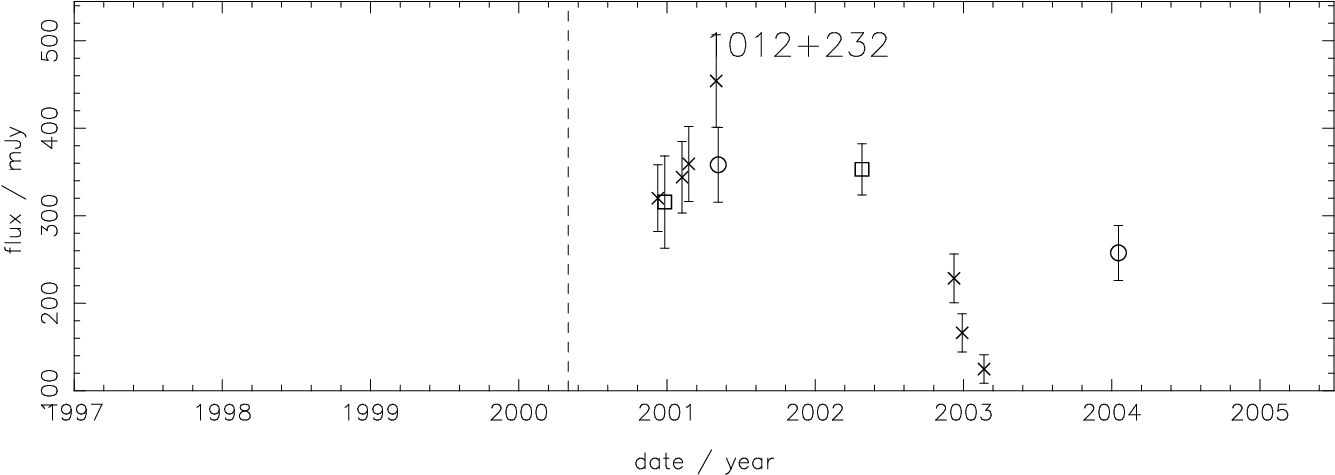}
\\
\includegraphics[angle=-90,width=76mm]{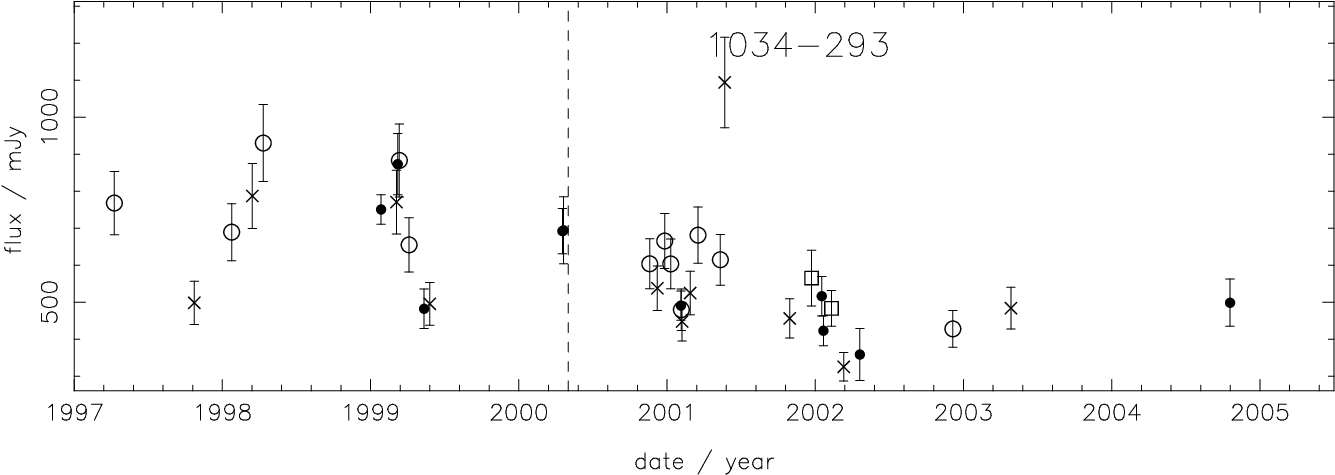}
&
\includegraphics[angle=-90,width=76mm]{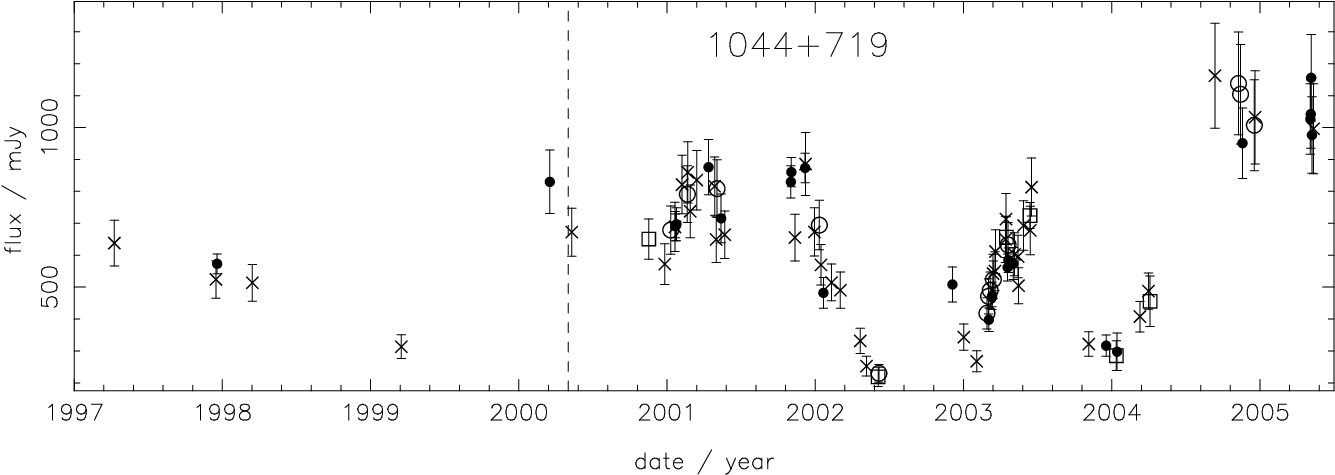}
\\
\includegraphics[angle=-90,width=76mm]{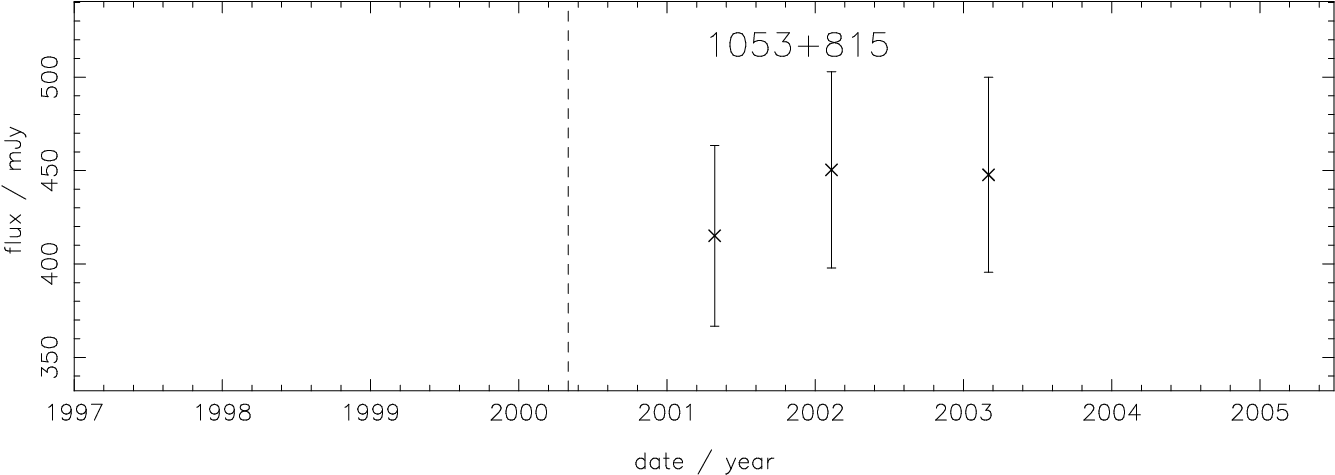}
&
\includegraphics[angle=-90,width=76mm]{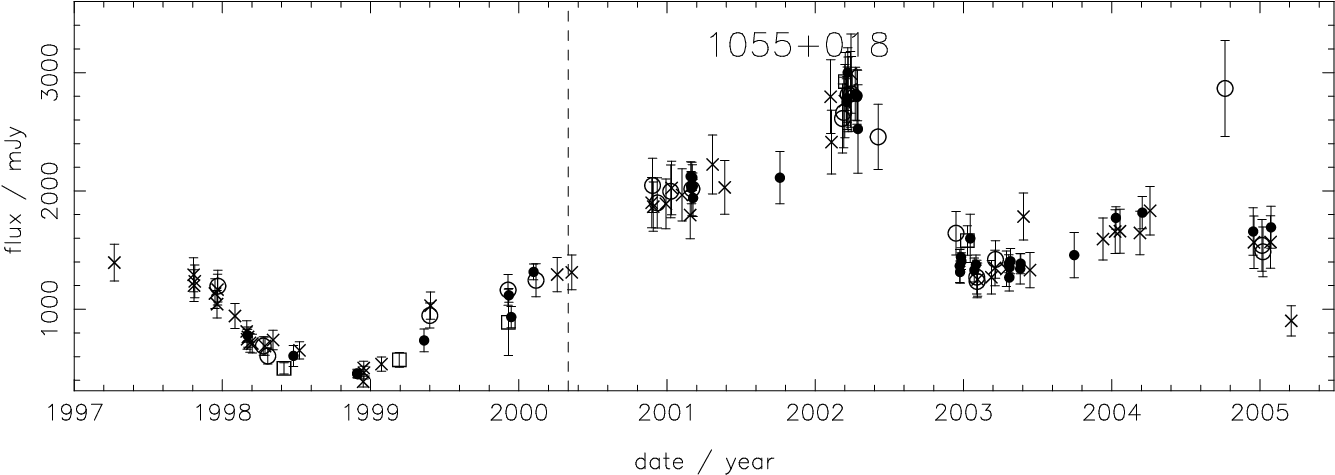}
\\
\includegraphics[angle=-90,width=76mm]{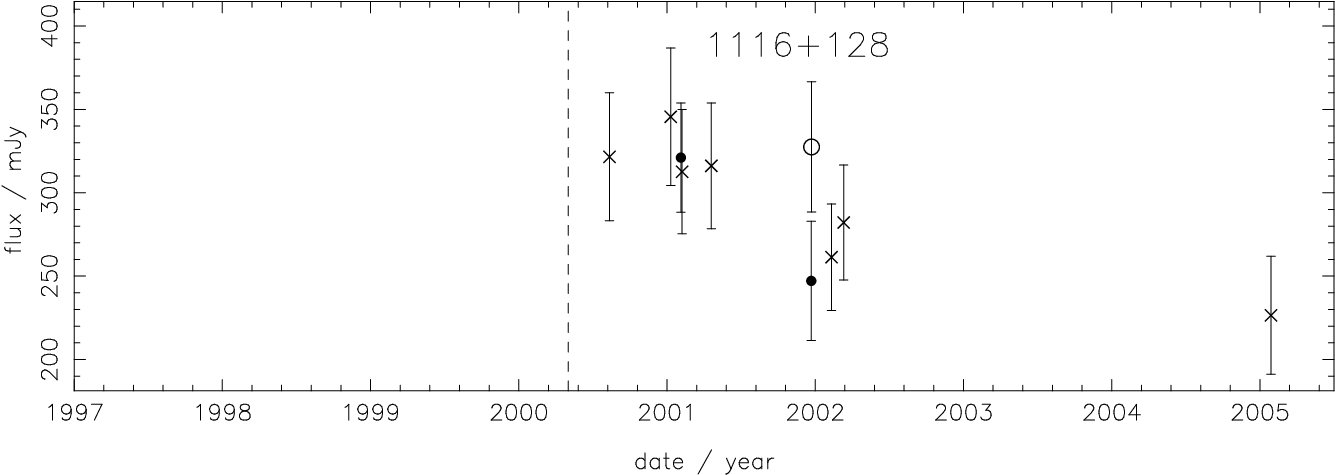}
&
\includegraphics[angle=-90,width=76mm]{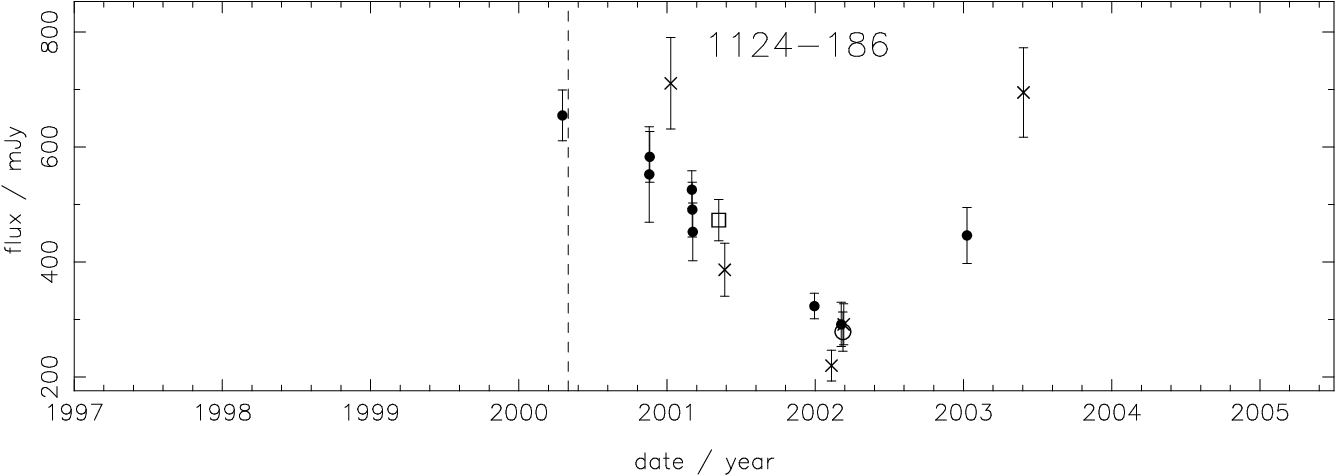}
\\
\end{tabular}
\contcaption{}
\label{fig:light-curves4}
\end{figure*}
\clearpage
\begin{figure*}
\begin{tabular}{cc}
\includegraphics[angle=-90,width=76mm]{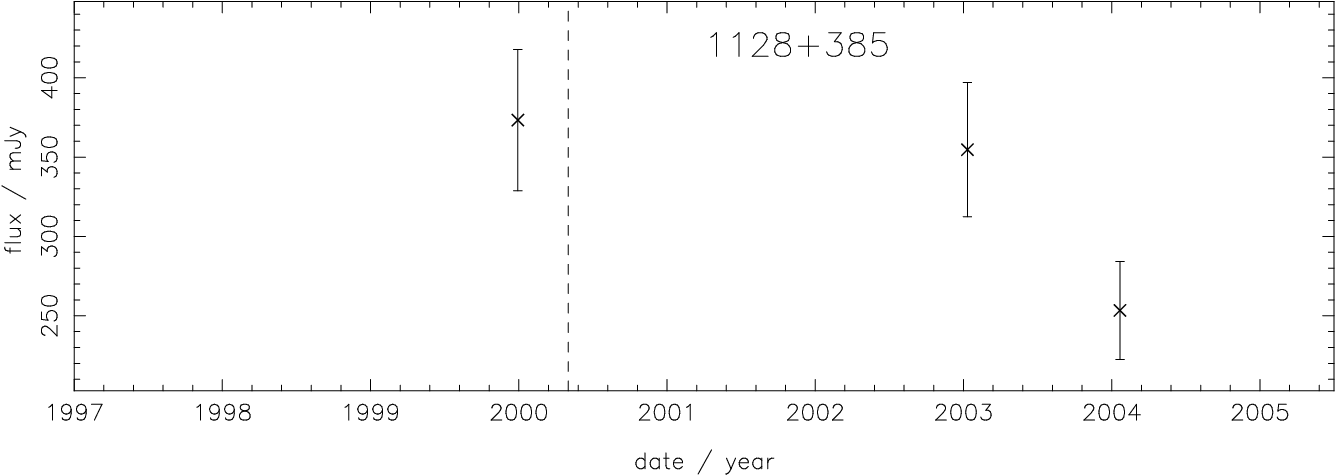}
&
\includegraphics[angle=-90,width=76mm]{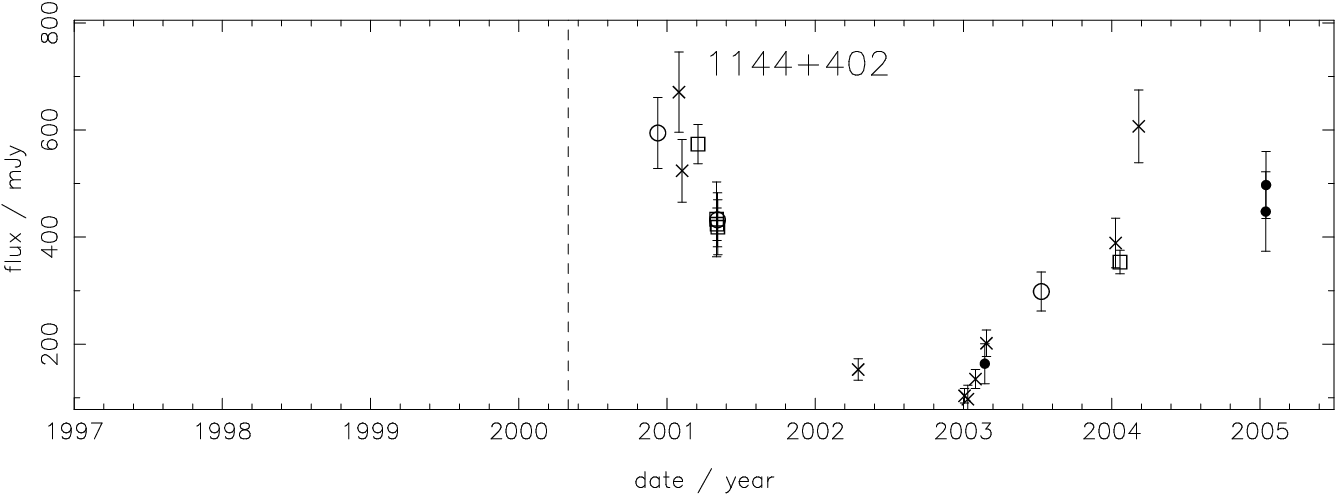}
\\
\includegraphics[angle=-90,width=76mm]{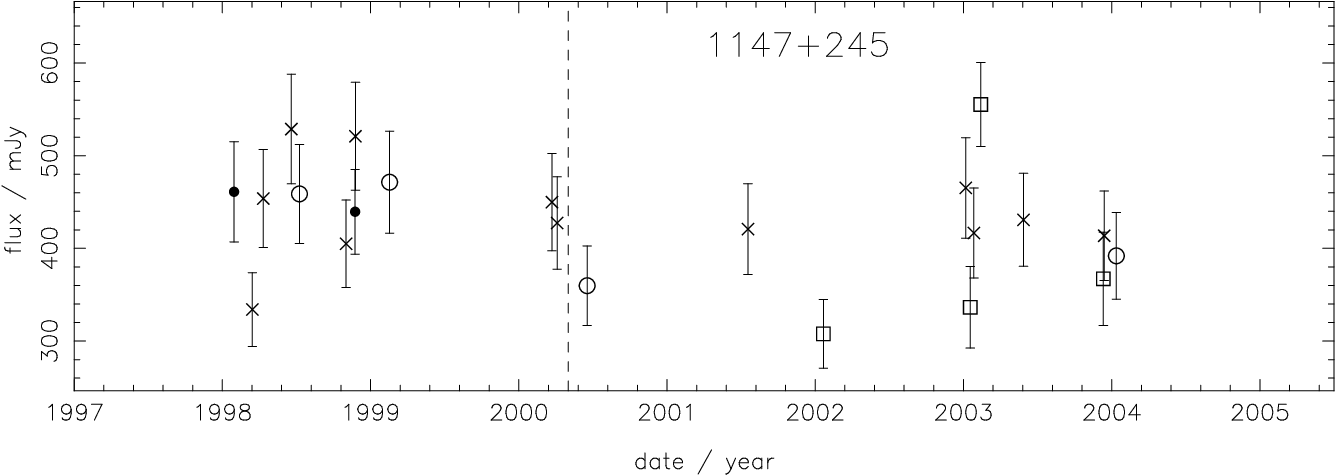}
&
\includegraphics[angle=-90,width=76mm]{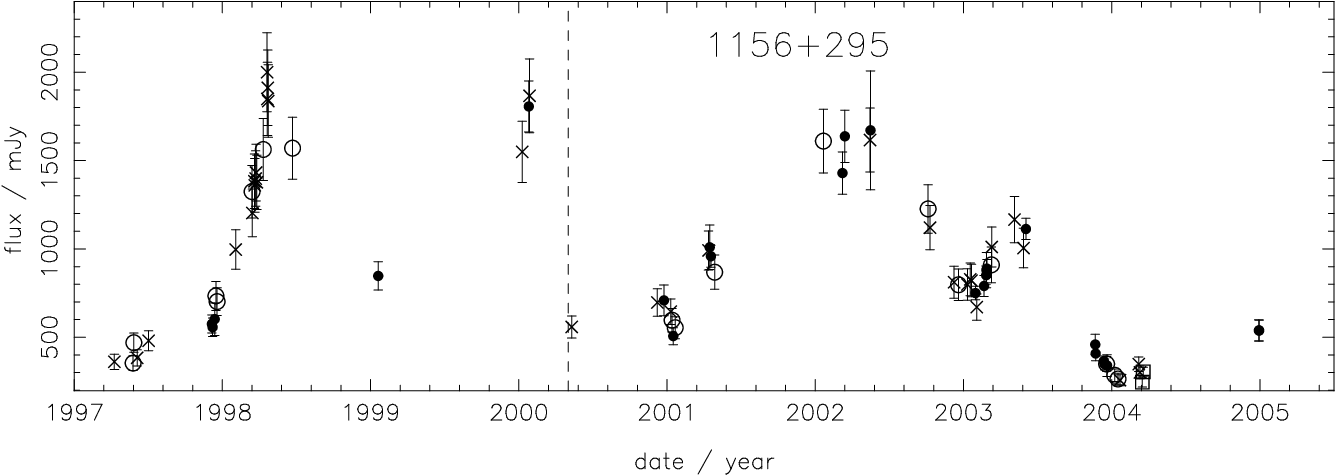}
\\
\includegraphics[angle=-90,width=76mm]{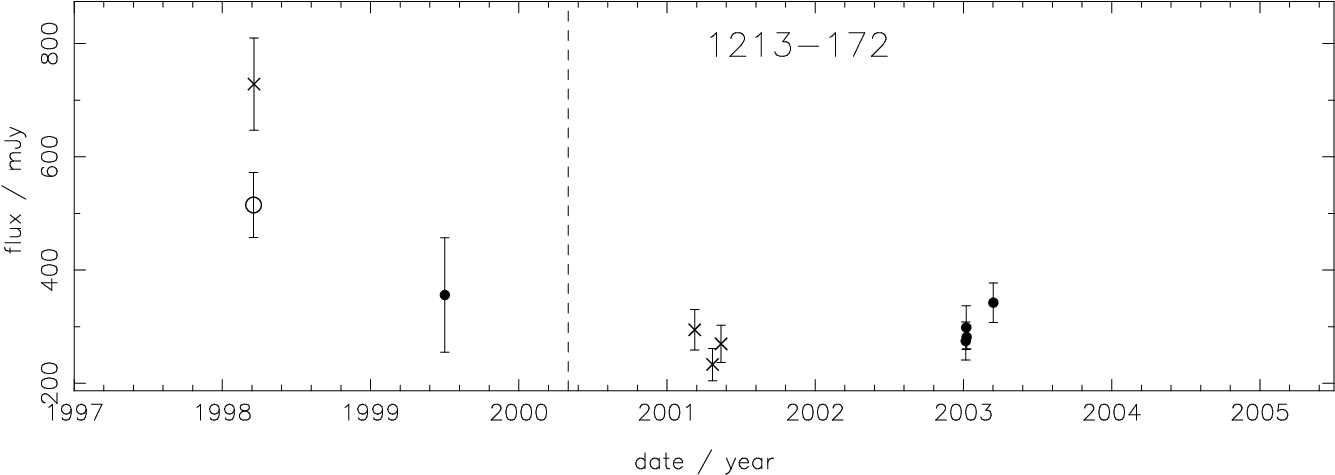}
&
\includegraphics[angle=-90,width=76mm]{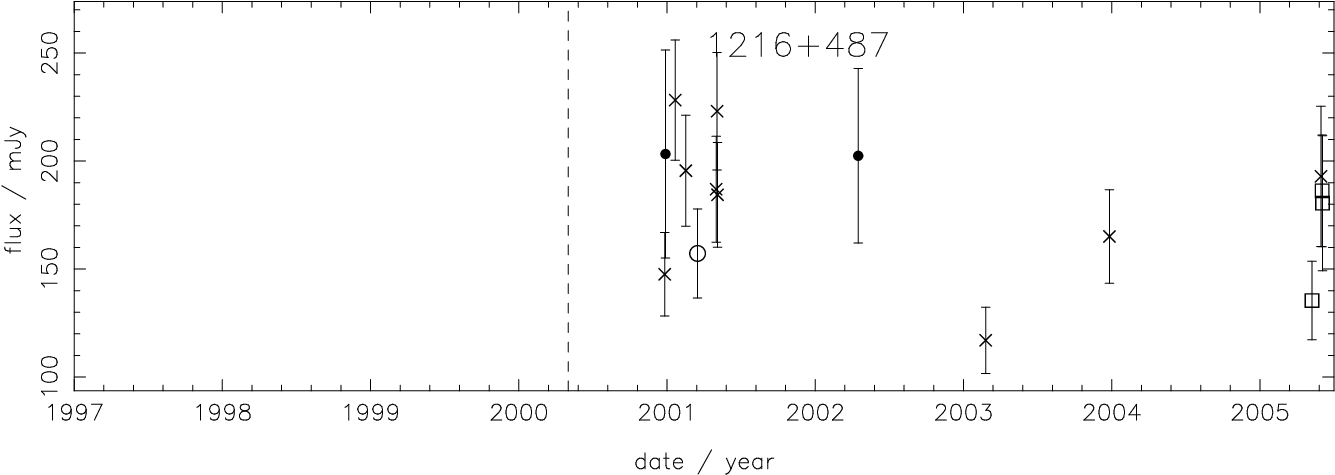}
\\
\includegraphics[angle=-90,width=76mm]{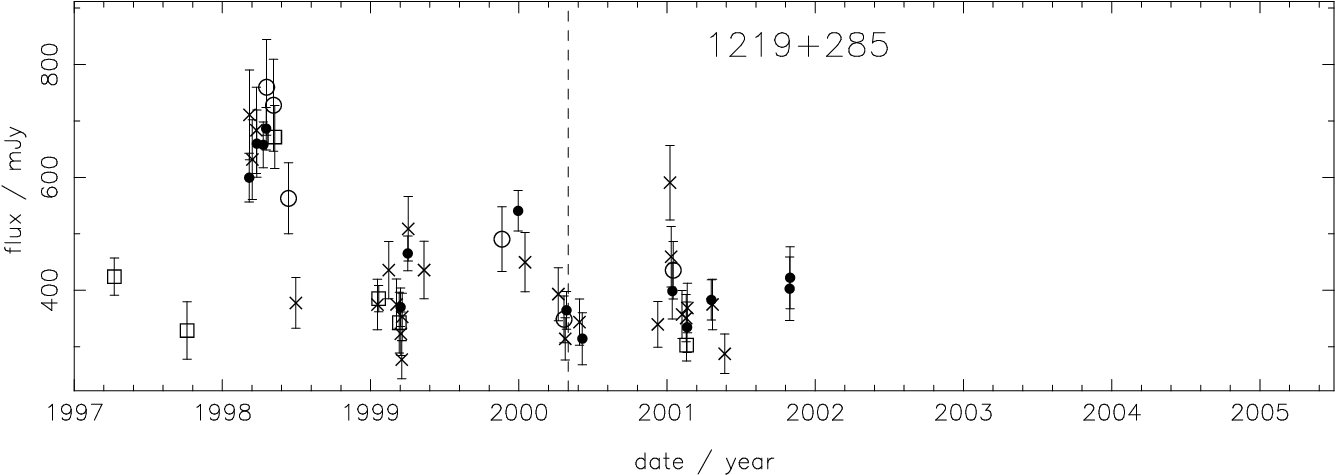}
&
\includegraphics[angle=-90,width=76mm]{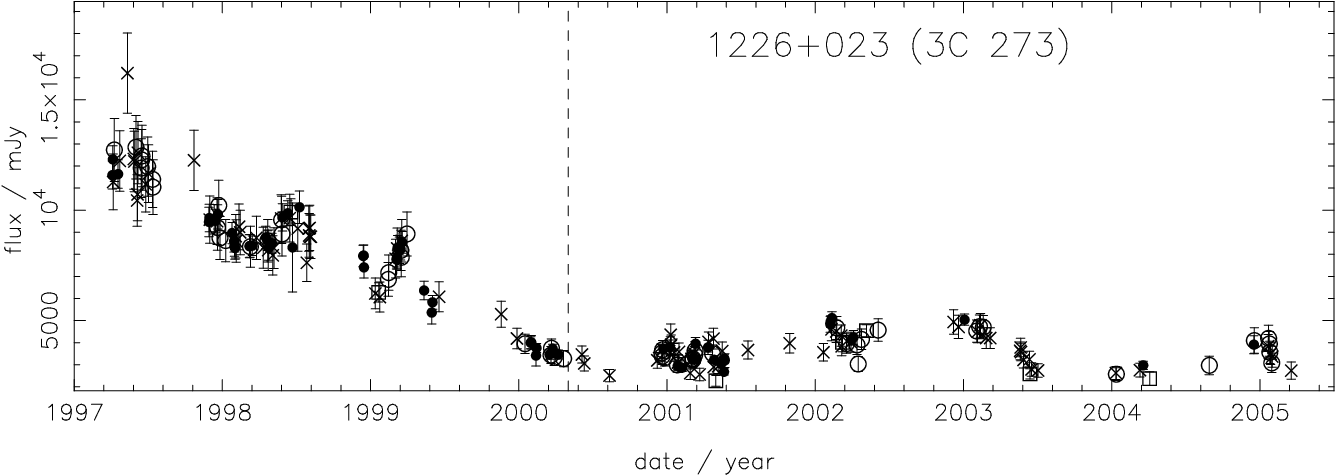}
\\
\includegraphics[angle=-90,width=76mm]{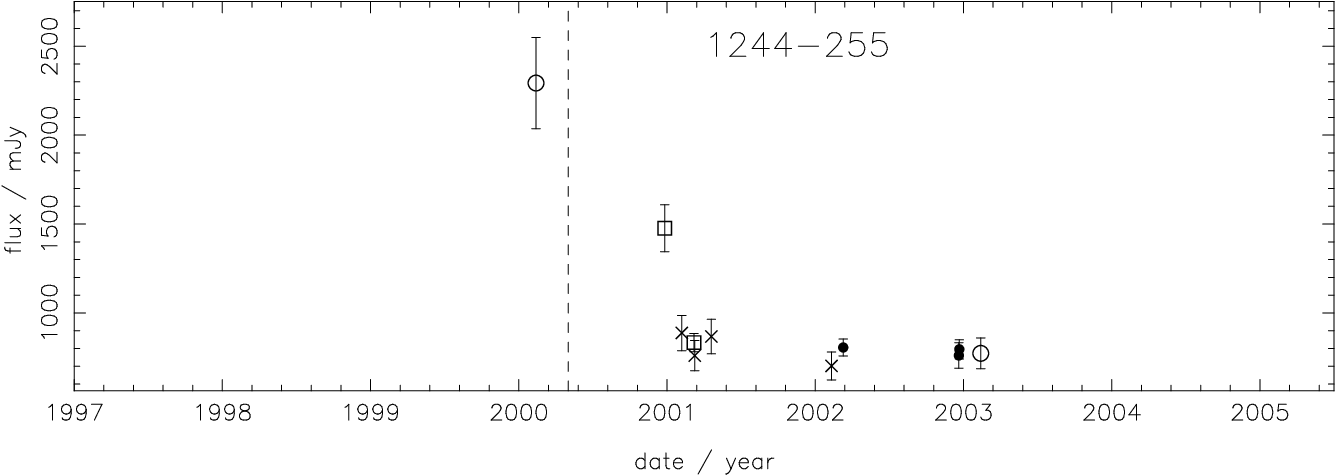}
&
\includegraphics[angle=-90,width=76mm]{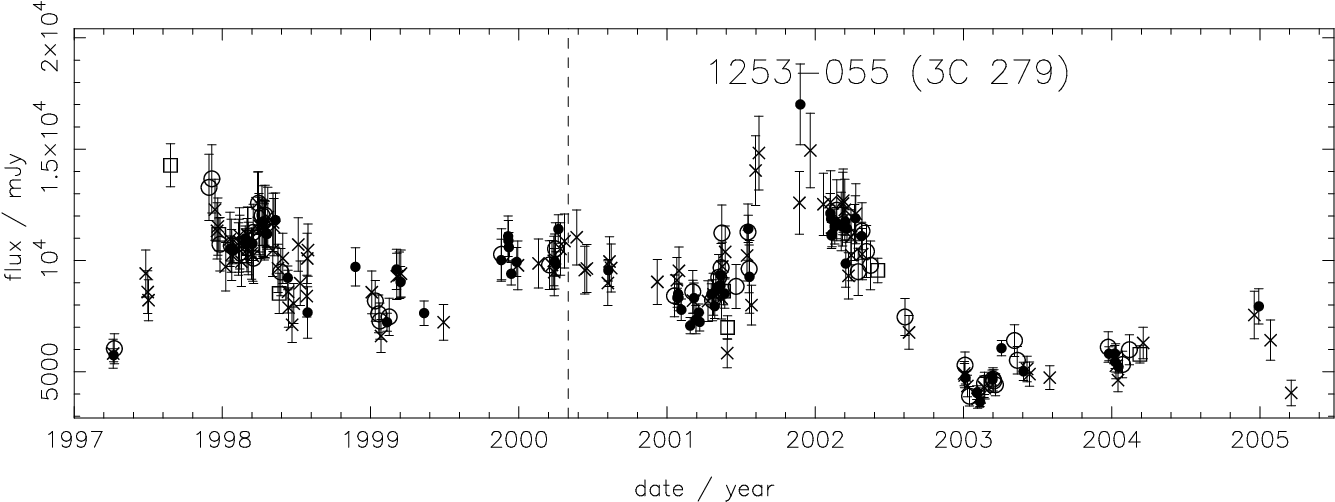}
\\
\includegraphics[angle=-90,width=76mm]{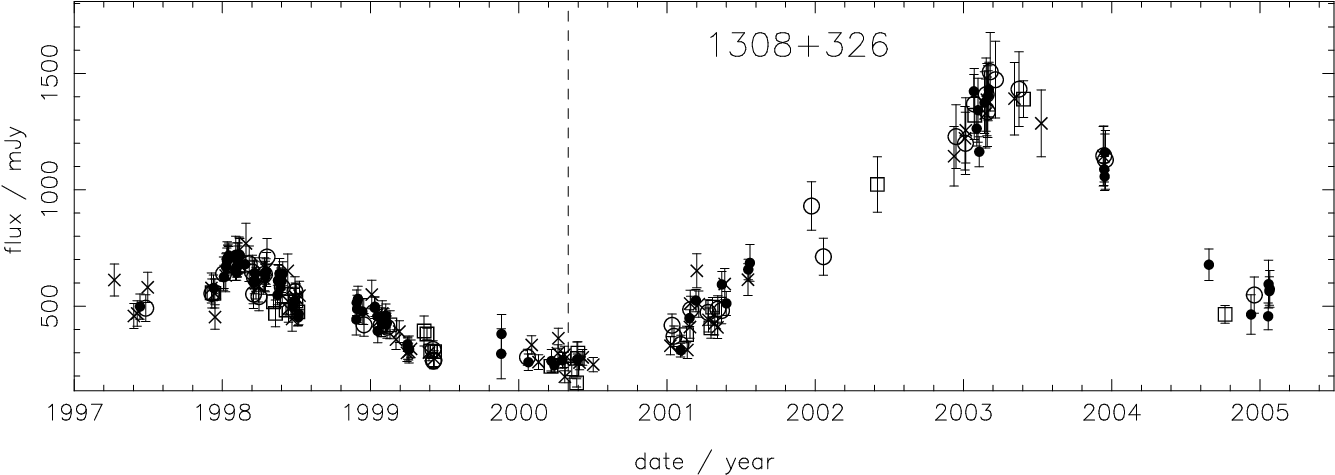}
&
\includegraphics[angle=-90,width=76mm]{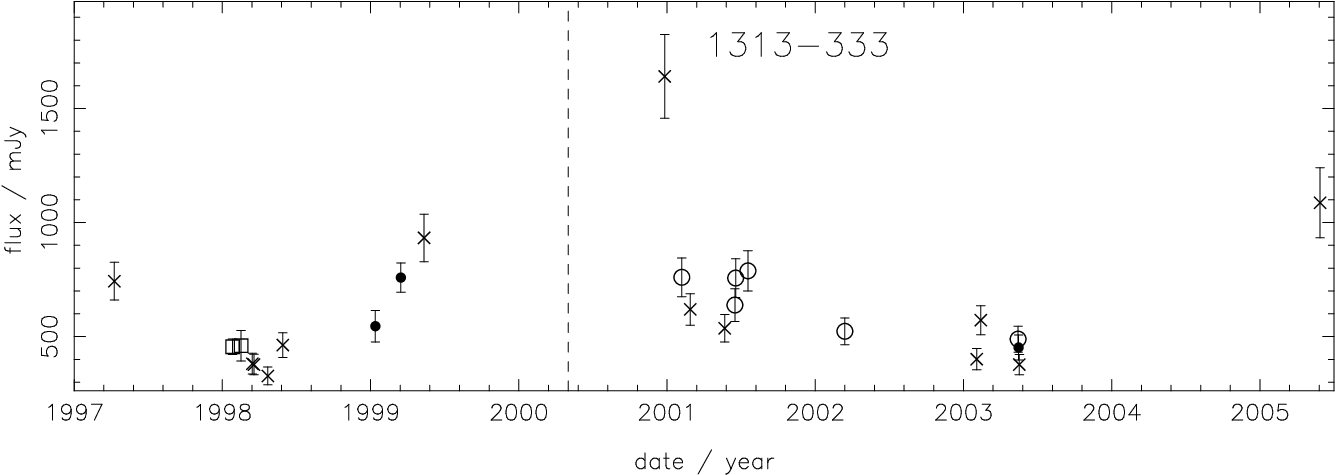}
\\
\includegraphics[angle=-90,width=76mm]{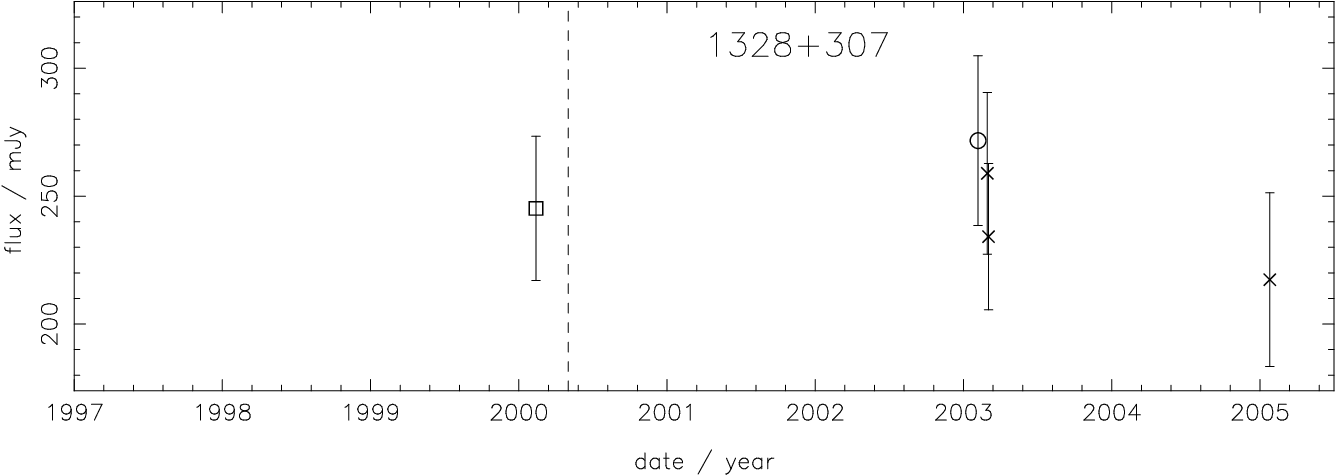}
&
\includegraphics[angle=-90,width=76mm]{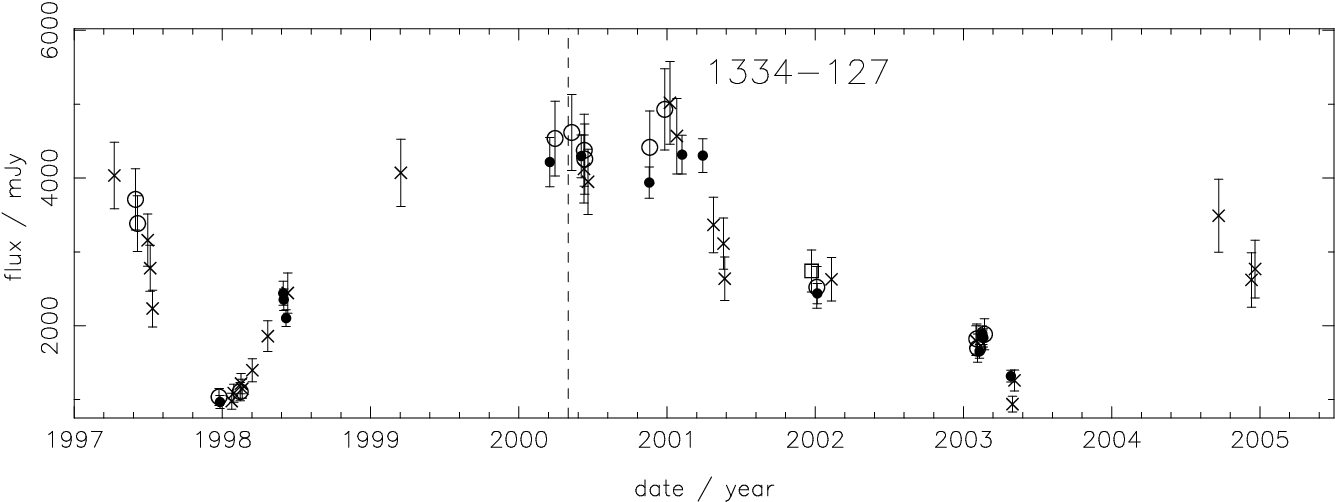}
\\
\end{tabular}
\contcaption{}
\label{fig:light-curves5}
\end{figure*}
\clearpage
\begin{figure*}
\begin{tabular}{cc}
\includegraphics[angle=-90,width=76mm]{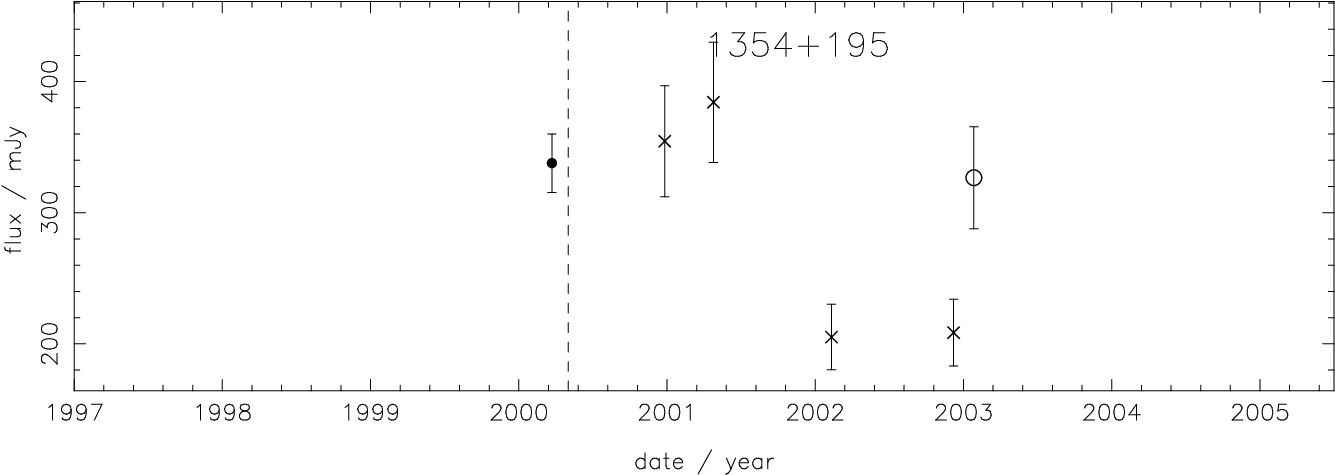}
&
\includegraphics[angle=-90,width=76mm]{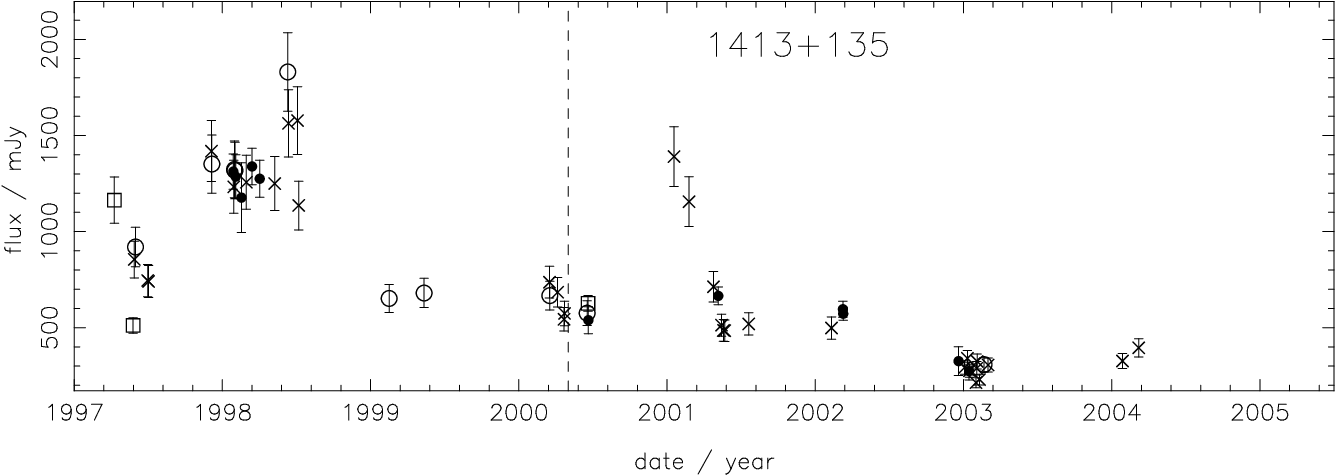}
\\
\includegraphics[angle=-90,width=76mm]{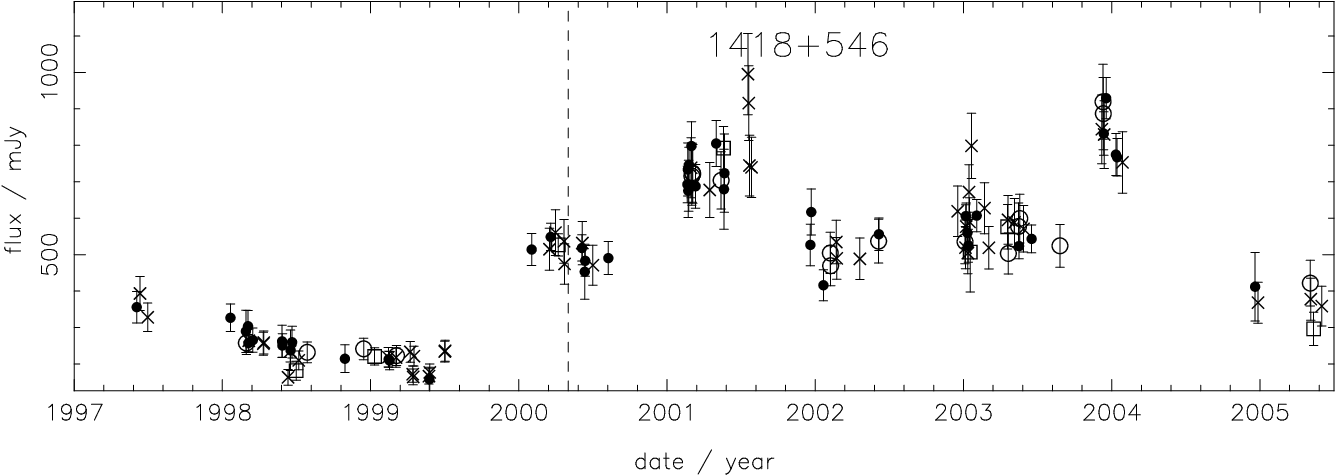}
&
\includegraphics[angle=-90,width=76mm]{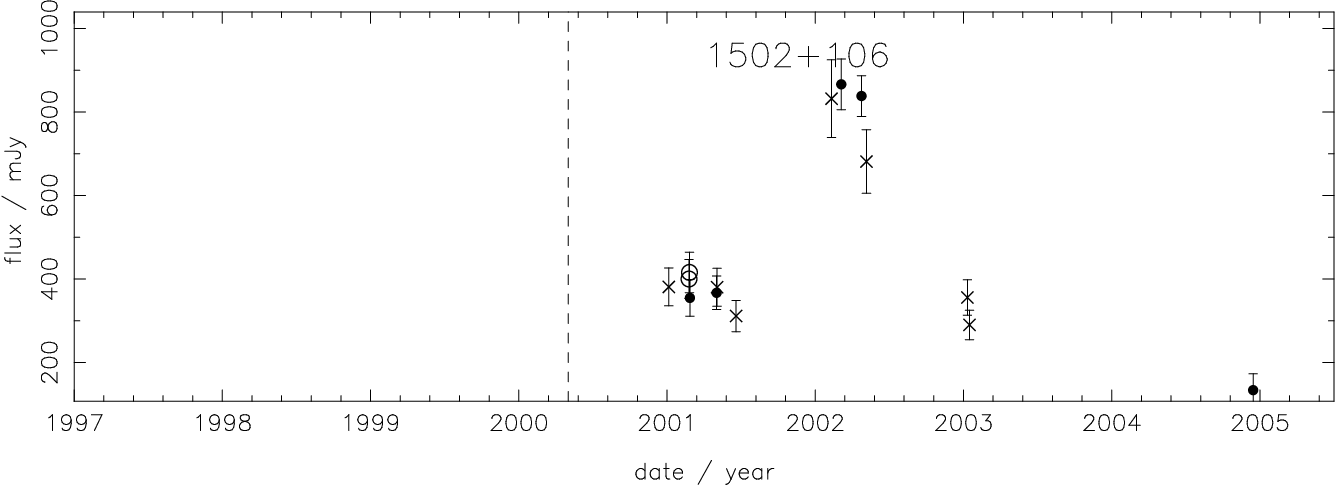}
\\
\includegraphics[angle=-90,width=76mm]{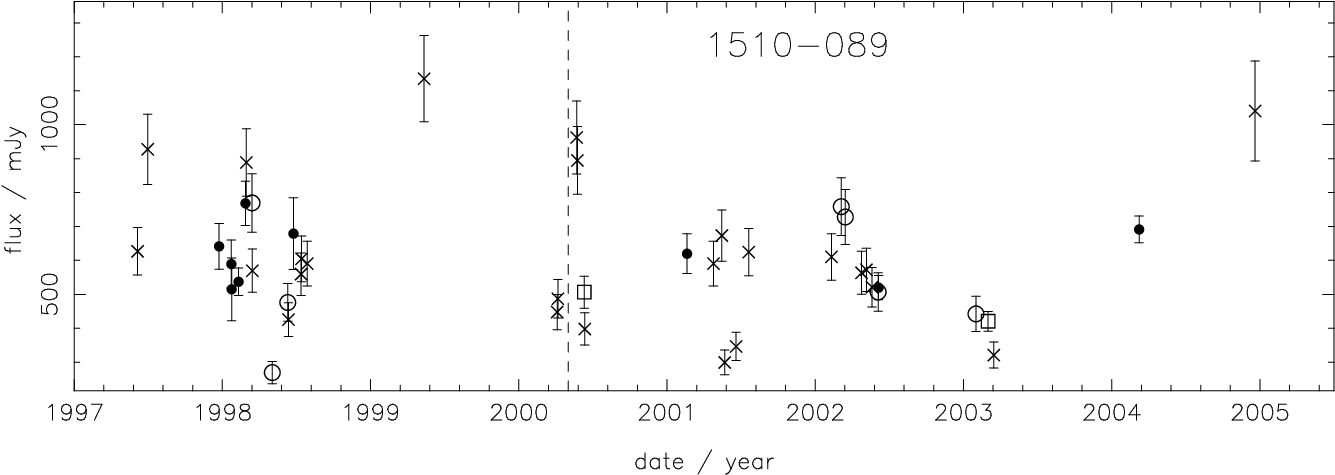}
&
\includegraphics[angle=-90,width=76mm]{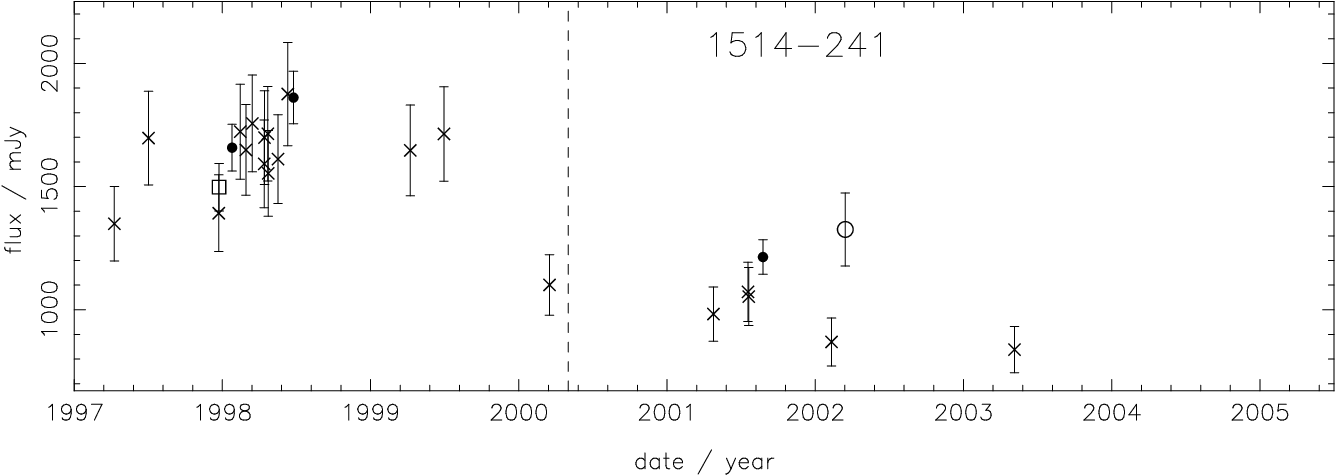}
\\
\includegraphics[angle=-90,width=76mm]{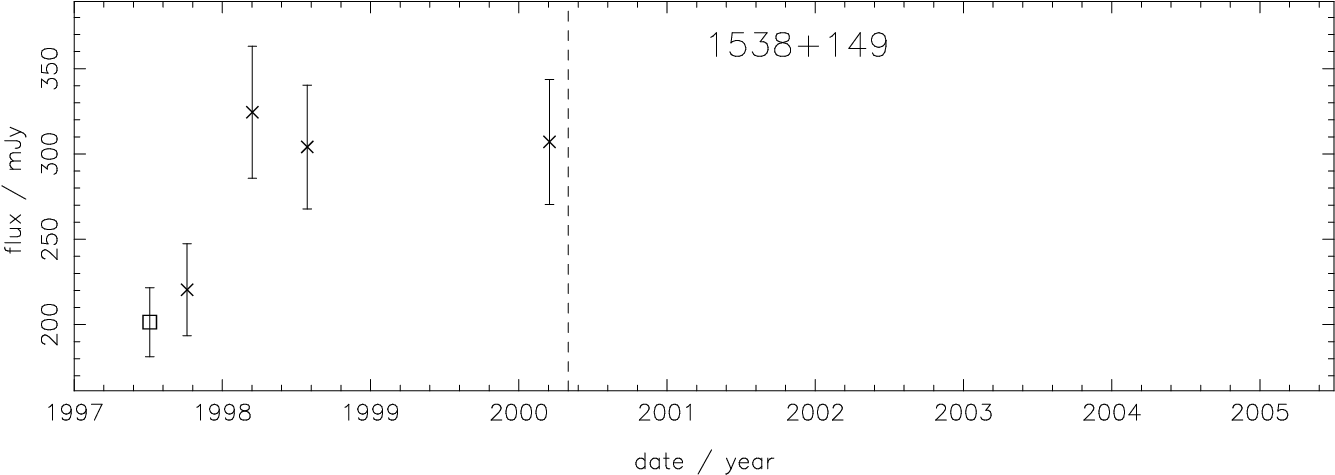}
&
\includegraphics[angle=-90,width=76mm]{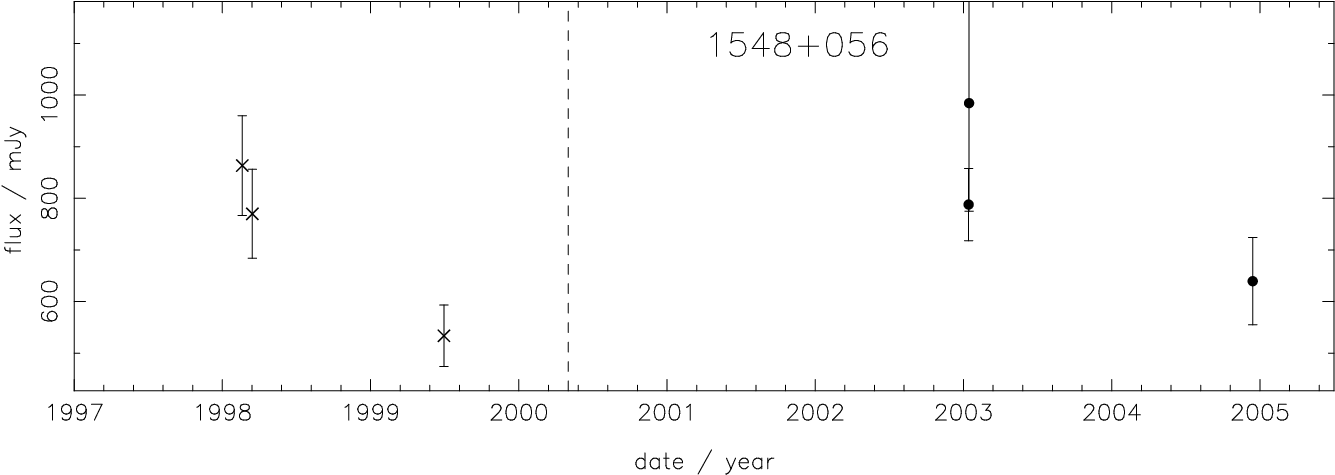}
\\
\includegraphics[angle=-90,width=76mm]{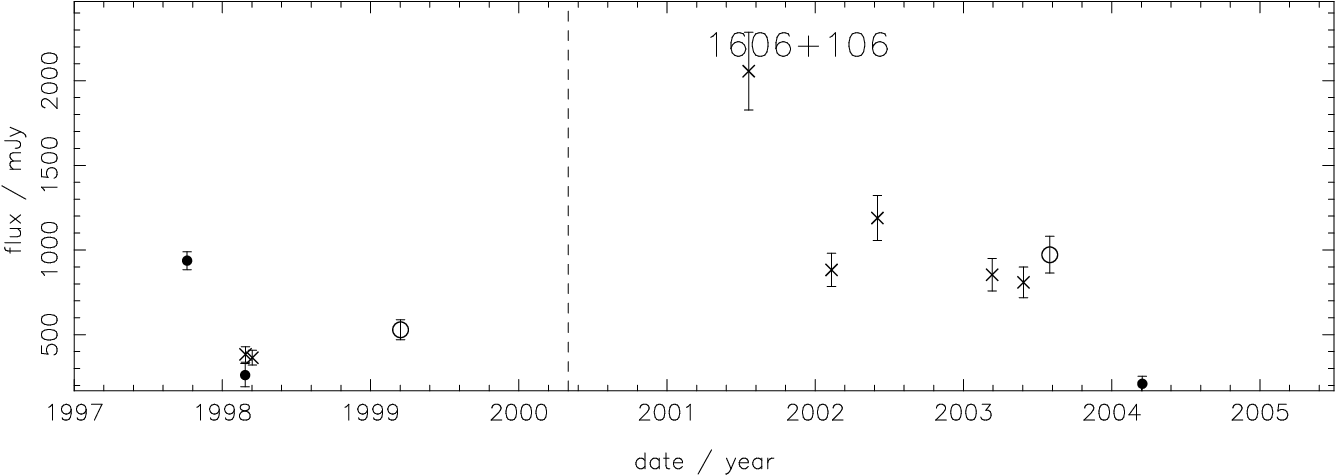}
&
\includegraphics[angle=-90,width=76mm]{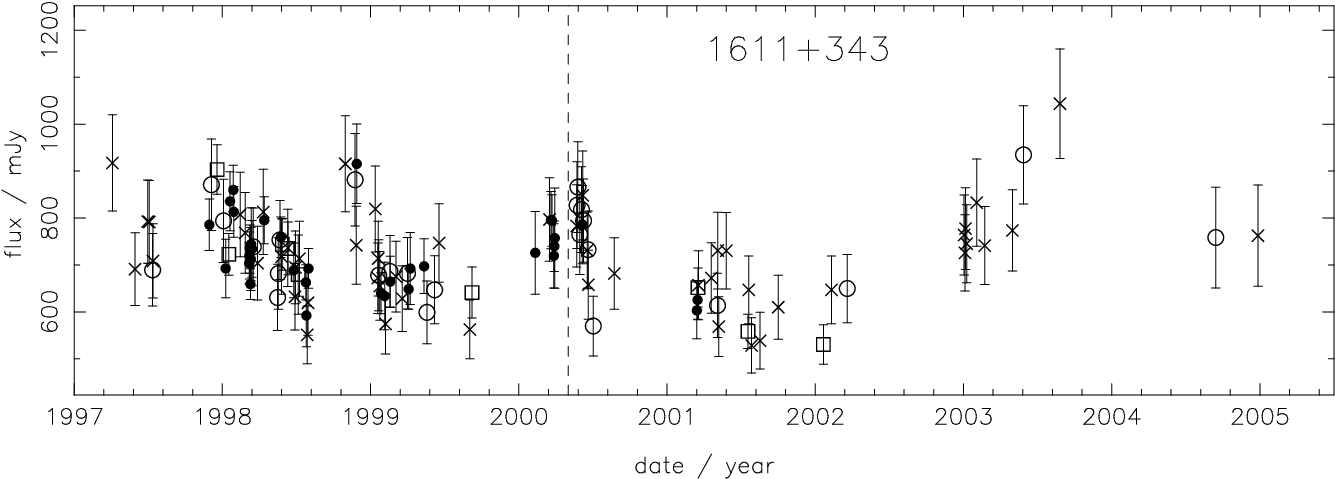}
\\
\includegraphics[angle=-90,width=76mm]{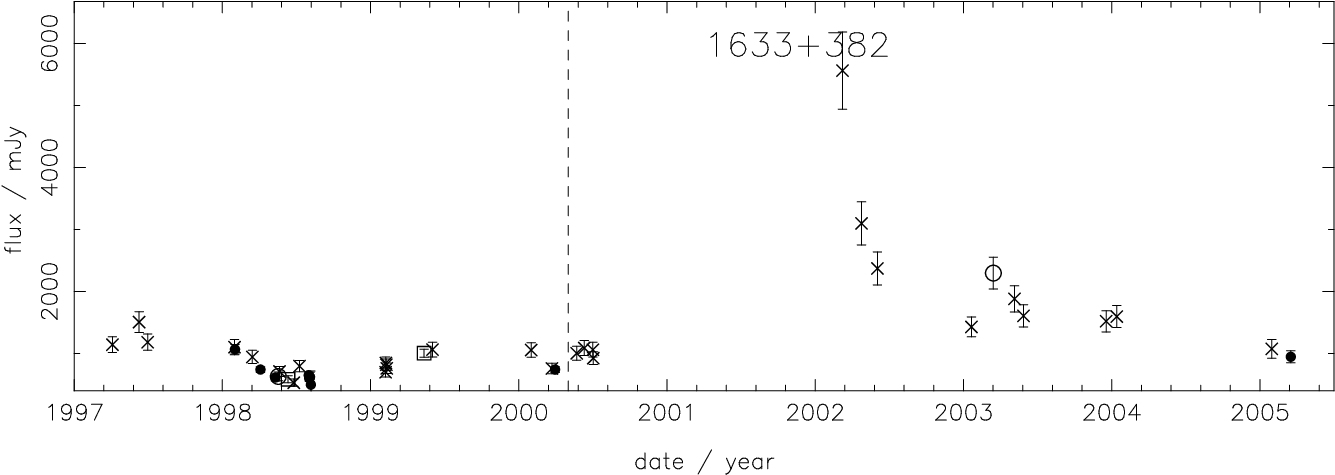}
&
\includegraphics[angle=-90,width=76mm]{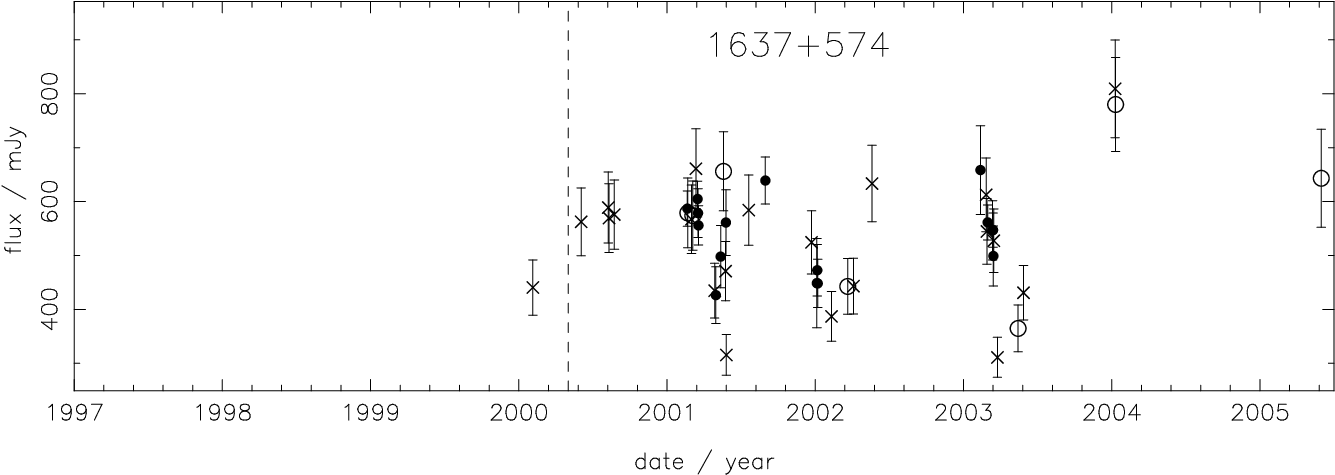}
\\
\includegraphics[angle=-90,width=76mm]{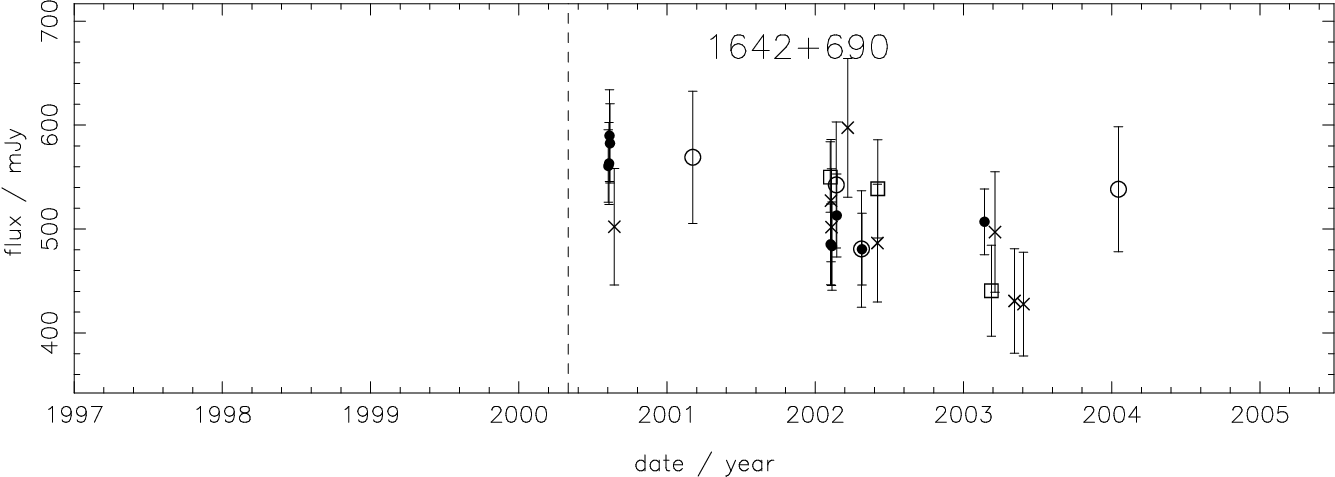}
&
\includegraphics[angle=-90,width=76mm]{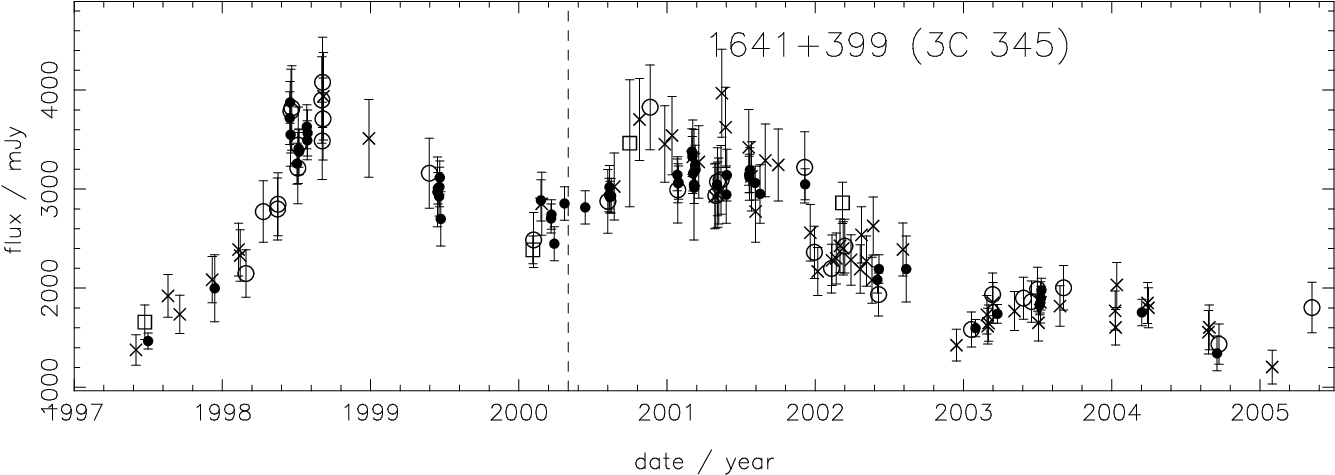}
\\
\end{tabular}
\contcaption{}
\label{fig:light-curves6}
\end{figure*}
\clearpage
\begin{figure*}
\begin{tabular}{cc}
\includegraphics[angle=-90,width=76mm]{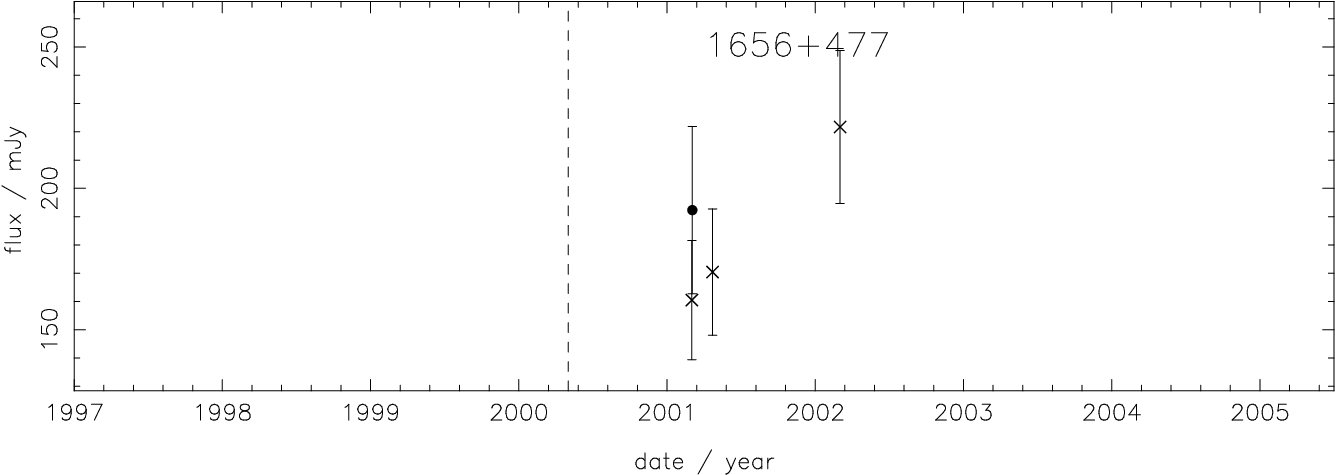}
&
\includegraphics[angle=-90,width=76mm]{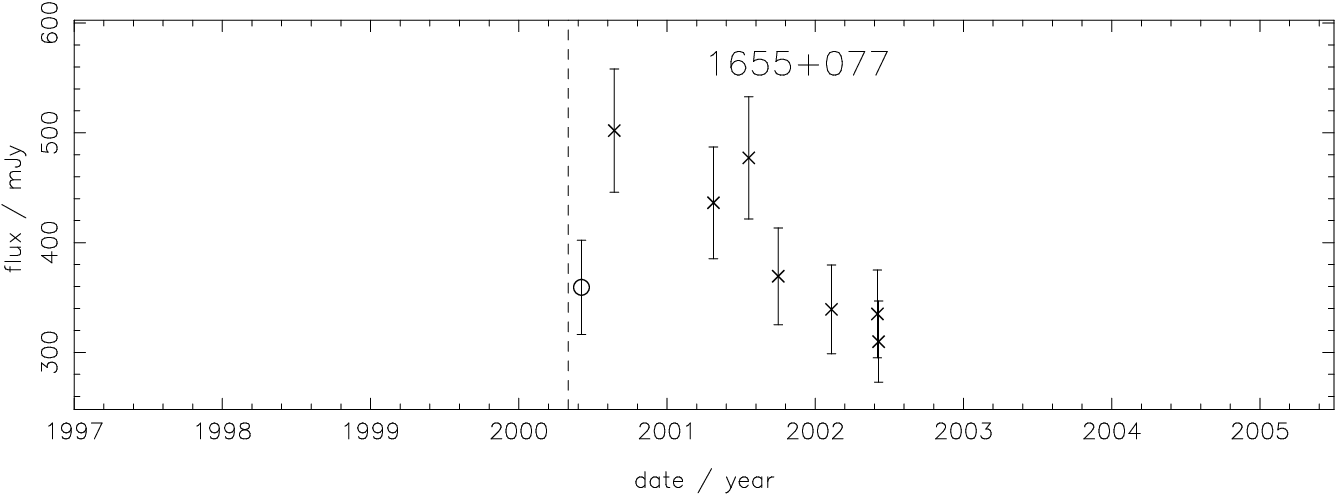}
\\
\includegraphics[angle=-90,width=76mm]{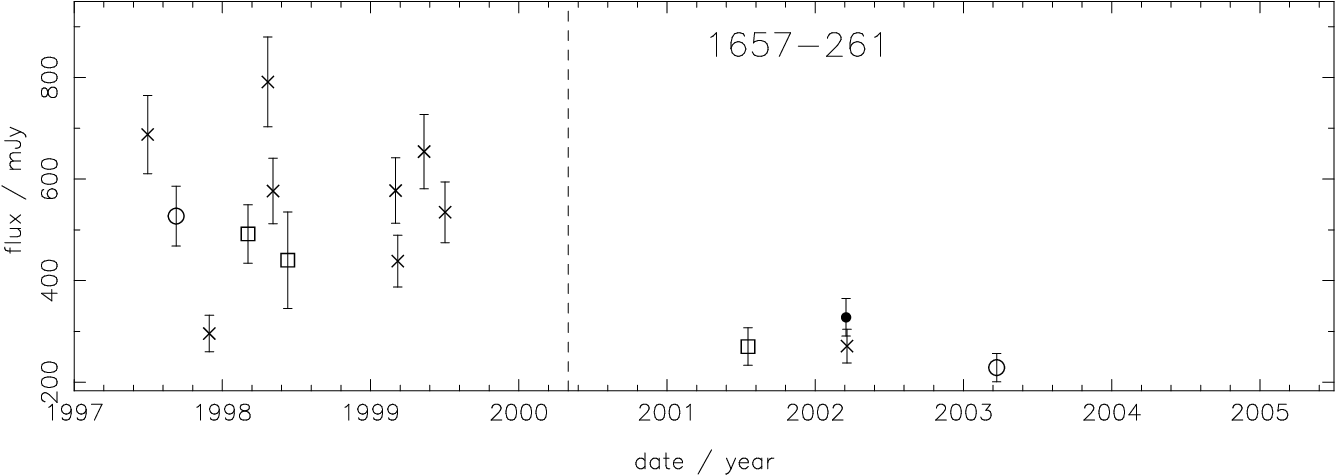}
&
\includegraphics[angle=-90,width=76mm]{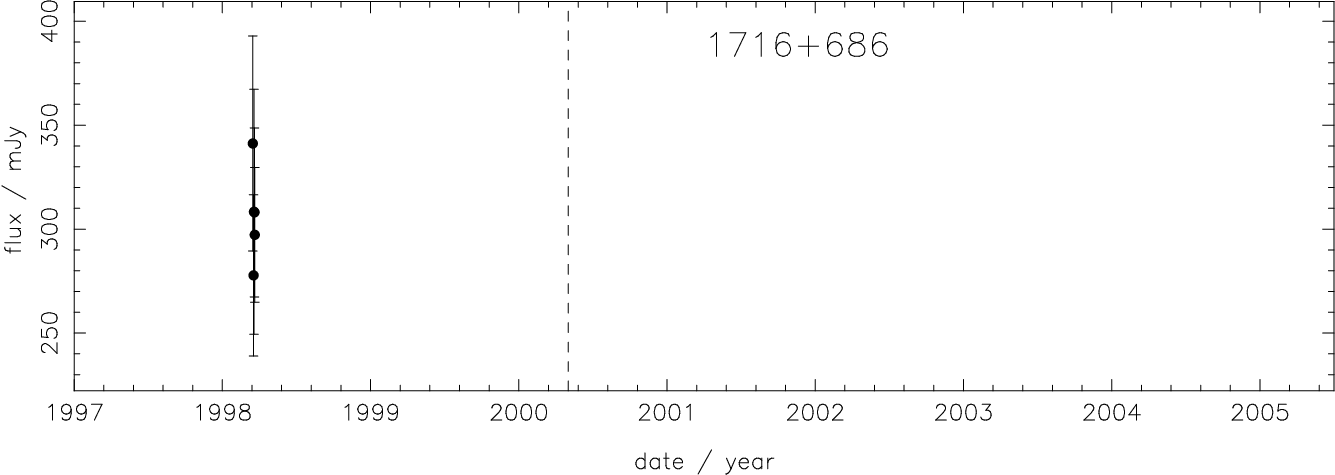}
\\
\includegraphics[angle=-90,width=76mm]{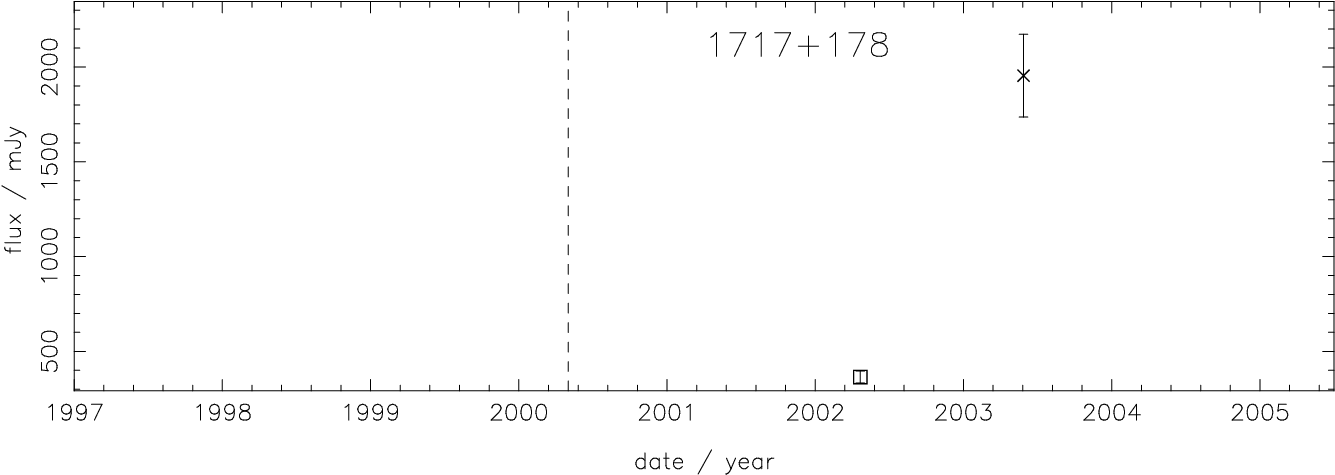}
&
\includegraphics[angle=-90,width=76mm]{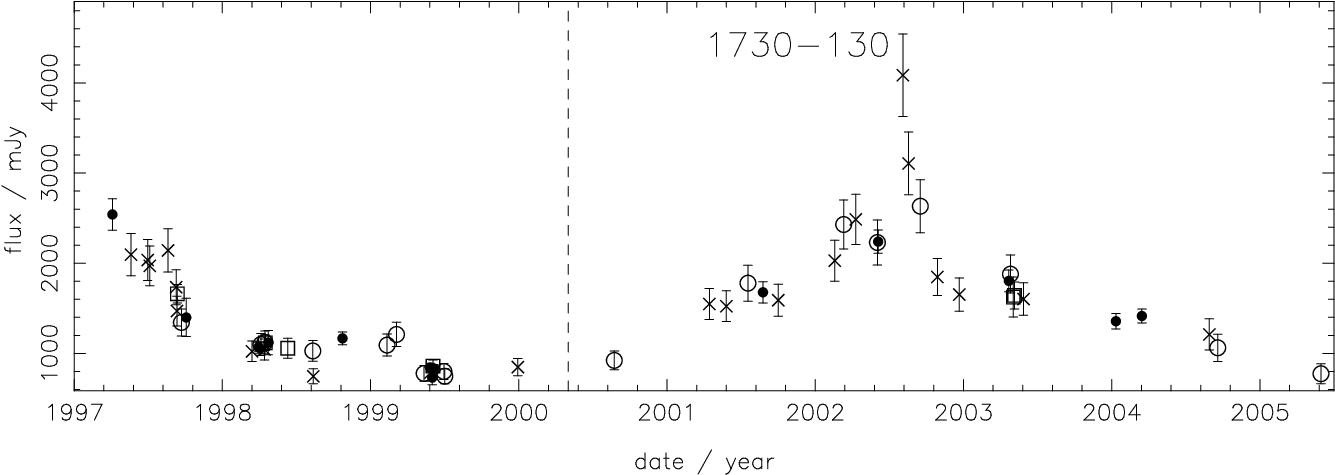}
\\
\includegraphics[angle=-90,width=76mm]{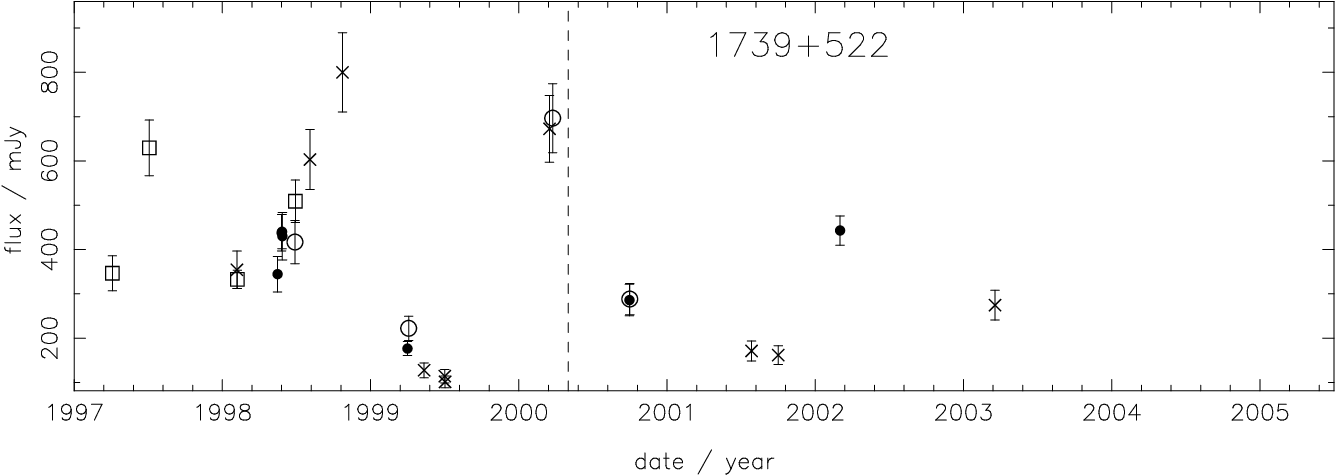}
&
\includegraphics[angle=-90,width=76mm]{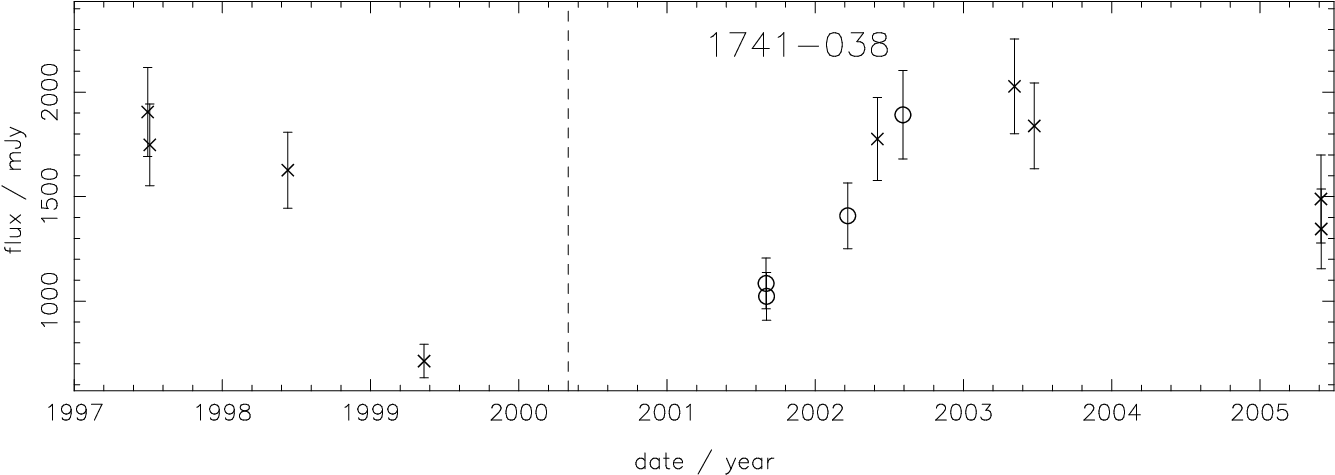}
\\
\includegraphics[angle=-90,width=76mm]{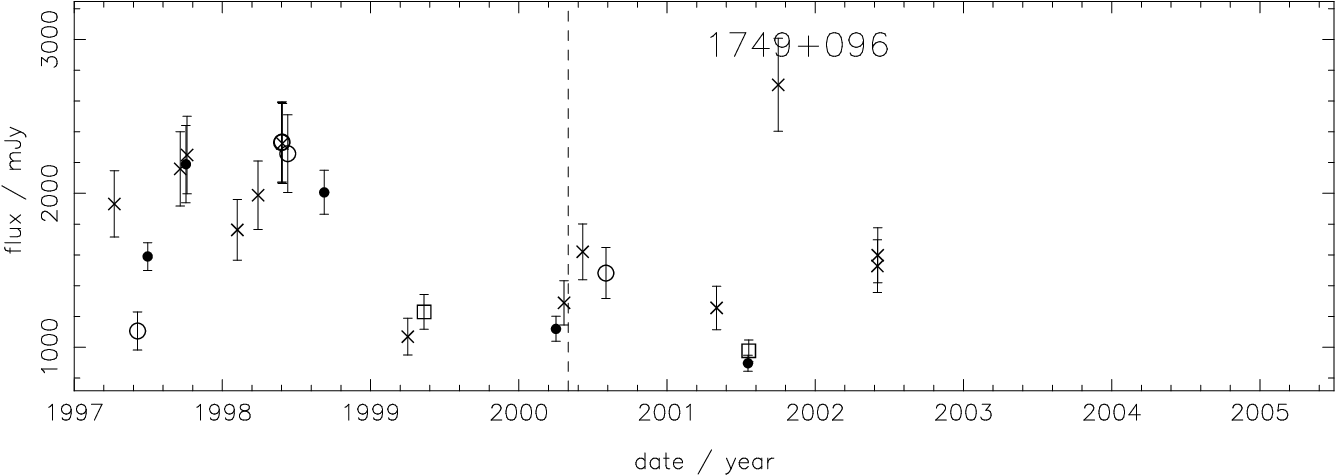}
&
\includegraphics[angle=-90,width=76mm]{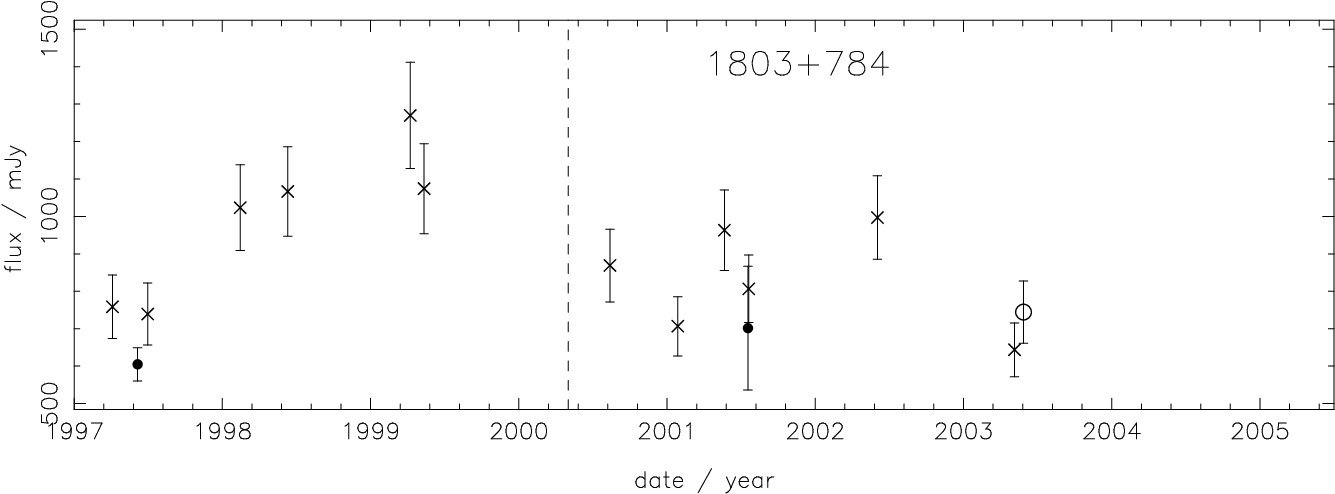}
\\
\includegraphics[angle=-90,width=76mm]{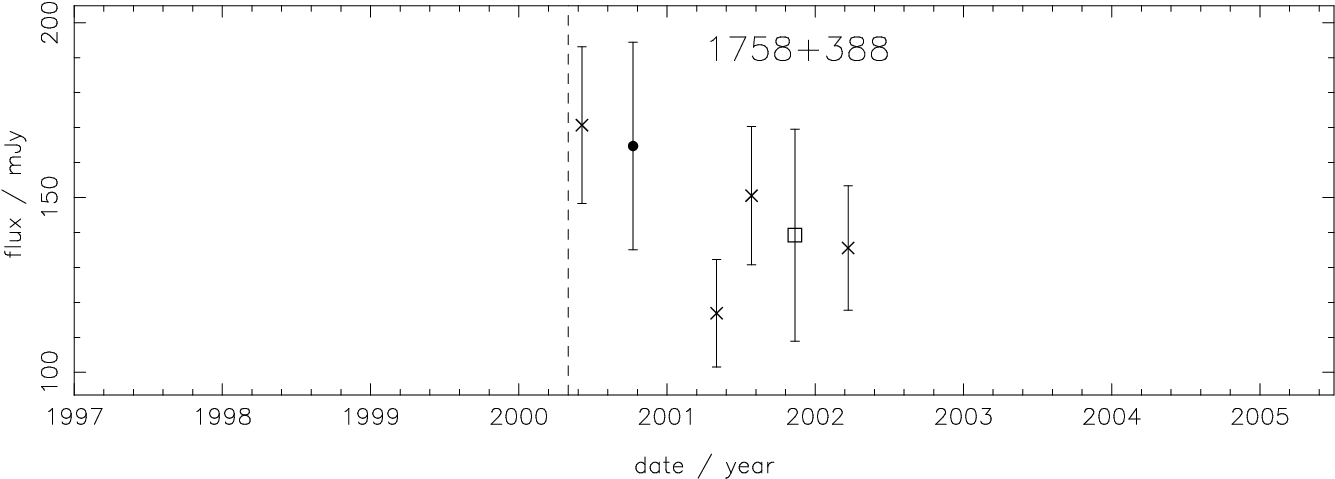}
&
\includegraphics[angle=-90,width=76mm]{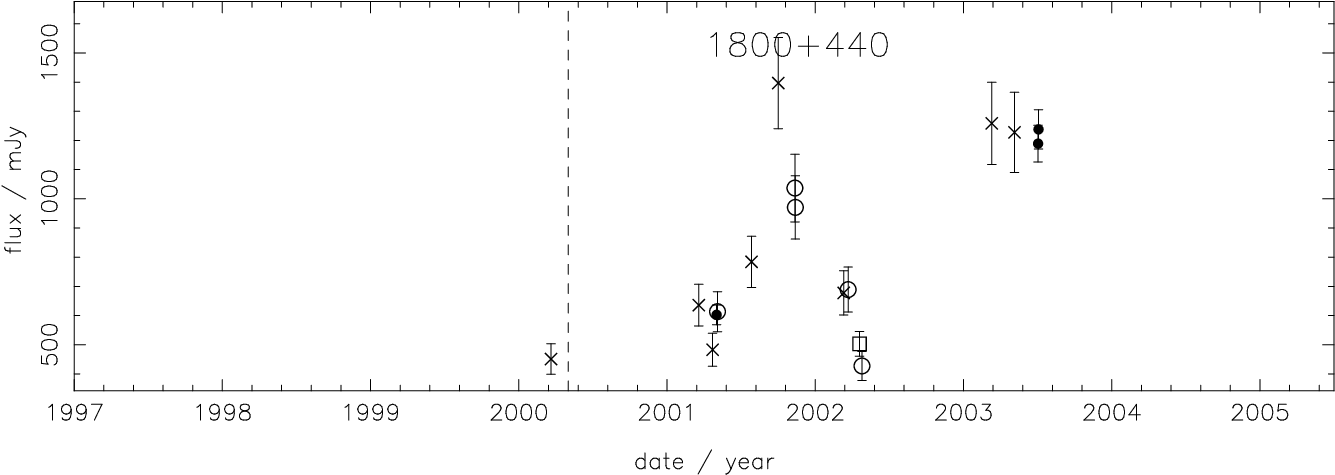}
\\
\includegraphics[angle=-90,width=76mm]{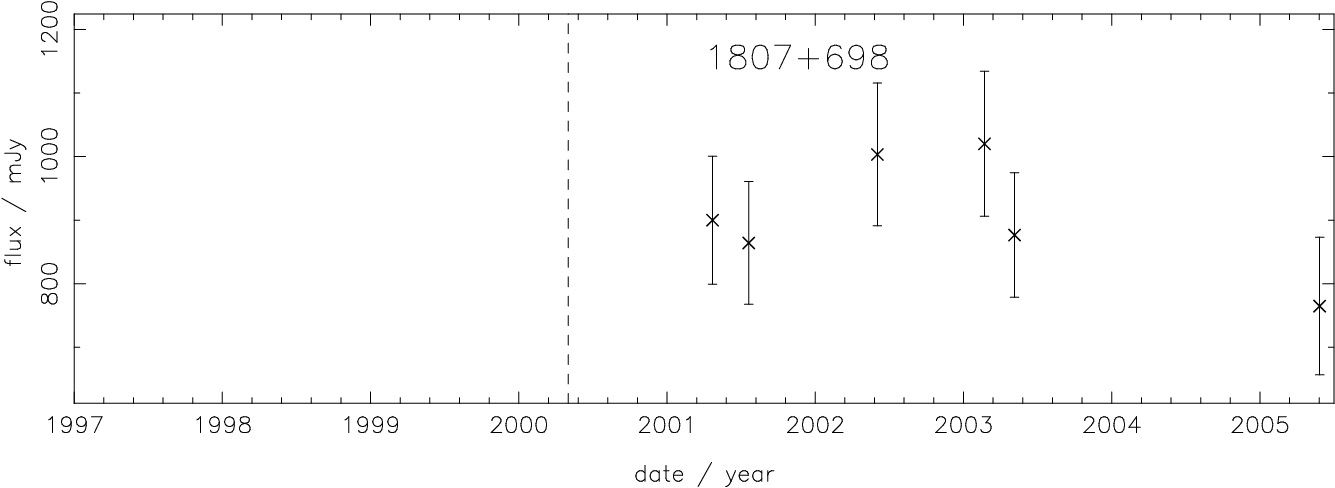}
&
\includegraphics[angle=-90,width=76mm]{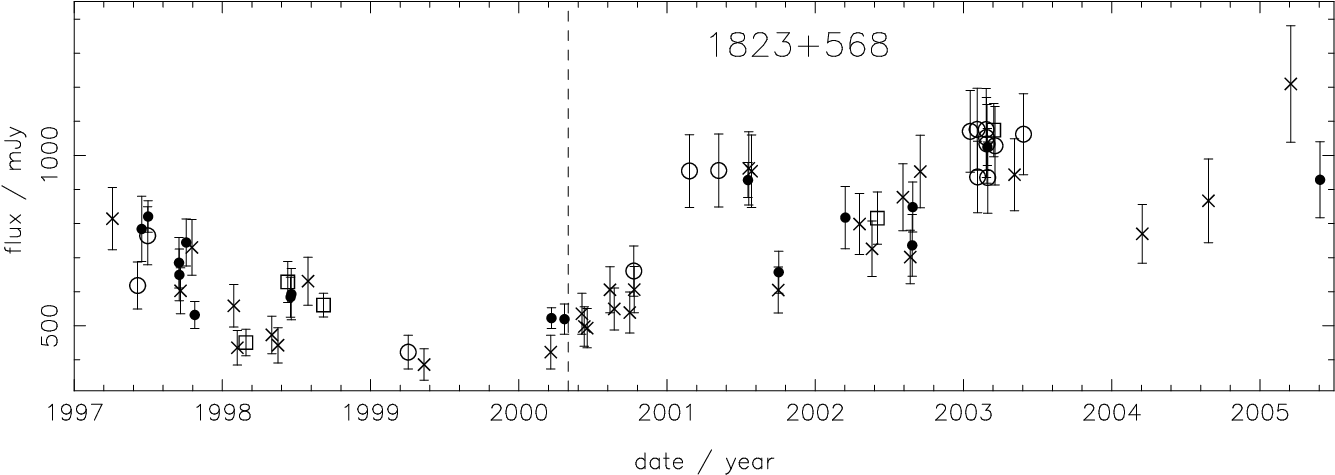}
\\
\end{tabular}
\contcaption{}
\label{fig:light-curves7}
\end{figure*}
\clearpage
\begin{figure*}
\begin{tabular}{cc}
\includegraphics[angle=-90,width=76mm]{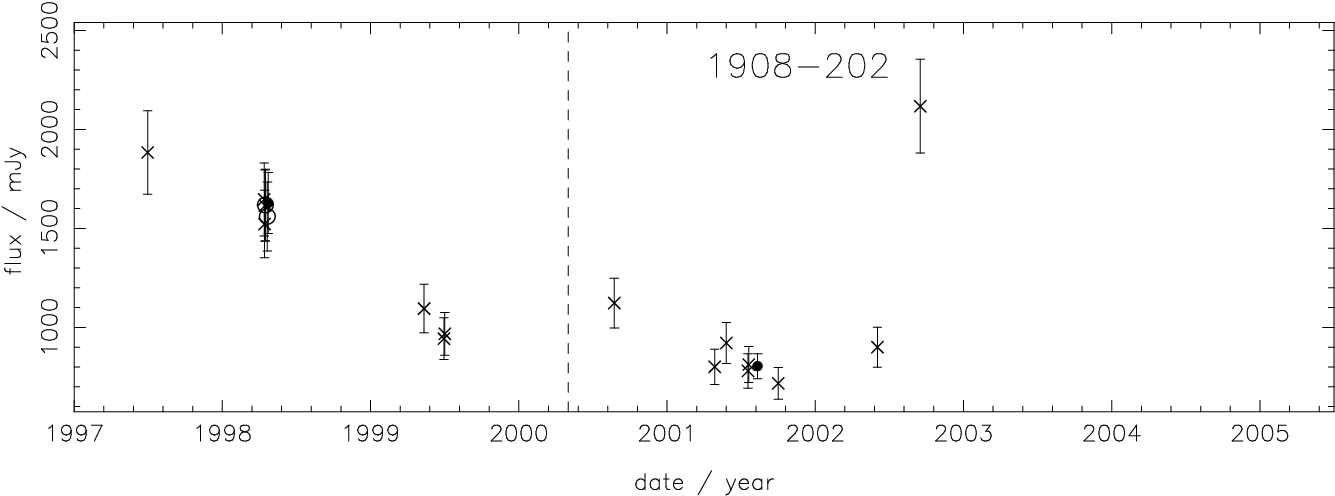}
&
\includegraphics[angle=-90,width=76mm]{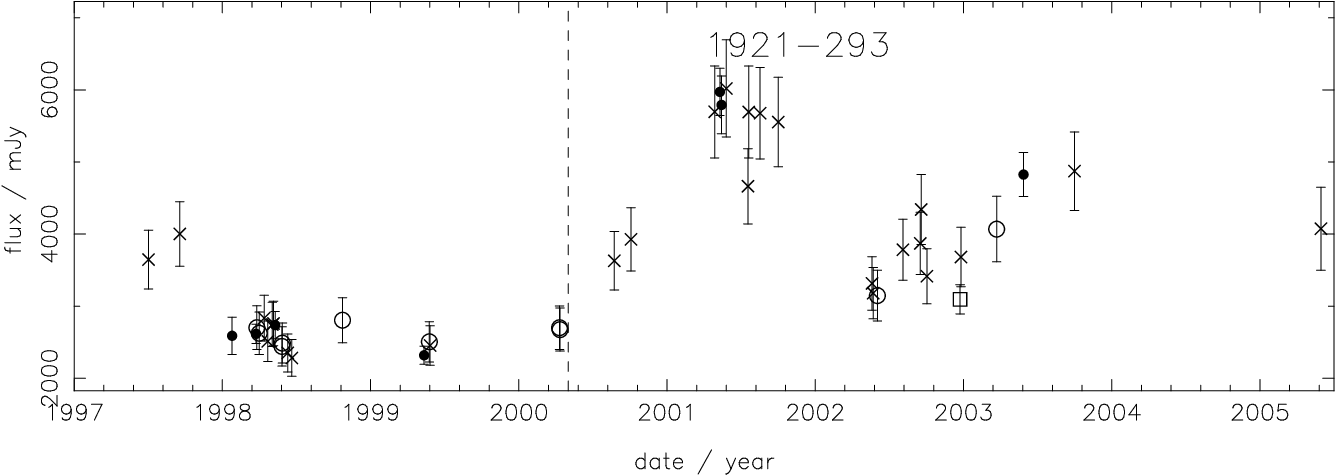}
\\
\includegraphics[angle=-90,width=76mm]{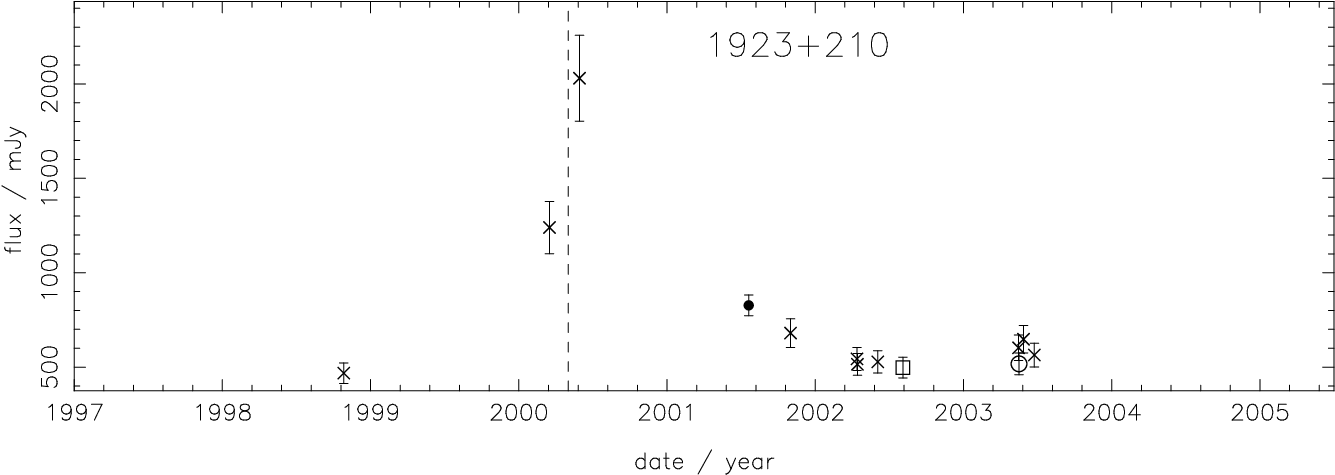}
&
\includegraphics[angle=-90,width=76mm]{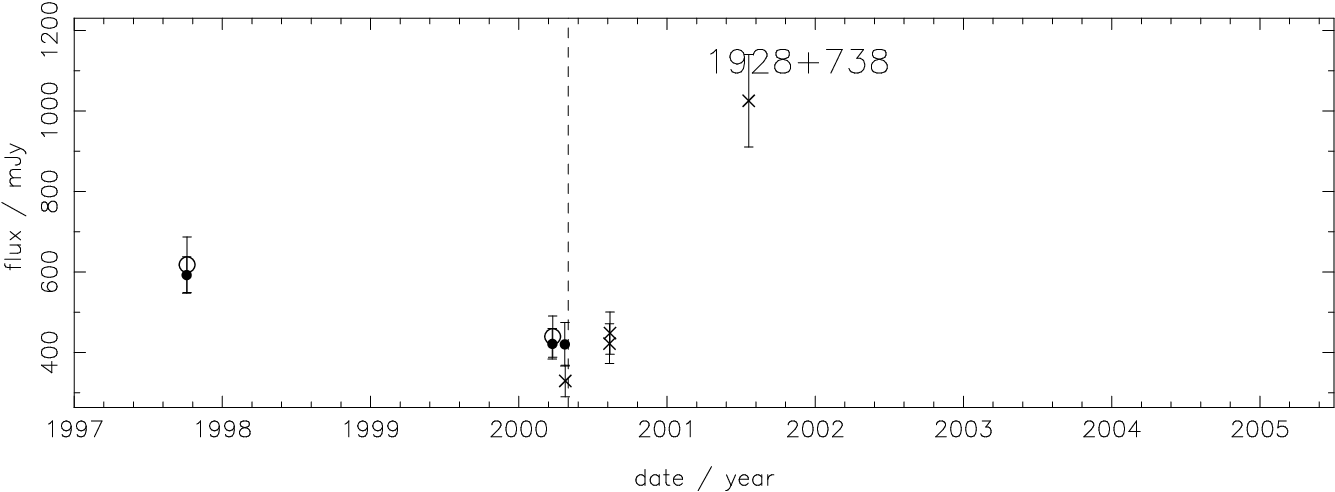}
\\
\includegraphics[angle=-90,width=76mm]{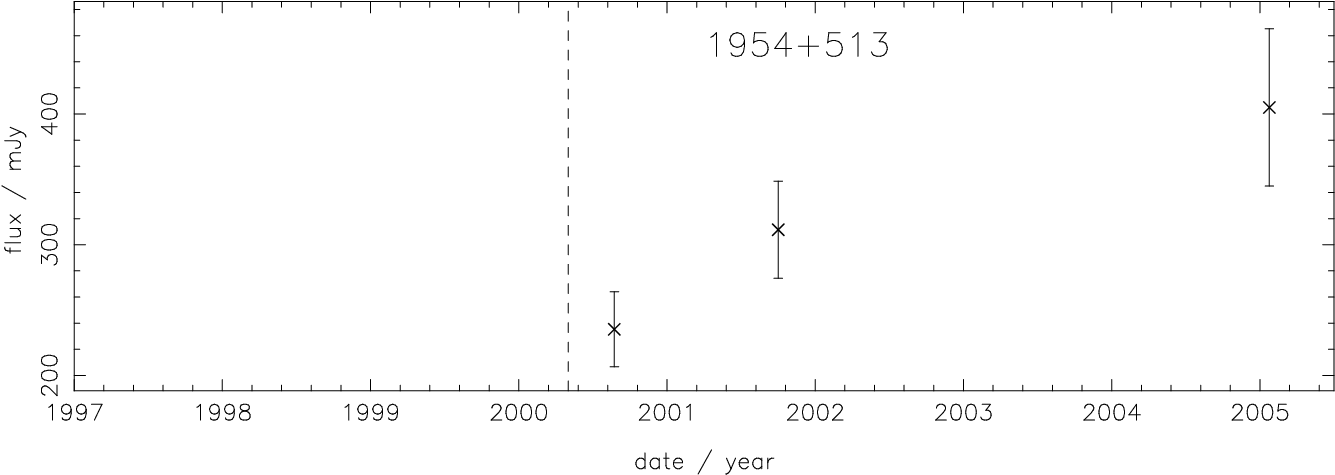}
&
\includegraphics[angle=-90,width=76mm]{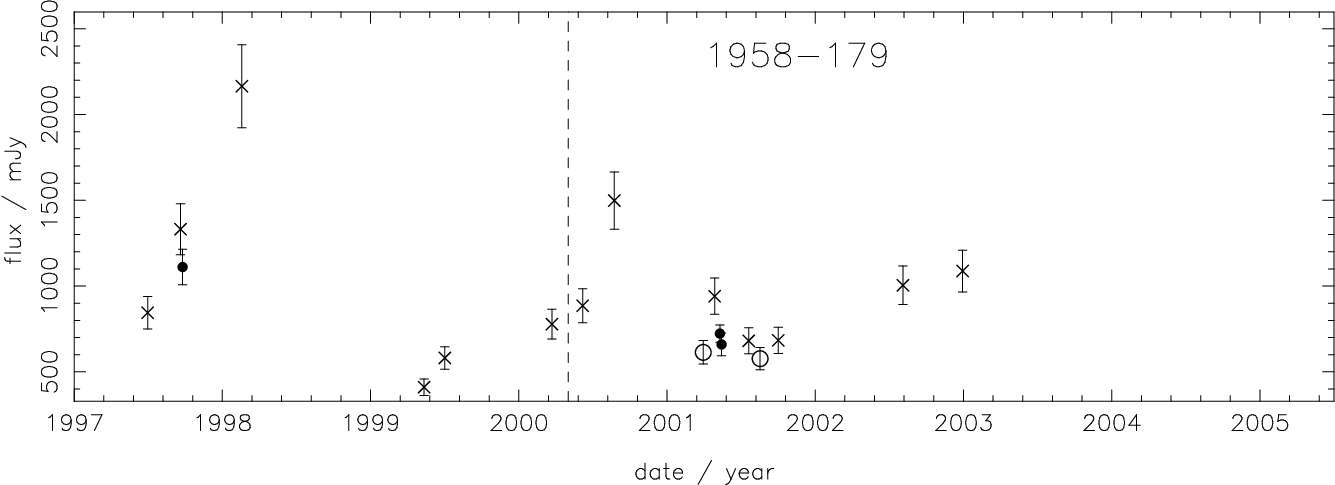}
\\
\includegraphics[angle=-90,width=76mm]{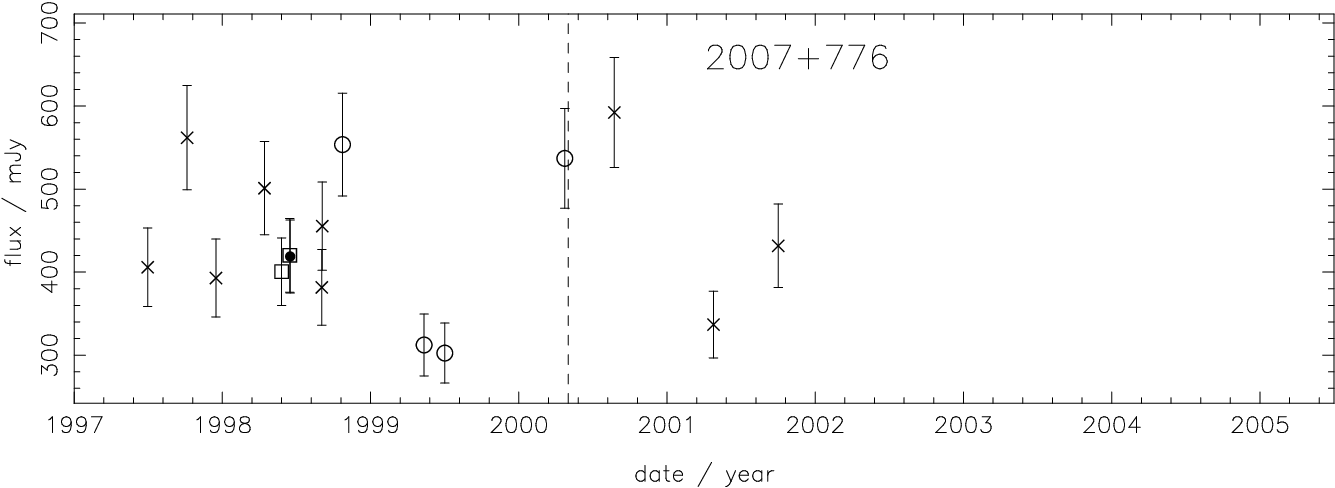}
&
\includegraphics[angle=-90,width=76mm]{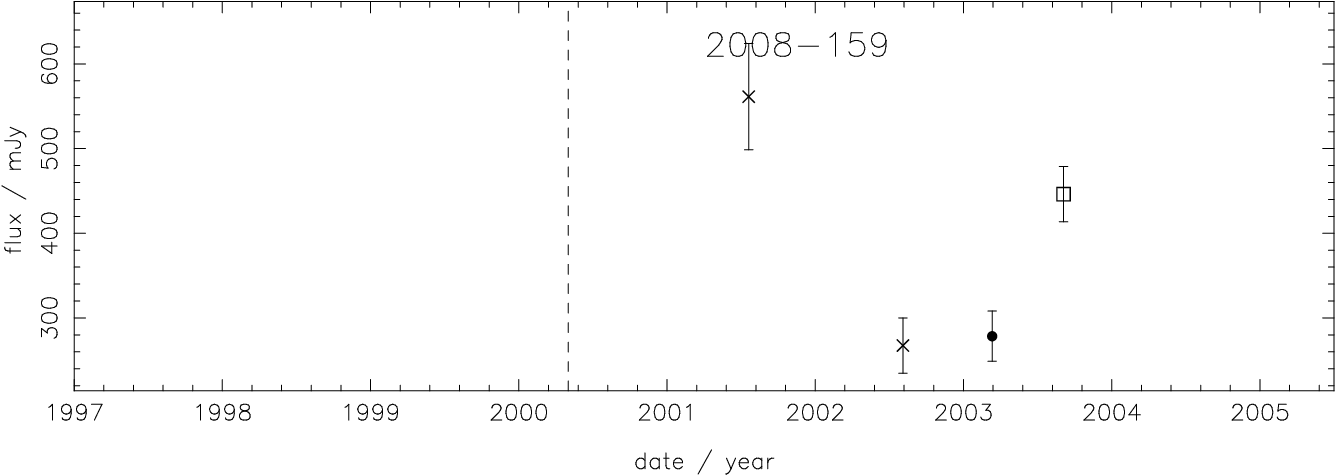}
\\
\includegraphics[angle=-90,width=76mm]{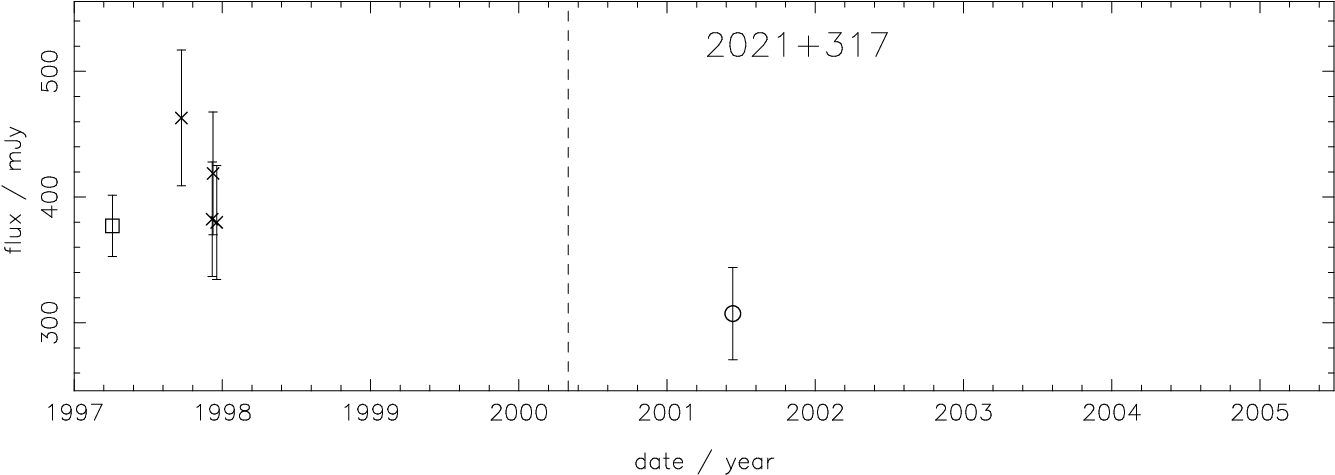}
&
\includegraphics[angle=-90,width=76mm]{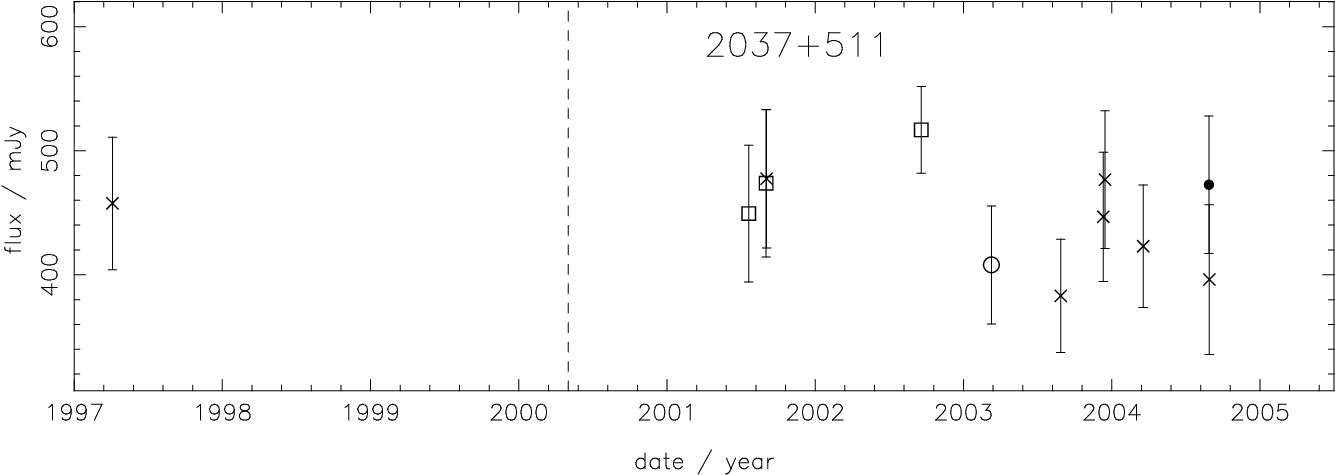}
\\
\includegraphics[angle=-90,width=76mm]{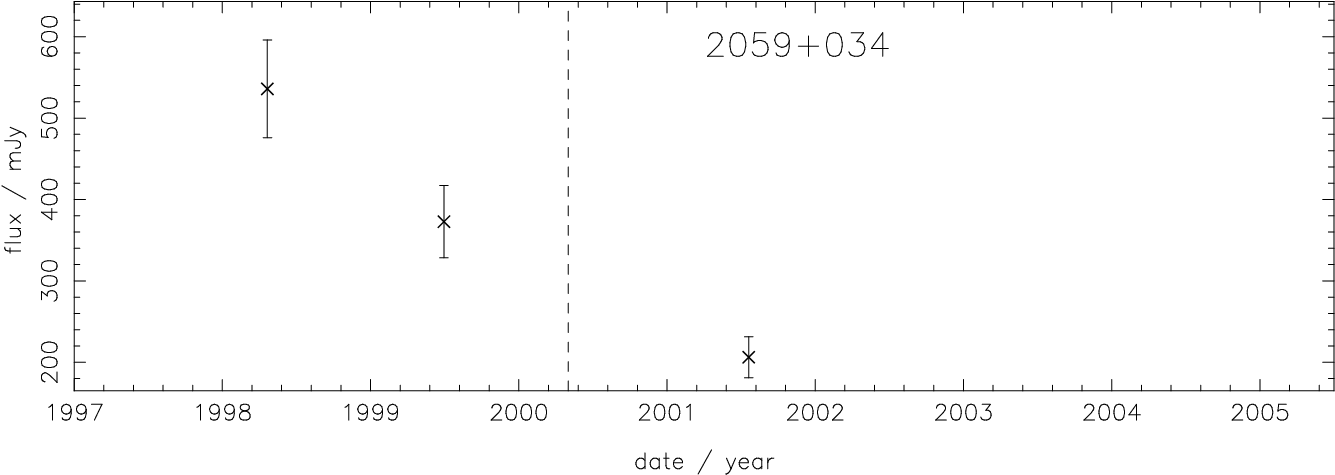}
&
\includegraphics[angle=-90,width=76mm]{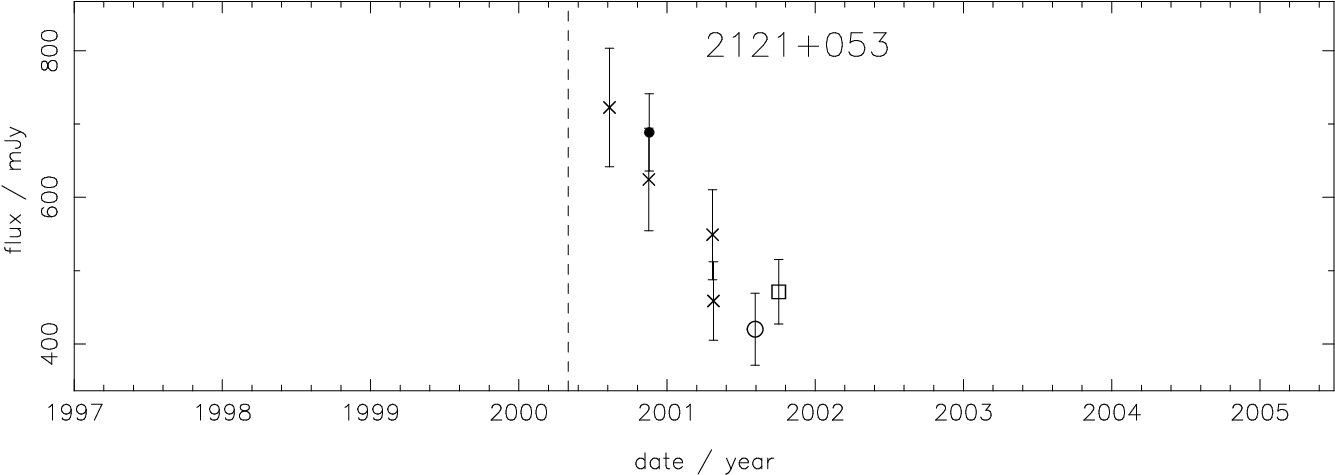}
\\
\includegraphics[angle=-90,width=76mm]{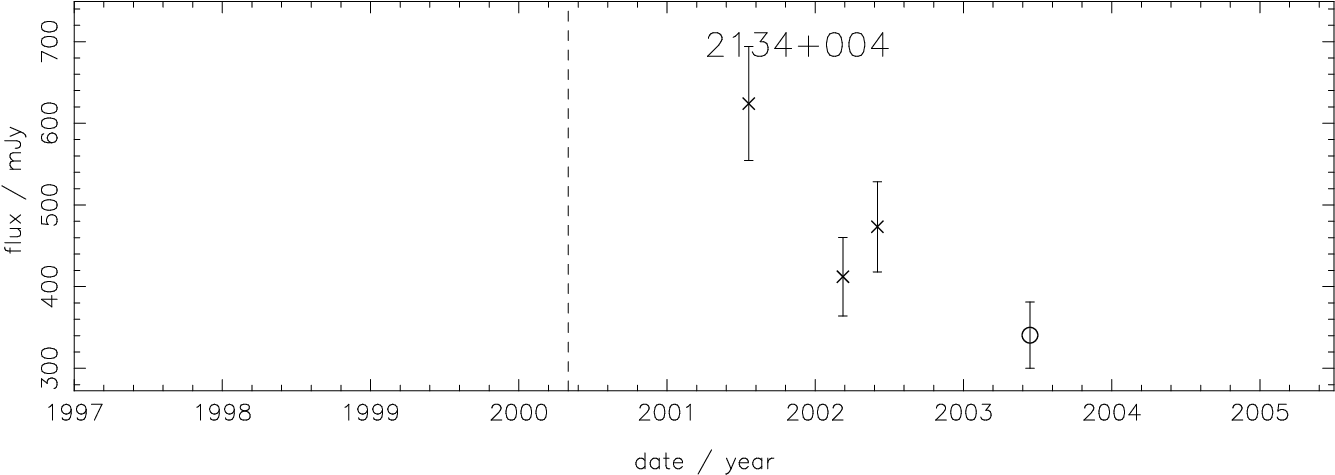}
&
\includegraphics[angle=-90,width=76mm]{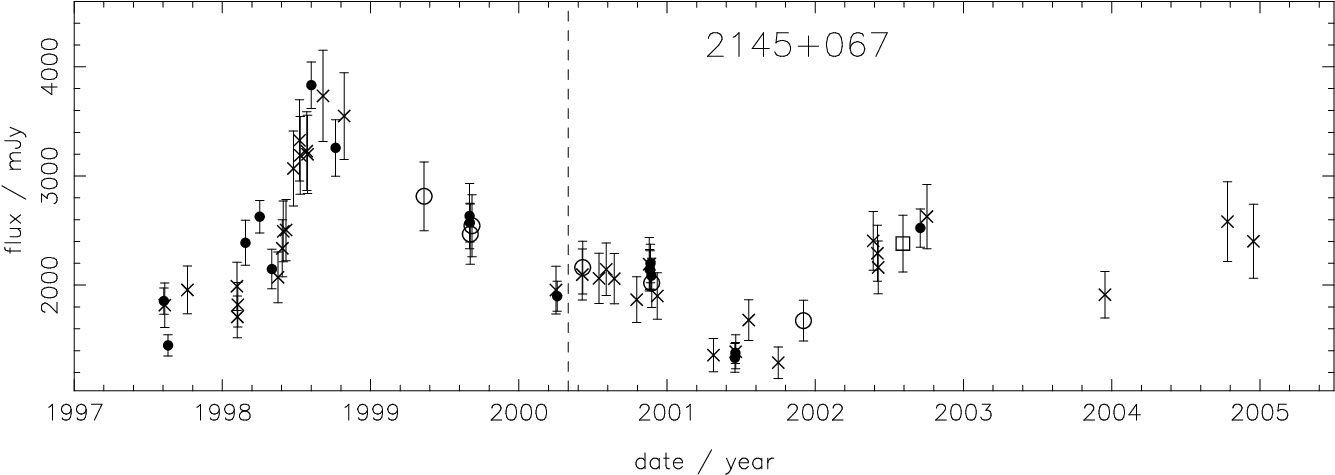}
\\
\end{tabular}
\contcaption{}
\label{fig:light-curves8}
\end{figure*}
\clearpage
\begin{figure*}
\begin{tabular}{cc}
\includegraphics[angle=-90,width=76mm]{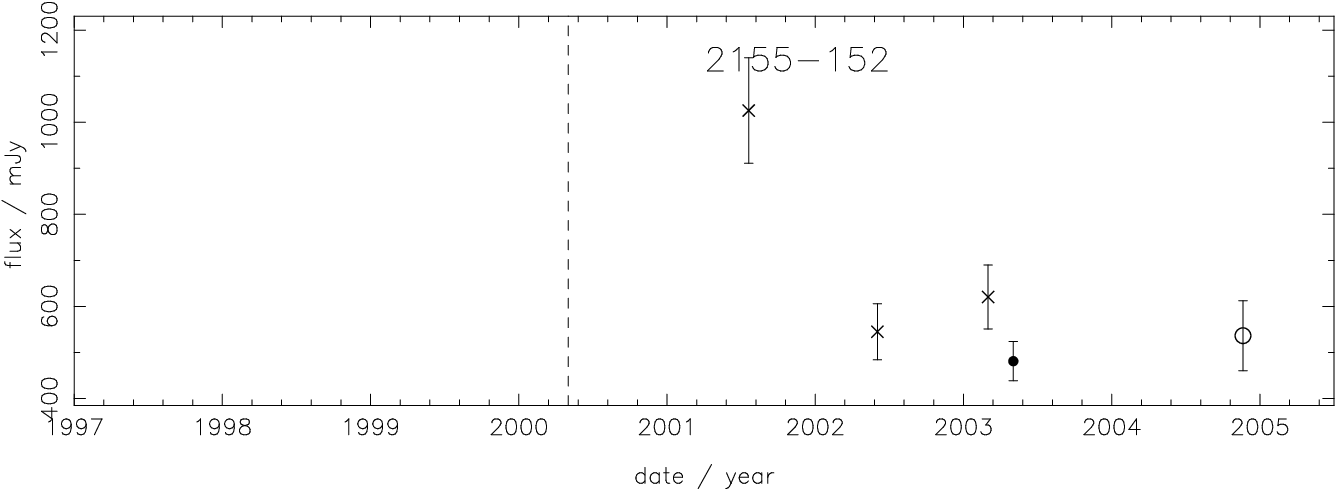}
&
\includegraphics[angle=-90,width=76mm]{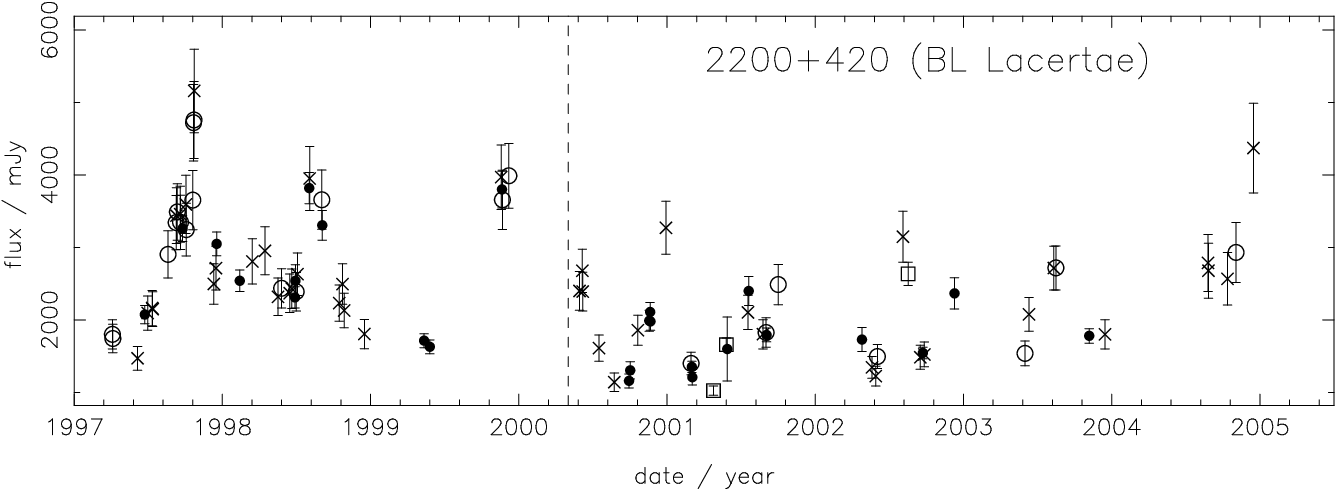}
\\
\includegraphics[angle=-90,width=76mm]{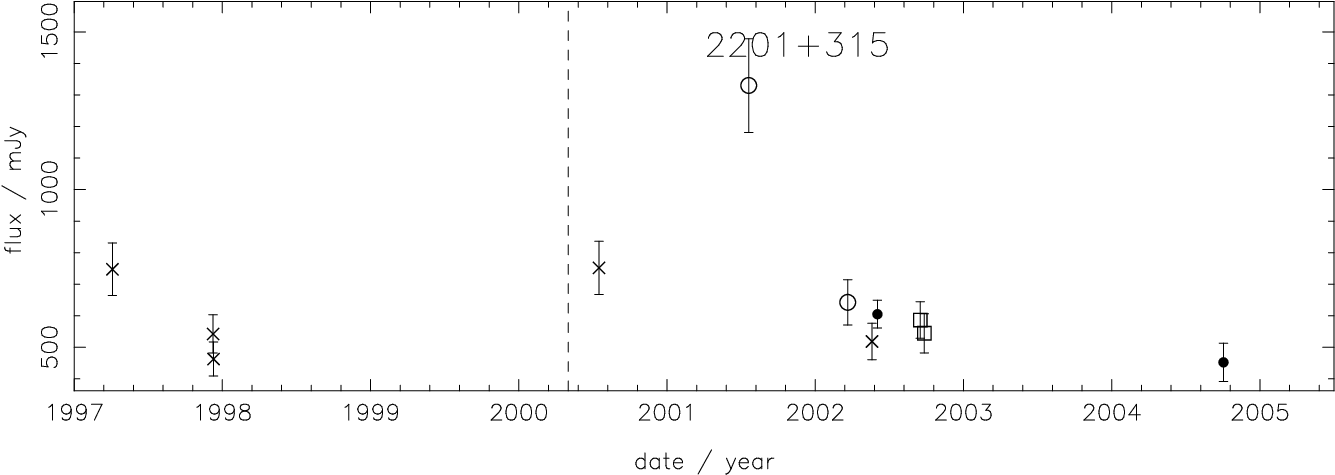}
&
\includegraphics[angle=-90,width=76mm]{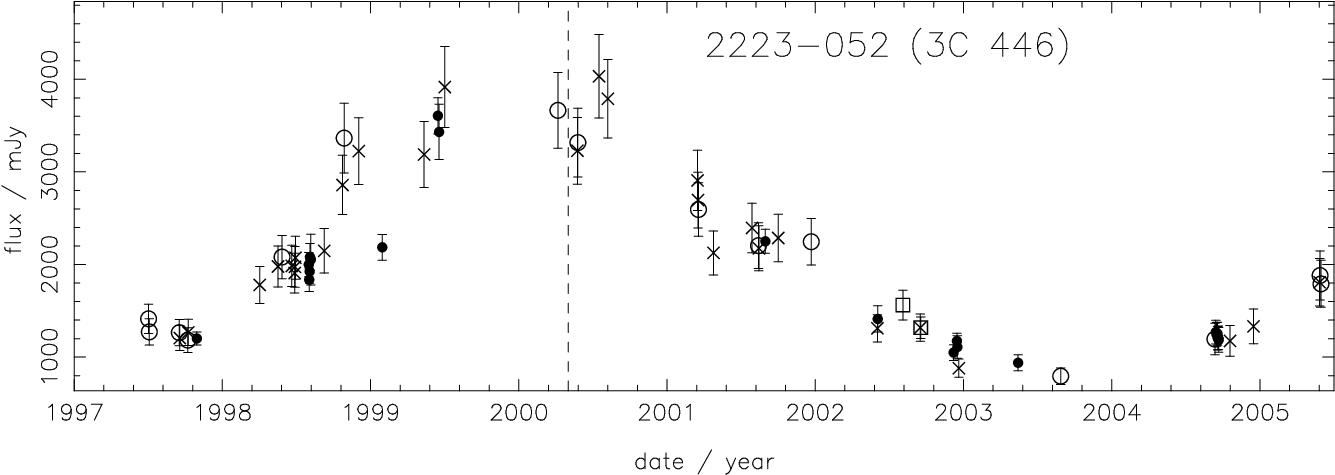}
\\
\includegraphics[angle=-90,width=76mm]{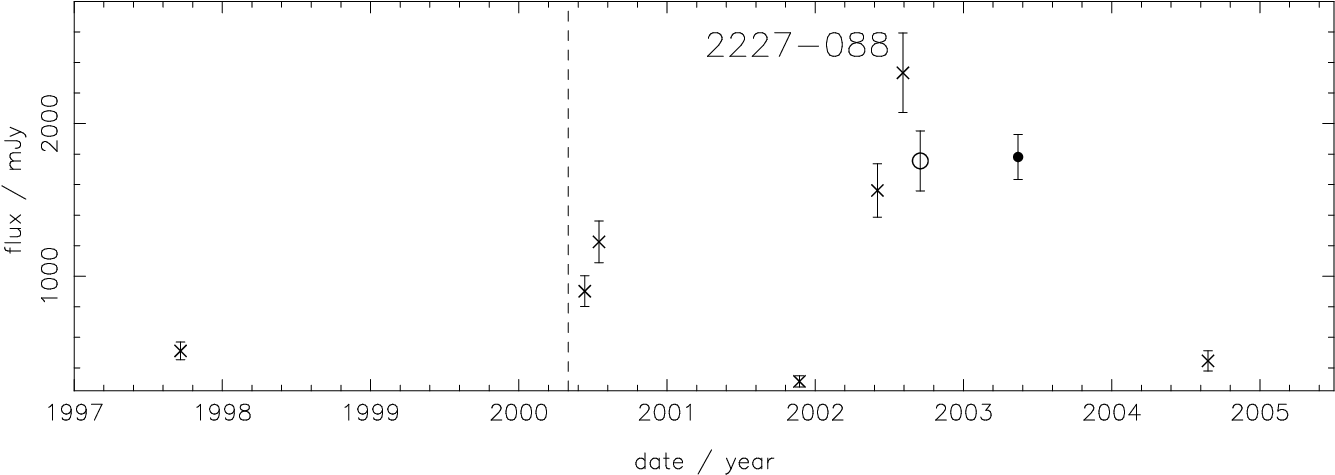}
&
\includegraphics[angle=-90,width=76mm]{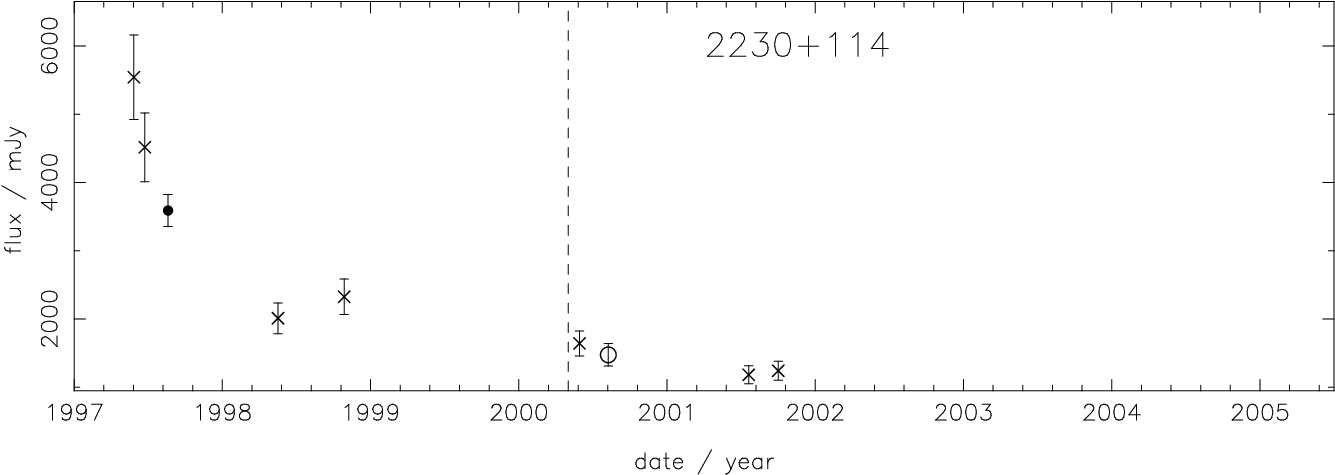}
\\
\includegraphics[angle=-90,width=76mm]{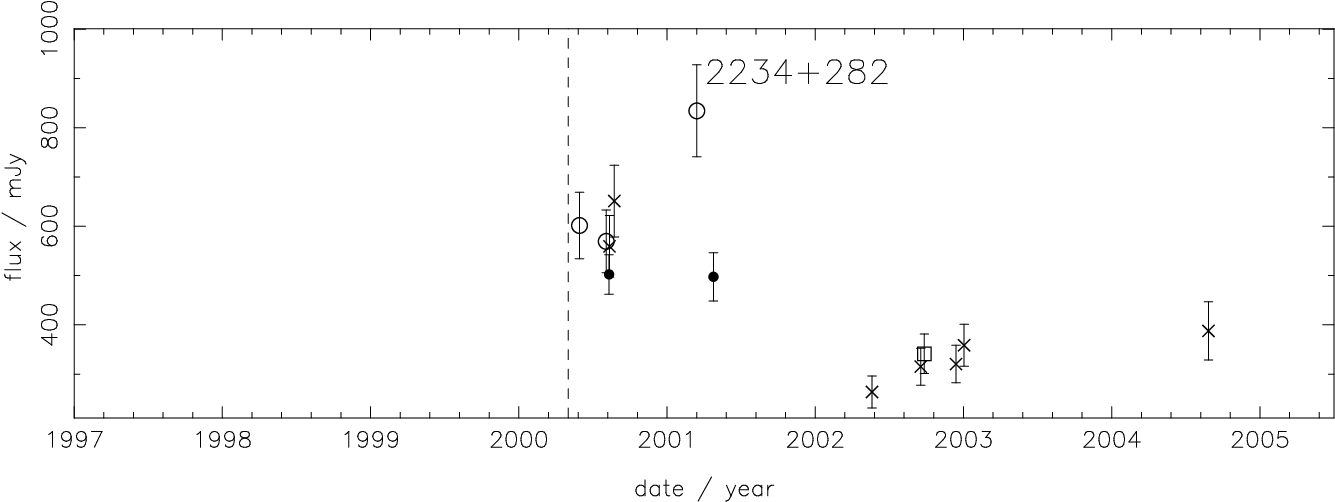}
&
\includegraphics[angle=-90,width=76mm]{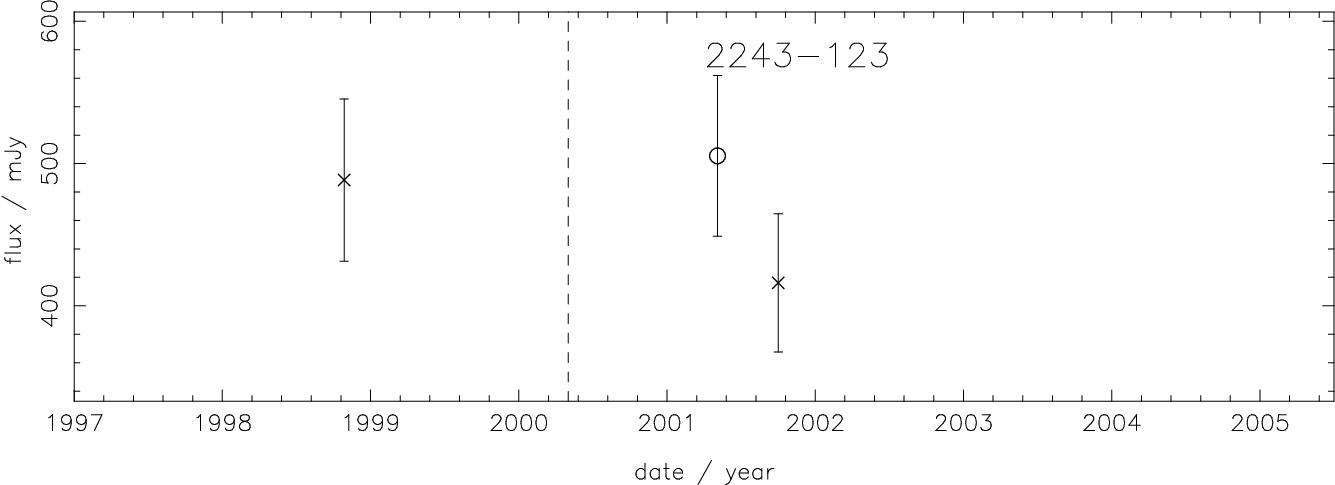}
\\
\includegraphics[angle=-90,width=76mm]{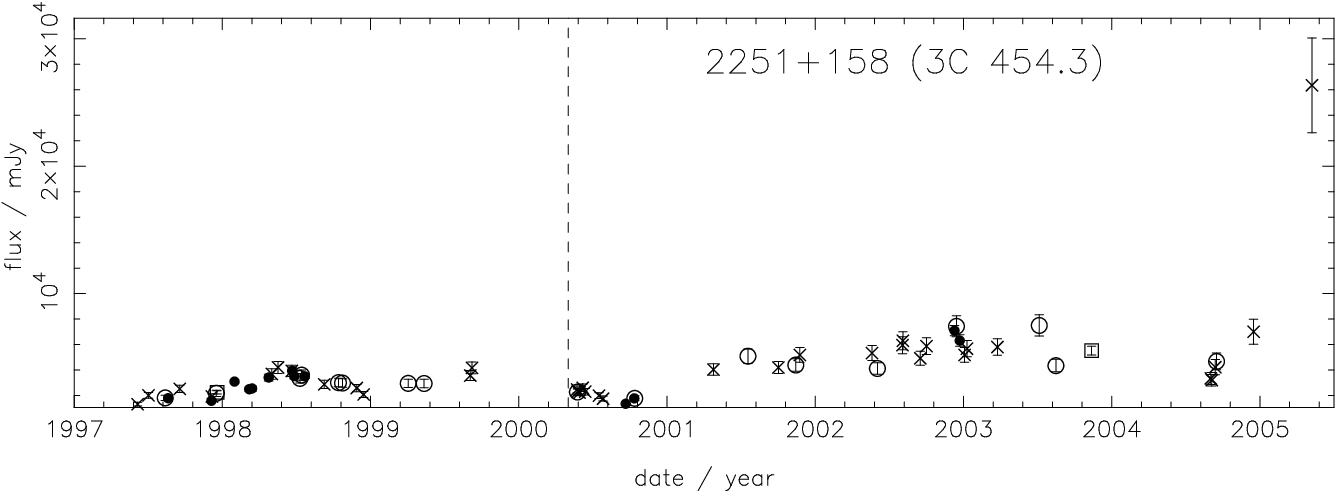}
&
\includegraphics[angle=-90,width=76mm]{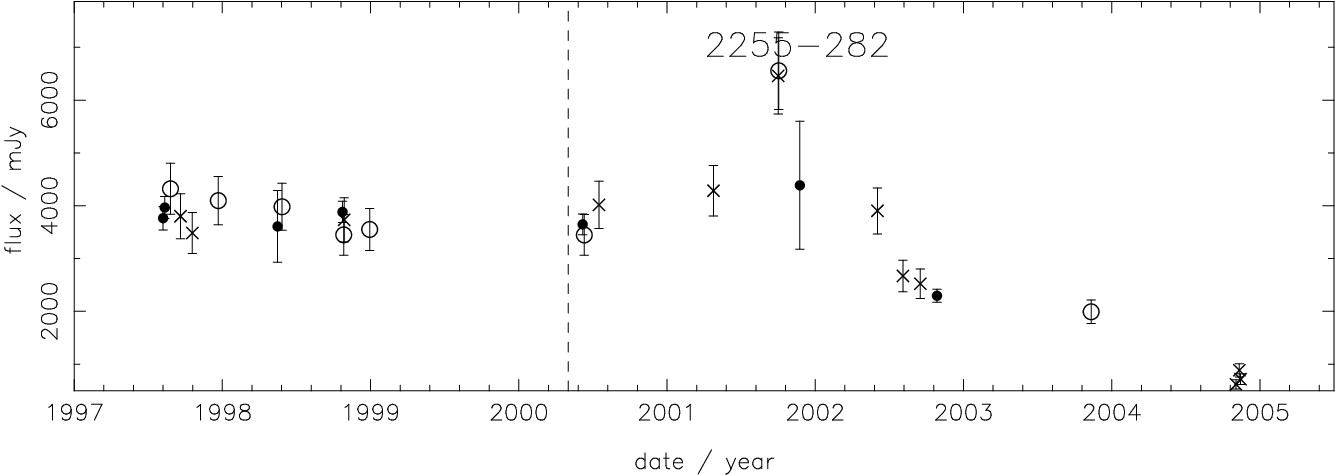}
\\
\includegraphics[angle=-90,width=76mm]{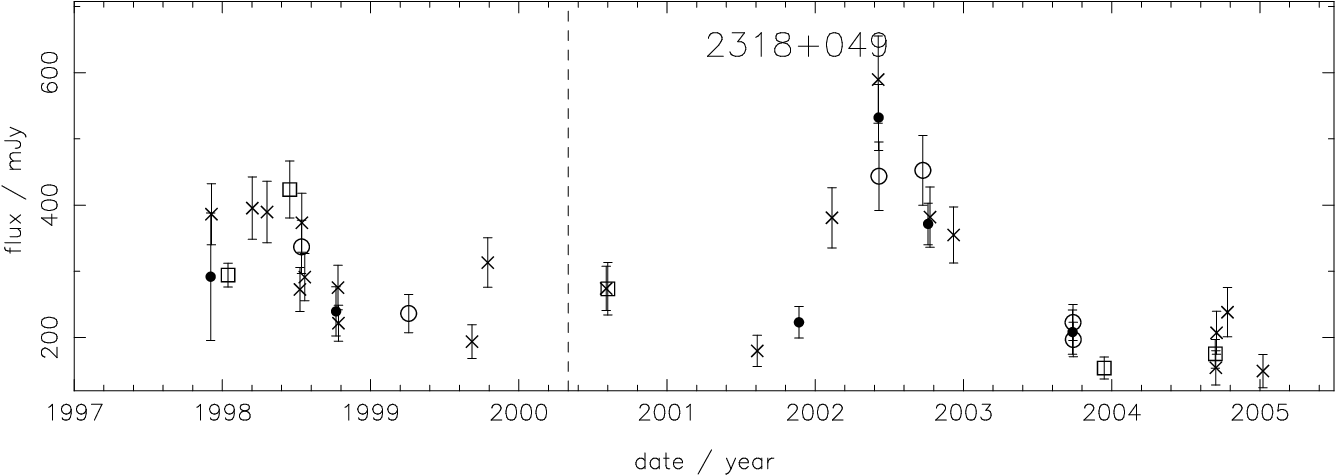}
&
\includegraphics[angle=-90,width=76mm]{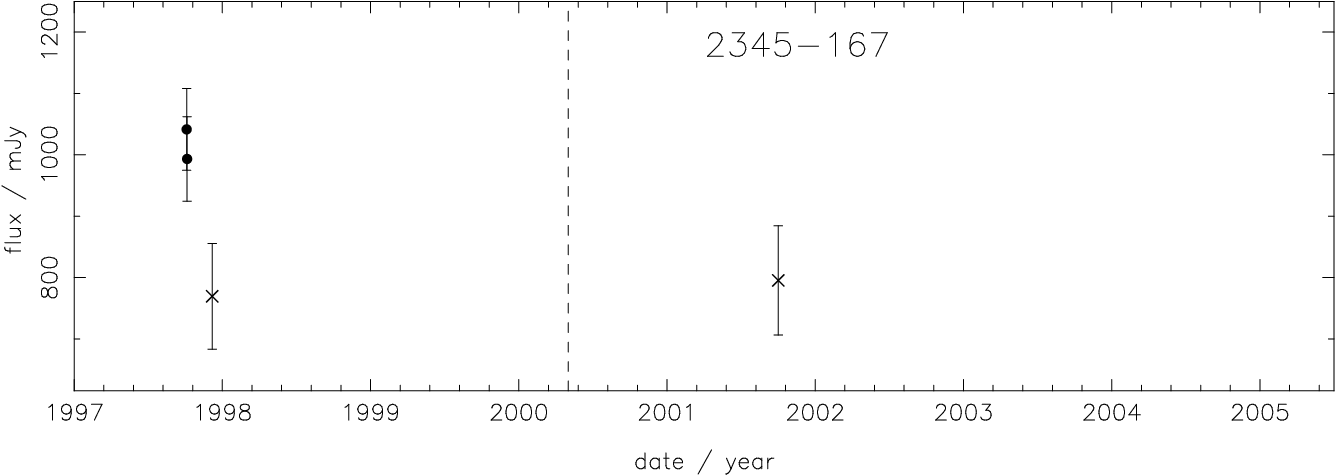}
\\
\end{tabular}
\contcaption{}
\label{fig:light-curves9}
\end{figure*}

\section*{ACKNOWLEDGMENTS}

The James Clerk Maxwell Telescope is operated by the Joint Astronomy
Centre in Hilo, Hawaii on behalf of the parent organisations the
Science and Technology Facilities Council in the United Kingdom, the
National Research Council of Canada and The Netherlands Organisation
for Scientific Research.

\bibliography{blazars}

\begin{thebibliography}{}

\bibitem[\protect\citeauthoryear{{Archibald} et~al.,}{{Archibald}
  et~al.}{2002}]{2002MNRAS.336....1A}
{Archibald} et~al., 2002, \mnras, 336, 1

\bibitem[\protect\citeauthoryear{{Feng}, {Shen}, {Cai}, {Chen}, {Lu} \&
  {Huang}}{{Feng} et~al.}{2006}]{2006A&A...456...97F}
{Feng} S.-W.,  {Shen} Z.-Q.,  {Cai} H.-B.,  {Chen} X.,  {Lu} R.-S.,    {Huang}
  L.,  2006, \aap, 456, 97

\bibitem[\protect\citeauthoryear{{Greaves} et~al.,}{{Greaves}
  et~al.}{2003}]{2003MNRAS.340..353G}
{Greaves} et~al., 2003, \mnras, 340, 353

\bibitem[\protect\citeauthoryear{{Holland} et~al.,}{{Holland}
  et~al.}{1999}]{1999MNRAS.303..659H}
{Holland} et~al., 1999, \mnras, 303, 659

\bibitem[\protect\citeauthoryear{{Jenness}, {Stevens}, {Archibald}, {Economou},
  {Jessop} \& {Robson}}{{Jenness} et~al.}{2002}]{2002MNRAS.336...14J}
{Jenness} T.,  {Stevens} J.~A.,  {Archibald} E.~N.,  {Economou} F.,  {Jessop}
  N.~E.,    {Robson} E.~I.,  2002, \mnras, 336, 14

\bibitem[\protect\citeauthoryear{Jessop, , Coulson, Greaves, Holland \&
  Jenness}{Jessop et~al.}{2000}]{2000JCMTNews15NJ}
Jessop N.,   Coulson I.,  Greaves J.~S.,  Holland W.~S.,    Jenness T.,  2000,
  in No.~15, The JCMT Newsletter

\bibitem[\protect\citeauthoryear{{Jorstad} et~al.,}{{Jorstad}
  et~al.}{2007}]{2007AJ....134..799J}
{Jorstad} et~al., 2007, \aj, 134, 799

\bibitem[\protect\citeauthoryear{{Lindfors} et~al.,}{{Lindfors}
  et~al.}{2006}]{2006A&A...456..895L}
{Lindfors} et~al., 2006, \aap, 456, 895

\bibitem[\protect\citeauthoryear{{Nartallo}, {Gear}, {Murray}, {Robson} \&
  {Hough}}{{Nartallo} et~al.}{1998}]{1998MNRAS.297..667N}
{Nartallo} R.,  {Gear} W.~K.,  {Murray} A.~G.,  {Robson} E.~I.,    {Hough}
  J.~H.,  1998, \mnras, 297, 667

\bibitem[\protect\citeauthoryear{{Robson}, {Stevens} \& {Jenness}}{{Robson}
  et~al.}{2001}]{2001MNRAS.327..751R}
{Robson} E.~I.,  {Stevens} J.~A.,    {Jenness} T.,  2001, \mnras, 327, 751

\bibitem[\protect\citeauthoryear{{Siebenmorgen}, {Freudling}, {Kr{\"u}gel} \&
  {Haas}}{{Siebenmorgen} et~al.}{2004}]{2004A&A...421..129S}
{Siebenmorgen} R.,  {Freudling} W.,  {Kr{\"u}gel} E.,    {Haas} M.,  2004,
  \aap, 421, 129

\bibitem[\protect\citeauthoryear{{Soldi} et~al.,}{{Soldi}
  et~al.}{2008}]{2008A&A...486..411S}
{Soldi} et~al., 2008, \aap, 486, 411

\bibitem[\protect\citeauthoryear{{Stevens}, {Robson} \& {Holland}}{{Stevens}
  et~al.}{1996}]{1996ApJ...462L..23S}
{Stevens} J.~A.,  {Robson} E.~I.,    {Holland} W.~S.,  1996, \apjl, 462, L23+

\bibitem[\protect\citeauthoryear{{Stirling}, {Spencer}, {Cawthorne} \&
  {Paragi}}{{Stirling} et~al.}{2004}]{2004MNRAS.354.1239S}
{Stirling} A.~M.,  {Spencer} R.~E.,  {Cawthorne} T.~V.,    {Paragi} Z.,  2004,
  \mnras, 354, 1239

\end{thebibliography}
\bibliographystyle{mn2e}

\bsp

\end{document}